\documentclass[11pt]{JHEP3}
\newcommand{\Comment}[1]{{}}
\usepackage{amsfonts,amsthm,amsmath,amssymb,slashed}
\newcommand\ignore[1]{}
\def\one{{\,\hbox{1\kern-.8mm l}}}

\newcommand{\tr}{\operatorname{tr}}

\def\Tr{{\rm Tr\, }}

\newcommand{\SO}{\mathrm{SO}} 
\newcommand{\SU}{\mathrm{SU}} \newcommand{\U}{\mathrm{U}}
\newcommand{\ie}{\emph{i.e.}\:}
\newcommand{\eg}{\emph{e.g.}\;} \newcommand{\pd}{\partial}

\def\a{\alpha}\def\b{\beta}

\def\l{\lambda}

\def\d{\partial}

\newcommand{\zd}{z^\dagger}

\newcommand{\be}{\begin{equation}}
\newcommand{\bea}{\begin{eqnarray}}

\newcommand{\ee}{\end{equation}}
\newcommand{\eea}{\end{eqnarray}}

\newcommand{\nn}{\nonumber}

\title{Higgsing M2 to D2 with gravity: $\mathcal{N}=6$ chiral supergravity from topologically gauged ABJM theory}
\author{Xiaoyong Chu${}^{a,}$\footnote{E-mail address: {\tt xiaoyong.chu@ulb.ac.be}}\;, Horatiu Nastase${}^{b,}$\footnote{E-mail address: {\tt
    nastase@ift.unesp.br}}\;, Bengt E.W. Nilsson${}^{c,}$\footnote{E-mail address: {\tt
    tfebn@chalmers.se}}\, and Constantinos Papageorgakis${}^{d,}$\footnote{E-mail address:
                                 {\tt costis.papageorgakis@kcl.ac.uk}}\\~\\\it
  $^a$ Service de Physique Th\'eorique, CP225, Universit\'e Libre de Bruxelles,\\
  ~ Boulevard du Triomphe (Campus de la Plaine), 1050 Bruxelles,
Belgium \\~\\
  $^b$ Instituto de F\'{i}sica Te\'{o}rica, UNESP-Universidade Estadual Paulista,\\ ~ R. Dr. Bento T. Ferraz 271, Bl. II, Sao Paulo 01140-070, SP, Brazil \\~\\   $^c$ Fundamental Physics, Chalmers University of Technology,\\ ~ SE-412 96 G\"{o}teborg, Sweden
  \\~\\ $^d$ Department of Mathematics, King's College London,\\ ~ The Strand, London WC2R 2LS, London, UK
}

\abstract{We present the higgsing of three-dimensional ${\cal N}=6$ superconformal ABJM type   theories coupled to conformal supergravity, so called topologically gauged ABJM theory, thus providing a gravitational extension of previous   work on the relation between $N$ M2 and $N$ D2-branes.  The resulting ${\cal N}=6$ supergravity   theory appears at a chiral point similar to that of three-dimensional chiral gravity introduced   recently by Li, Song and Strominger, but with the opposite sign for the Ricci scalar term in the   lagrangian. We identify the supersymmetry in the broken phase as a particular linear combination of the supersymmetry and special conformal supersymmetry in the original topologically gauged ABJM theory. We also discuss the higgsing procedure in detail paying special attention to the role   played by the $\U(1)$ factors in the original ABJM model and the $\U(1)$ introduced in the   topological gauging.}

\begin{document}

\section{Introduction and Overview}

The models of Aharony, Bergman, Jafferis and Maldacena (ABJM) \cite{Aharony:2008ug}, describing the physics of $N$ M2-branes on $\mathbb C^4/\mathbb Z_k$, has received a lot of attention recently. One important step in unravelling the physics of these models is to analyze their relation to D2-branes through a certain Higgs effect connected to a reduction of the background theory from eleven to ten dimensions. Such a higgsing procedure was introduced by Mukhi and Papageorgakis  \cite{Mukhi:2008ux} in the context of the $\mathcal N=8$ BLG theory \cite{Bagger:2006sk,Bagger:2007jr,Bagger:2007vi, Gustavsson:2007vu}. This approach  was developed further in a series of consecutive papers, see \eg \cite{Li:2008ya, Pang:2008hw} and references therein, where also $\mathcal N=6$ ABJM-type theories were analyzed. The end result contains in many cases Yang-Mills theories appropriate for an interpretation in terms of D2-branes.

The purpose of this paper is to generalize this higgsing procedure to the topologically gauged ABJM-type theories which were recently constructed by Chu and Nilsson in \cite{Chu:2009gi} following an earlier attempt for the BLG case in \cite{Gran:2008qx}.  These topologically gauged theories involve M2-branes whose global symmetries have been gauged by coupling them to superconformal gravity given by a set of Chern-Simons (CS) terms, one for each of the dreibein, Rarita-Schwinger and R-symmetry gauge fields.  As will become evident later, the higgsing leads to theories that have some unusual features if viewed as theories of D2-branes. One particularly interesting feature is that the theory, as noted already in \cite{Chu:2009gi},  ends up at a chiral point in a sense similar to that of Li, Song and Strominger \cite{Li:2008dq}. 
We also find that the six supersymmetries of this theory arise as linear combinations of the supersymmetry and special superconformal symmetries of the original topologically gauged ABJM theory. However, a number of questions concerning the interpretation of these theories need to be answered. Some of these issues will be addressed in this paper but further study will be required in order to  determine their precise role in string/M-theory.

The coupling of superconformal M2-brane theories to three-dimensional gravity also provides an arena where many other interesting questions can be addressed. Gravitational theories in three dimensions have been extensively studied in a variety of different forms in the past.  Examples are for instance topologically massive gravity (TMG) \cite{Deser:1981wh} and `chiral gravity' \cite{Li:2008dq}; for some recent work in these directions and further references, see \cite{Bergshoeff:2010mf, Maloney:2009ck, Skenderis:2009nt, Deser:2010df}. In these specific cases many of the key features depend on the presence of both an Einstein-Hilbert and a gravitational Chern-Simons term in the lagrangian, the latter of which is of third order in derivatives. Including also an AdS-type cosmological constant term makes these theories exhibit interesting properties related to their spectra of physical states and their boundary conformal field theories. Interestingly enough, whether the propagating modes or the black holes are physical (have positive energy) depends crucially on the sign of the Einstein-Hilbert term in the lagrangian; for a recent discussion of this issue see \cite{Deser:2010df}.

As mentioned above, in this paper our starting point will instead be a very particular class of models, namely the ones that arise from the topological gauging of the ABJM theory \cite{Chu:2009gi}, which provide gravitationally coupled theories preserving superconformal invariance. These models, which have no propagating gravitational degrees of freedom (see \eg \cite{Nilsson:2008ri}), have some features in common with the more familiar Polyakov action in string theory as already pointed out in \cite{Gran:2008qx}. In analogy with that case, coupling worldvolume gravity to M2 theories at the conformal fixed point should be done while preserving the conformal invariance, even though this leads to a dependence on the internal metric that in turn prevents its elimination through its field equations.\footnote{This is also the case for the worldsheet metric in the Polyakov action   once a Tseytlin term is added. One may also compare this to the situation for D3 branes   discussed in \cite{Liu:1998bu}.}  Of course, within string theory there is no independent internal gravity on the branes,\footnote{Like there is for instance in the GDP model \cite{Dvali:2000hr}.} so it must  be possible to either eliminate the metric through its field equations, or for the metric to be related to the spacetime or worldvolume fields already present in the theory.

Our main goal will be to gain a better understanding of these models by performing the higgsing of \cite{Mukhi:2008ux}, which turns them into theories for stacks of D2-branes coupled to supergravity in a very special way. The  supergravity found here is at a `chiral point' similar to that discussed in the bosonic case by Li, Song and Strominger \cite{Li:2008dq} and in the supersymmetric case by Becker, Bruillard and Downes \cite{Becker:2009mk}. Thus, as we will see in detail later, the TMG and, in particular, chiral-point gravity models are closely related to the present work as was first noticed in \cite{Chu:2009gi}. The fact that the propagating modes have been argued to disappear in chiral gravity \cite{Li:2008dq,Maloney:2009ck} ties well in with the fact that the topologically gauged BLG/ABJM theories have no propagating gravitational modes and  provides additional incentive for the investigation of the topologically gauged M2-brane theories carried out in this paper. One should, however, be aware of the controversial aspects of these models connected to logarithmic gravity and log-CFT on the boundary, see \eg~\cite{Grumiller:2010rm, Grumiller:2010tj}, that might have implications also for the unhiggsed topologically gauged BLG/ABJM theories.

\subsection{Some comments on our results}

As we will demonstrate below, when reducing the topologically gauged M2-brane theory to D2-branes through the Higgs mechanism of \cite{Mukhi:2008ux}, it is clear that one will end up with a lagrangian that also contains a standard Einstein-Hilbert gravity term: When ABJM is coupled to conformal gravity such a higgsing procedure will turn the conformal coupling term of the Ricci scalar to the ABJM scalars, $R|Z|^2$, into the usual EH kinetic term. Giving a VEV to one of the ABJM scalars breaks conformal invariance and generates dynamical gauge fields, resulting in a super Yang-Mills (SYM) theory with a tunable coupling.  In addition, the new terms in the scalar potential arising from the topological gauging also produce a non-trivial AdS cosmological constant term. Thus after higgsing the models considered here are precisely of the kind discussed in the contexts of TMG and chiral gravity alluded to above.

However, because we are starting from a conformal theory and turning an interaction term into the leading gravitational kinetic term, the resulting $\mathcal N=6$ theory will have some unusual properties. Especially intriguing is the fact, as was noticed in \cite{Chu:2009gi},  that this theory automatically sits at a chiral point in the sense of  \cite{Li:2008dq}, albeit with an opposite sign for the EH term.  That the rather involved new sixth order potential terms that arise in the topological gauging gives exactly the correct cosmological constant for the chiral point theory is somewhat non-trivial.  

After the higgsing there are a number of parameters which will play an important role. We will therefore present two slightly different sets of limits with different virtues and problems. The physical interpretation of these two limits should probably be kept apart and will not be discussed much. Furthermore, in order to simplify the presentation, we discuss only one of the two sets in the bulk of the paper. In that limit the Newton constant is identified with the inverse VEV, while also giving rise to a background with a constant curvature which does not scale with the VEV. When the VEV becomes large gravity ends up interacting very weakly with matter. We postpone the introduction of the second limit, together with a comparison of the two, until Section \ref{summary}.

An important open question that could perhaps be addressed in this context is whether it is possible to couple these BLG/ABJM models (currently formulated only in a kind of static gauge) to general eleven dimensional supergravity backgrounds, in a fashion similar to that done for a single M2-brane in \cite{Bergshoeff:1987cm}. A step in this direction might be to couple BLG/ABJM to an M2 worldvolume metric, which one then could imagine as being the worldvolume restriction of a more general spacetime metric. Exactly how such a picture should be related to the case considered in this paper is however not clear. Attempts to couple  M2-brane theories to non-gravitational background fields can be found for instance in \cite{Lambert:2009qw,Kim:2010hj}.

\subsection{Explicit results}\label{explicit}

The main result of this work can be summarized in the following expression for the $\mathcal N=6$ chiral supergravity/matter
 lagrangian, which we obtain after higgsing the topologically gauged  ABJM theory of
 \cite{Chu:2009gi} (but before taking the VEV 
$v \rightarrow \infty$):
\be\label{final}
L = \frac{1}{\kappa^2}L_{\rm CG} + \frac{1}{\kappa} L^1_{\rm Int}+ ( L_{\rm SYM} + L^2_{\rm Int}) + \mathcal O(\kappa)\;,
\ee
with $\kappa$ the gravitational coupling and where $L_{\rm CG}$ contains the original conformal supergravity terms:\footnote{After the higgsing the original $\SU(4)$ unbroken R-symmetry gauge field $B^A_{\mu\,B}$, $A = 1,...,4$, decomposes into an $\SU(3)$ part $B^{A'}_{\mu\,B'}$, $A'=1,...,3$, a  ${\bf 3}$ of $\SU(3)$ $B^4_{\mu\,A'}$,  and a singlet $B^4_{\mu\,4}=iB_{0\mu}$. The six supersymmetry generators are in the ${\bf 3}+\bar{\bf 3}$ of the R-symmetry group $\SU(3)$.}
\bea\label{finalsugra}
\frac{1}{\kappa^2}L_{\rm CG} &=& \frac{1}{\kappa^2}\Big[-e (R - 2\Lambda)+\frac{1}{2\mu}\varepsilon^{\mu\nu\lambda}\Tr_\alpha(\tilde\omega_\mu\pd_\nu\tilde \omega_\lambda+\frac{2}{3}\tilde\omega_\mu \tilde \omega_\nu \tilde\omega_\lambda)\nn\\
&&-\frac{1}{2\mu}\varepsilon^{\mu\nu\rho}C_\mu \d_\nu C_\rho
+8 |{B^{A'}_\mu}_4|^2 - 8(qC_{\mu}+B_{\mu})^2\cr
&& -\frac{2}{\mu}\varepsilon^{\mu\nu\rho}\Tr_{A'}
(B_{\mu}\partial_{\nu}B_{\rho}+\frac{2}{3}B_{\mu}B_{\nu}B_{\rho})
-\frac{4}{\mu}\varepsilon^{\mu\nu\rho}B^{A'}_{\mu\,4}\check D_{\nu}B^4_{\rho\,A'}+\frac{2}{\mu}\varepsilon^{\mu\nu\rho}B_{0\mu}\partial_{\nu}B_{0\rho}\cr
&&- \frac{ie}{\mu}\tilde D_\mu \bar\chi_\nu^{AB }\gamma^{\rho\sigma}\gamma^{\mu\nu}\tilde D_\rho\chi_{\sigma AB}+c.c.
+2i\varepsilon^{\mu\nu\rho}\bar \chi_\mu^{AB}\Big(\tilde D_\nu +\frac{1}{2\ell}\gamma_\nu\Big)\chi_{\rho A B}+c.c.\Big] \;.
\eea
In the above $\ell$ is the ${\rm AdS}_3$ radius ($\Lambda=-\tfrac{1}{\ell^2}$) and $\mu$ is a  parameter satisfying $\mu \ell =1$ as required for a theory sitting at the chiral point \cite{Li:2008dq}.  The covariant derivatives $\tilde D$ are in general with respect to the spin connection and the various gauge fields that appear in the theory, to be given in detail in due course, while 
\be
\check{D}_{\nu}B^4_{\rho\,A'} \equiv \pd_{\nu}B_\rho^{4}{}_{A'} + i B_{0\nu} B_\rho^{4}{}_{A'}-\frac{2}{3}B_\nu^{B'}{}_{A'}B_\rho^{4}{}_{B'}\;.
\ee

Note that the signs in front of the mass terms for the two vector fields in the second line of \eqref{finalsugra} is irrelevant for stability since they are linear in mass. We  want to emphasize that in the one of the two limiting procedures where the above expression is valid, the  lagrangian describes gravitational excitations around the ${\rm   AdS}_3 $ background and that the action is valid only for small $\kappa$, or low energy. Also note  that in the strict limit the matter and gravity sectors completely decouple.  On the other hand, in the second way of taking the limits the  parameters of the theory are all free and tunable, independent of the limiting value $v\to\infty$. We  discuss these two possibilities in more detail in Section~\ref{summary}.

The expression (\ref{finalsugra}) for the gravity part of the theory is valid in the case of $\SU(N)\times \SU(N)$ matter while for the perhaps more physically relevant case (with a possible gravity dual) $\U(N)\times \U(N)$ the mass term $(qC+B)^2$ is replaced by the interaction term
\be
-\frac{4\sqrt{N\mu}}{g_{YM}}\varepsilon^{\mu\nu\rho}(qC_{\mu}+i B_{0\mu})F_{\nu\rho}^{+0}.
\ee

The next set of terms in \eqref{finalsugra} contains matter-gravity interactions and is given by
\be
\frac{1}{\kappa} L^1_{\rm Int} = \frac{1}{\kappa}\Big[- e \Big(\frac{1}{2} R + 3 \mu^2\Big) \tilde X_0^4 -i\sqrt 2 e A  (\bar{\tilde \psi}_0^{A'} + i \bar{\tilde \psi}_0^{A'+4})\gamma^{\mu\nu}(\tilde D_{\mu}+\frac{1}{2\ell}\gamma_{\mu}) \chi_{\nu A'4} + c.c \Big]\;.
\ee
We comment on the presence of  terms linear in $\tilde X_0^4$ during their derivation in Section~\ref{bosonic}.

For the $\mathcal O (0)$ terms, $L_{\rm SYM}$ is the usual 2+1d $\mathcal N=8$ SYM action\footnote{The eight supersymmetries occur here for the same reasons as  in the higgsed ABJM theory without gravity \cite{Aharony:2008ug}.} (minimally coupled to supergravity)  with coupling $g_{YM}$, although the maximal supersymmetry is broken down to $\mathcal N=6$ in the full theory due to the coupling to supergravity. Finally, the second set of matter-gravity interaction terms are given by
\bea\label{interactions}
L^2_{\rm Int} &=& -2 \mu g_{YM} ef^{abc}X_a^{A'+4}X_b^{A'}X_c^8 -\frac{3}{2}e\mu^2\tilde X_0^4 \tilde X_0^4 -\frac{e}{8} \Big( \frac{1}{2} R+ 3\mu^2\Big) (\tilde X_0^I \tilde X_0^I + X_a^I X_a^I)\nn\\
&&+\frac{3i \mu}{2}e\Big[\bar{\tilde{\psi}}_0^{A'}\tilde\psi_0^{A'}+\bar{\tilde{\psi}}_0^{A'+4} \tilde\psi_0^{A'+4}\Big]+\frac{i \mu}{2}e\Big[\bar\psi_a^{A'}\psi_a^{A'}+\bar\psi_a^{A'+4}\psi_a^{A'+4}\Big]\nn\\
&&+\frac{i\mu}{2}e\Big[5(\bar{\tilde{\psi}}_0^4\tilde\psi_0^4+\bar{\tilde{\psi}}_0^8
\tilde\psi_0^8)-3(\bar\psi_a^4\psi_a^4+\bar\psi_a^8\psi_a^8)\Big]\;,
\eea
where $\tilde X^4_0$, $X^8_a$, $X^{A'}$ and $X^{A'+4}$ are real fields coming from the four complex bifundamental fields $Z^A$ in the ABJM model. The various coefficients in the lagrangian reproduced here depend on which of the two sets of limits is used and if the parameters are taken to their limiting values or not.

The rest of this paper is organized as follows: We begin in Section~\ref{3andbi} by rewriting the topologically gauged $\mathcal N=6$ model of \cite{Chu:2009gi} in terms of bifundamental fields in order to make contact with the original formulation of the ABJM theories. We present  the details of the Higgs mechanism in Section \ref{Higgsing}, where we first review the gauge theory results and include a discussion on how the mechanism affects the abelian factors, which seems to have been omitted in the literature. We then separately address the gravity and interaction sectors, as well as the supersymmetry transformations. In Section~\ref{interpretation} we discuss the resulting $\mathcal N=6$ supergravity theory coupled to SYM and compare it with the pure chiral  gravity of \cite{Li:2008dq} as well as its $\mathcal N=1$ relative discussed in  \cite{Becker:2009mk}. 

For readers who are interested in the main points of the higgsing, the process is summarized in Section~\ref{summary} where we sketch the relevant steps while skipping most of the technical details. We also postpone the discussion and detailed comparison of the two limits to the end, while using only one of them in the  bulk of the paper to increase readability. Some concluding comments are collected in Section~\ref{conclusions}. A number of useful identities and calculational results can be found in Appendices~\ref{expansion} and \ref{subleading}.

\section{From three-algebra to bifundamental formulation}\label{3andbi}

The lagrangian and transformation rules of the topologically gauged ABJM theory constitute the starting point for the discussion in the remaining sections. Therefore, we  begin in Section~\ref{original} with a brief account of this theory as constructed  in \cite{Chu:2009gi}.  This is followed in Section~\ref{gauging} by a detailed description  of the steps needed for turning the results of \cite{Chu:2009gi} into a form more suitable for the higgsing  which is the main goal of this paper. The  lagrangian and transformation rules we find after performing these steps are presented in Section~\ref{lagandsusy}. 

We would also like to establish some nomenclature for the different theories appearing throughout this paper. The theory of \cite{Chu:2009gi} called the ``topologically gauged ABJM theory'' actually refers to a wide class of models, namely all those constructed from a complex three-algebra according to the classification of \cite{Schnabl:2008wj, Palmkvist:2009qq}. Here we will restrict this class to those without $\U(1)$ factors and as a result our starting point will contain the $\SU(N)\times \SU(N)$ but not the $\U(N)\times \U(N)$ ABJM models. The latter is usually obtained from the former by performing an extra step where some $\U(1)$ factors are gauged, as already described in \cite{Aharony:2008ug, Lambert:2010ji}.\footnote{For a recent discussion on $\U(1)$ factors see \cite{Gustavsson:2010yr}.} In our case however this will be done after the topological gauging (which is gravitational) and will in fact be carried out in an entirely different way compared to ABJM. We will be precise about which of the two cases we are analyzing at each point.

\subsection{The topologically gauged three-algebra theory}\label{original}

The formulation of the ungauged ABJM type models presented in \cite{Aharony:2008ug} makes use of ordinary gauge fields and Lie algebras. However, in the original work of Bagger-Lambert \cite{Bagger:2006sk,Bagger:2007jr,Bagger:2007vi} and Gustavsson \cite{Gustavsson:2007vu} the construction of the $\mathcal N=8$ (BLG) theory is in terms of three-algebras.  Subsequently, an alternative formulation of the $\mathcal N=6$ ABJM type theories in terms of such three-algebras was given in \cite{Bagger:2008se,Nilsson:2008kq}. In \cite{Gran:2008qx, Chu:2009gi} the topologically gauged M2-brane models were all constructed in this three-algebra language while the higgsing, to be performed later in this paper, is more conveniently implemented in the ordinary gauge field and Lie algebra formulation. We will therefore make the effort of rewriting the topologically gauged ABJM type theories of \cite{Chu:2009gi} in this more conventional language before subjecting them to the higgsing procedure.

The formulation of ABJM theory  obtained in \cite{Bagger:2008se} (see also \cite{ Nilsson:2008kq}) in terms of four-indexed structure constants is based on complex three-algebra generators $T^a$  and their triple product. For gauge theories with $\mathcal N=6$ supersymmetry the  associated structure constants and the trace are defined by
\be
[T^a,T^b;T_c]={f^{ab}}_{cd}T^d\,,\qquad \Tr(T^a T_b) = \delta^a_b\;,
\ee
where the structure constants are antisymmetric in both the upper and lower pair of indices. 
All indices are moved up or down by complex conjugation. For further details on the three-algebra and the lagrangian reproduced below we refer the reader to \cite{Chu:2009gi}.

The  topologically gauged M2-brane theories \cite{Gran:2008qx, Chu:2009gi} are  gravitationally coupled theories  preserving the superconformal symmetry. All global symmetries of the M2 matter theories are gauged in the process. Those preserving $\mathcal N=6$ supersymmetry are given in terms of the following lagrangian \cite{Chu:2009gi}:
\begin{eqnarray}\label{BengtYong}
 L&=&L_{sugra}^{conf}+L_{ABJM}^{cov} +\tfrac{1}{2}\epsilon^{\mu\nu\rho}C_{\mu}\partial_{\nu}C_{\rho}\cr
  &&+iAe\bar\chi_{\mu}^{BA}\gamma^{\nu}\gamma^{\mu}\Psi_{Aa}
       (\tilde D_{\nu}\bar Z^a_B-\frac{i}{2} A\bar\chi_{\nu BC}\Psi^{Ca})+c.c.\cr
  &&+i \epsilon ^{\mu\nu\rho} (\bar\chi_{\mu}^{AC}\chi_{\nu BC}) Z^B_a
       \tilde D_{\rho} \bar Z^a_A+c.c.\cr
 &&-i A(\bar f^{\mu AB}\gamma_{\mu}\Psi_{Aa}\bar Z_B^a+
 \bar f^{\mu}_{AB}\gamma_{\mu}\Psi^{Aa}Z^B_a)\cr
 &&-\tfrac{e}{8}\tilde R   |Z| ^2\cr
 &&+\tfrac{i}{2}  |Z|  ^2\bar f^{\mu}_{AB}\chi_{\mu}^{AB}\cr
 &&+2ieA f^{ab}{}_{cd}(\bar \chi_{\mu AB}\gamma^{\mu}\Psi^{d[B})Z^{D]}_a Z^A_b \bar Z^c_D+c.c.\cr
 &&-i\epsilon^{\mu\nu\rho}(\bar \chi_{\mu AB}\gamma_{\nu}\chi_{\rho}^{CD})(Z^A_a Z^B_b\bar Z^c_C \bar Z^d_D)f^{ab}{}_{cd}\cr
 &&+\tfrac{i}{4}\epsilon^{\mu\nu\rho}(\bar\chi_{\mu AB}\gamma_{\nu}\chi_{\rho}^{AB})
 (Z^C_a Z^D_b \bar Z^c_C \bar Z^d_D)f^{ab}{}_{cd}\cr
 &&-\tfrac{i}{16}e \epsilon^{ABCD}(\bar\Psi_{Aa}\Psi_{Bb})\bar Z^a_C\bar Z^b_D+c.c.\cr
 &&+\tfrac{i}{16} e(\bar\Psi_{Db}\Psi^{Db})  |Z| ^2-\frac{i}{4}e(\bar\Psi_{Db}\Psi^{Bb})\bar Z^a_B Z^D_a\cr
 &&+\tfrac{i}{8}e(\bar \Psi_{Db}\Psi^{Da})\bar Z^b_B Z^B_a
   +\tfrac{3i}{8}e (\bar \Psi_{Db}\Psi^{Ba})\bar Z^b_B Z^D_a \cr
 &&-\tfrac{i}{16}e A(\bar \chi_{\mu AB} \gamma^{\mu}\Psi^{Bb})   |Z| ^2 Z_b^A
 -\tfrac{i}{4}e A(\bar \chi_{\mu AB} \gamma^{\mu}\Psi^{Db}) Z^A_a Z^B_b \bar Z^a_D+c.c\cr
 &&-\tfrac{i}{4}\epsilon^{\mu\nu\rho}(\bar \chi_{\nu AB} \gamma_{\rho}\chi_{\mu}^{CD}) Z^A_a Z^B_b \bar Z^a_C \bar Z^b_D
  +\frac{i}{64}\epsilon^{\mu\nu\rho}(\bar \chi_{\nu AB} \gamma_{\rho}\chi_{\mu}^{AB})  |Z| ^4 \cr
 &&+\tfrac{1}{8}ef^{ab}{}_{cd}   |Z| ^2 Z^C_a Z^D_b \bar Z^c_C \bar
          Z^d_D+\frac{1}{2}e f^{ab}{}_{cd}Z^B_a Z^C_b Z^D_e \bar Z^e_B \bar Z^c_C \bar Z^d_D\cr
 &&+\tfrac{5}{12 \cdot 64}e  |Z| ^6 -\frac{1}{32}e   |Z| ^2 Z^A_b Z^C_a \bar Z^b_C \bar Z^a_A+
 \tfrac{1}{48}e Z^A_a Z^B_b Z^C_d \bar
          Z^b_A \bar Z^d_B \bar Z^a_C,
\end{eqnarray}
where 
\bea\label{ABJM}
L^{cov}_{ABJM}&=& -e(\tilde D_\mu Z^{A}{}_a)(\tilde D^\mu \bar{Z}_A{}^a) - \frac{1}{2}(ie \bar \Psi^{Aa} \gamma^\mu  \tilde D_\mu  \Psi_{Aa}+
ie \bar \Psi_{Aa} \gamma^\mu \tilde D_\mu  \Psi^{Aa})
 \cr
 &&  - ie f^{ab}{}_{cd}\bar {\Psi}^{Ad}
 \Psi_{Aa}Z^B{}_b \bar{Z}_{B}{}^{c}+
 2ie f^{ab}{}_{cd}\bar {\Psi}^{Ad}  \Psi_{Ba}Z^B{}_b \bar{Z}_{A}{}^{c}
  \cr
 &&
 -\tfrac{i}{2}e \epsilon_{ABCD} f^{ab}{}_{cd}\bar {\Psi}^{Ac}  \Psi^{Bd} Z^C{}_a Z^D{}_b
  -\tfrac{i}{2}e \epsilon^{ABCD} f^{cd}{}_{ab}\bar {\Psi}_{Ac}  \Psi_{Bd}\bar{Z}_{C}{}^a \bar{Z}_{D}{}^{b}
\cr
 && -eV +\tfrac{1}{2}\epsilon^{\mu\nu\lambda}(
f^{ab}{}_{cd}A_{\mu}{}^d{}_{b}
\partial_\nu A_{\lambda}{}^c{}_{a}+ \tfrac{2}{3} f^{bd}{}_{gc} f^{gf}{}_{ae}
A_{\mu}{}^a{}_{b}  A_{\nu}{}^c{}_{d} A_{\lambda}{}^e{}_{f})\,,
\\
V &=& \tfrac{2}{3} \Upsilon^{CD}{}_{Bd}\bar\Upsilon_{CD}{}^{Bd}\,,\\
\Upsilon^{CD}{}_{Bd}&=& f^{ab}{}_{cd} Z^C{}_a{Z}^D{}_b \bar{Z}_B{}^c
  + f^{ab}{}_{cd}\delta^{[C}{}_B Z^{D]}{}_a{Z}^E{}_b \bar{Z}_{E}{}^{c}
\eea
and
\bea
L^{conf}_{sugra} &=&\frac{1}{2}\varepsilon^{\mu\nu\rho}
\Tr_{\alpha}(\tilde\omega_{\mu}\partial_{\nu}\tilde\omega_{\rho}+
\frac{2}{3}\tilde\omega_{\mu}\tilde\omega_{\nu}\tilde\omega_{\rho})
-2\varepsilon^{\mu\nu\rho}\Tr_A
(B_{\mu}\partial_{\nu}B_{\rho}+\frac{2}{3}B_{\mu}B_{\nu}B_{\rho})\\\nn
&&-i e^{-1} \varepsilon^{\alpha\mu\nu}\epsilon^{\beta\rho\sigma}(\tilde
D_{\mu}\bar{\chi}^{AB}_{\nu}\gamma_{\beta}\gamma_{\alpha}\tilde
D_{\rho}\chi_{\sigma AB}).
\eea
Here  $B_{\mu}^A{}_B$ is an $\SO(6)$ R-symmetry gauge field, $A = \pm \sqrt 2$ and   $|Z|^2$ stands for
\be
Z^A_a\bar Z_A^a=\Tr(\bar Z_A,Z^A)\;,
\ee
while the action of the $\tilde D$-covariant derivative on a spinor is given by\footnote{Here we follow the three-algebra literature conventions, with the gauge fields being anti-hermitian.}
\be
\tilde D_{\mu}\Psi^{Aa}=\partial_{\mu}\Psi^{Aa}+\tfrac{1}{4}\tilde
\omega_{\mu \alpha\beta}\gamma^{\alpha\beta}\Psi^{Aa}+B_{\mu
B}^A\Psi^{Ba}+\tilde A_{\mu
b}^a\Psi^{Ab}+qC_{\mu} \Psi^{Aa}.\label{Ccovderivative}
\ee
From the work of \cite{Chu:2009gi} we know that $q^2=\tfrac{1}{16}$ in order for the theory to possess six local special conformal as well as  ordinary supersymmetries. 

 In the above formulas we have also used the definitions
\bea
f^{\mu}&=&\frac{1}{2}\varepsilon^{\mu\nu\rho}\tilde D_{\nu}\chi_{\rho}\,,\\
\tilde\omega_{\mu \a\b}&=&\omega_{\mu \a\b}(e)-\frac{i}{2}(\bar\chi_\mu \gamma_\b \chi_\a-\bar\chi_\mu \gamma_\a\chi_\b-\bar \chi_\a\gamma_\mu\chi_\b),
\eea
\ie $\tilde\omega_{\mu\a\b}=\tilde\omega_{\mu\a\b}(e,\chi)$ is the usual spin connection of supergravity in the second order formalism. 

The novel features of this lagrangian  are the conformal coupling term between the curvature scalar and the scalar fields $-\tfrac{e}{8}R|Z|^2$, as well as the new terms in the sixth order  scalar potential (\ie the last two lines in the lagrangian \eqref{BengtYong} above). These terms will play an important role later.

\subsection{Gauging of  $\U(1)$'s and  turning three-algebras into Lie algebras}\label{gauging}

Having given the three-algebra formulation of the topologically gauged theory, we will now switch to a Lie algebra formulation in order to facilitate the higgsing procedure presented in the next section. We also give a detailed discussion of the various ways $\U(1)$ factors appear in the theories analyzed here.

To find the Lie algebra version of these theories we need to replace the triple product and trace of the abstract three-algebra by equivalent expressions in terms of matrices. We will  follow 
the procedure presented in  \cite{Bagger:2008se}.
 As an example, consider the following  term already analyzed in \cite{Bagger:2008se}:
\be\label{prereplace}
-i {f^{ab}}_{cd}\bar \Psi^{Ad} \Psi_{Aa}Z^B_b \bar Z^c_{B}\rightarrow\
-i\Tr(\bar \Psi^A, {f^{ab}}_{c d}\Psi_{Aa}Z^B_b \bar Z^c_{B} T^d)
\rightarrow -i \Tr(\bar \Psi^A, [\Psi_A, Z^B;\bar Z_B])\;,
\ee
where fields without three-algebra indices have been contracted by an element from the algebra.
We will finally express everything in terms of the following $N\times N$ matrix realization of the three-algebra
\be
\Tr(\bar X,Y)=\tr(X^\dagger Y)\;,\qquad\qquad [X,Y;\bar Z]=\lambda (XZ^\dagger Y-YZ^\dagger X)\label{replace}\;,
\ee
where $\lambda = \frac{2\pi}{k}$. Thus, for example,

\be
Z^A_a\bar Z_A^a=\Tr(\bar Z_A,Z^A)=\tr(Z^\dagger_A Z^A)\;,
\ee

Applying the above procedure on one of the new interaction terms of the topologically gauged theory we get:
\bea
\frac{1}{8}ef^{ab}{}_{cd}   |Z| ^2 Z^C_a Z^D_b \bar Z^c_C \bar
          Z^d_D&\rightarrow&\frac{1}{8}e|Z|^2 \Tr(\bar Z_D,[Z^C,Z^D;\bar Z_C])\cr
&\rightarrow & \frac{1}{8}\lambda e|Z|^2\tr(Z^\dagger _DZ^CZ^\dagger_CZ^D-Z^\dagger_DZ^DZ^\dagger_C Z^C)\nn \;.
\eea
In the next subsection we will use the definition of the  trace to introduce one more
parameter which will then show up in the last form of the interaction term above.

It is important to note that the procedure of \cite{Bagger:2008se} only leads to an $\SU(N)\times \SU(N)$ gauge group \cite{Lambert:2010ji}. Hence one needs to introduce two abelian factors in order to turn $\SU(N)\times \SU(N)$ into $\U(N)\times \U(N)$, while also keeping the $\U(1)$ associated with $C_\mu$ and preserving $\mathcal N=6$ supersymmetry. This $\U(1)$-extended formulation will require that the terms in the  lagrangian related to the $C_{\mu}$ field be accompanied by terms containing two abelian fields with opposite CS-level $k'$ (independent of the one in the non-abelian sector), denoted $\hat{\hat A}^L$ and $\hat{\hat A}^R$. The introduction of $\hat{ \hat A}^L, \hat{\hat A}^R$ means that at the end these fields can be combined with the nonabelian $\tilde {A^a}_b$ to give the desired $\U(N)\times\U(N)$ gauge fields leading to the topologically gauged $\U(N)\times\U(N)$ ABJM model.

Instead of repeating the rather extensive calculations of \cite{Chu:2009gi} for this new situation, we will employ a simpler method to introduce additional abelian Chern-Simons fields without destroying any of the six local supersymmetries. We first  split $C_\mu$ as
$C_{\mu}= E_\mu + \tfrac{\a}{q} P_\mu$ and note that nothing changes in the check of supersymmetry if we define the variations of $E_\mu, P_\mu$ as
$\delta E_\mu =\delta C_\mu$ and $\delta P_\mu=0$. Replacing $C_\mu$ in the covariant derivative gives
\begin{eqnarray}
\tilde D_{\mu}\psi^{Aa}=\partial_{\mu}\psi^{Aa}+\tfrac{1}{4}\tilde
\omega_{\mu \alpha\beta}\gamma^{\alpha\beta}\psi^{Aa}+B_{\mu
B}^A\psi^{Ba}+\tilde A_{\mu
b}^a\psi^{Ab}+qE_{\mu}\psi^{Aa}+ \a P_{\mu}\psi^{Aa}\;.
\end{eqnarray}
Eventually we also want to replace the CS term for $C_\mu$  without affecting its supersymmetry variation. Hence we consider 
\be
\delta(\frac{1}{2}\varepsilon^{\mu\nu\rho}C_\mu \d_\nu C_\rho)=
\delta(\frac{1}{2}\varepsilon^{\mu\nu\rho}E_\mu \d_\nu E_\rho)+\frac{\a}{q}\varepsilon^{\mu\nu\rho}P_\mu\d_\nu (\delta E_\rho)\label{variation}
\ee
and redefine $P_\mu$ by setting $P_\mu=\hat{\hat A}_\mu^L-\hat{\hat A}_\mu^R$. 
Recalling that $\delta P^\mu=0$, any $\delta \hat{\hat A}^L_\mu=\delta \hat{\hat A}^R_\mu$ will do since it does not appear explicitly in the supersymmetry variations. The requirement that when we switch to the Lie algebra formulation the $\hat{\hat A}^L_\mu, \hat{\hat A}^R_\mu$ terms appear in the same way in the covariant derivative as the $\SU(N)$ pieces  fixes $\a=1$. We then replace the CS term for the $C_\mu$ field, $\frac{1}{2}\epsilon^{\mu\nu\rho}C_\mu \d_\nu C_\rho$, with 
\be
\frac{1}{2}\varepsilon^{\mu\nu\rho}E_\mu \d_\nu E_\rho+\frac{k'}{4\pi}\varepsilon^{\mu\nu\rho}\left(\hat{\hat A}_\mu^L\d_\nu \hat{\hat A}^L_\rho-
\hat{\hat A}^R_\mu\d_\nu \hat{\hat A}^R_\rho\right)=\frac{1}{2}\varepsilon^{\mu\nu\rho}E_\mu \d_\nu E_\rho+\frac{k'}{4\pi}\varepsilon^{\mu\nu\rho}
P_\mu\d_\nu (\hat{\hat A}^L_\rho+\hat{\hat A}^R_\rho)\,.
\ee
In order to keep the supersymmetry invariance, we equate its variation with (\ref{variation}), obtaining 
\bea
&&-i\delta \hat{\hat A}^L_\mu=-i\delta \hat{\hat A}^R_\mu = \frac{2\pi }{k'}\frac{\a}{q}\delta E\cr
\cr
&=&\frac{2\pi}{k'}(\bar\epsilon_{AB}\gamma_{\mu}\Psi^{Aa}Z^B_a-
 \bar\epsilon^{AB}\gamma_{\mu}\Psi_{Aa}\bar Z_B^a)
    +\frac{4\pi}{k'}(\bar\epsilon_g^{AD}\chi_{\mu BD}-\bar\epsilon_{gBD}\chi_{\mu}^{AD})Z^B_a\bar
    Z^a_A\;.
\eea
Finally, we rename $E_\mu$ as $C_\mu$, and note that in effect, we have introduced the two new fields $\hat{\hat A}^L_\mu, \hat{\hat A}^R_\mu$, 
on top of the $C_\mu$ field.

Thus in the case corresponding to $\U(N)\times \U(N)$ we will make use of the covariant derivative
\be
\tilde D_{\mu}\Psi^{Aa}=\partial_{\mu}\Psi^{Aa}+\tfrac{1}{4}\tilde
\omega_{\mu \alpha\beta}\gamma^{\alpha\beta}\Psi^{Aa}+B_{\mu
B}^A\Psi^{Ba}+qC_{\mu}\Psi^{Aa}+\tilde A_{\mu b}^a\Psi^{Ab}+(\hat{ \hat A}_\mu^L - \hat{ \hat A}_\mu^R)\delta_b^a \Psi^{Ab},\label{ALRcovderivative}
\ee
 and add a related Chern-Simons term to the lagrangian
\be
 \frac{k'}{4\pi} \varepsilon^{\mu\nu\lambda} (\hat{\hat A}^L_\mu\partial_\nu \hat{\hat A}^L_\lambda
-\hat{\hat A}^R_\mu\partial_\nu \hat{ \hat A}^R_\lambda ) \frac{1}{N}\tr(\one_{N\times N})\;.
\ee
Note that here we have introduced a factor $1=\tfrac{1}{N}\tr(\one_{N\times N})$ that will be useful later when combining the abelian fields with the $\SU(N)$ ones to form a $\U(N)\times \U(N)$ ABJM theory.

A further comment on notation is in order: When we next move to the Lie algebra formulation we will switch from the three-algebra conventions with anti-hermitian generators to the more familiar ones with hermitian generators, in order to connect with the usual ABJM analysis. We will treat the $\U(N)\times \U(N)$ gauge fields coming from the three-algebra, and the corresponding part of the covariant derivative separately, but for the $\SU(4)$ and $\U(1)$ fields ${B^A}_B, C,\hat{\hat A}^L,\hat{\hat A}^R$ this corresponds to  multiplying with an extra factor of $i$.

\subsection{The topologically gauged ABJM theory in terms of Lie algebras}\label{lagandsusy}

The higgsing procedure will involve taking limits in several parameters at the same time. One of these is the CS-level parameter $\lambda = \frac{2\pi}{k}$. However, there is also a natural way to introduce a (super-)gravitational coupling, while retaining  the supersymmetry of the lagrangian.  We will call this coupling $g_M$, keeping in mind that it could come from string theory after reduction to 10d. This can be achieved by multiplying all traces by $g_M^2$ and then rescaling the whole lagrangian by $\tfrac{1}{g_M^2}$, thus ensuring that the pure ABJM part remains unmodified.  Alternatively, one could just  insert a $g_M^2$ on the RHS of the trace identification (\ref{replace}) and then rescale the lagrangian by $\tfrac{1}{g_M^2}$. In this manner one obtains
\bea
\frac{1}{8}ef^{ab}{}_{cd}   |Z| ^2 Z^C_a Z^D_b \bar Z^c_C \bar
          Z^d_D&\rightarrow&\frac{1}{8}e|Z|^2 \Tr(\bar Z_D,[Z^C,Z^D;\bar Z_C])\cr
&\rightarrow & \frac{1}{8}g_M^2 \lambda e|Z|^2\tr(Z^\dagger _DZ^CZ^\dagger_CZ^D-Z^\dagger_DZ^DZ^\dagger_C Z^C)\nn \;,
\eea
where in the last expression $|Z|^2$ refers to the matrix trace $\tr$.

Note that while $L^{cov}_{ABJM}$ involves a  single  trace over gauge indices, $L^{conf}_{sugra}$ has no trace and hence the interaction terms are no-trace as well as  single-, double- and triple-trace. This will have important implications for the results  of the higgsing procedure and in particular for how the various limits affect the terms in the lagrangian.

\subsubsection{The lagrangian}

Following the procedure described so far, the resulting Lie algebra lagrangian is
\bea
L&=&\tfrac{1}{g_M^2}L_{sugra}^{conf}+L_{ABJM}^{cov}+
\tfrac{1}{2g_M^2}\epsilon^{\mu\nu\rho}C_{\mu}\partial_{\nu}C_{\rho} \label{uncoupled}\\
&&+iAe\bar \chi^{BA}_\mu \gamma^\nu\gamma^\mu \tr(\Psi
_{A}(\tilde{D}_\nu Z^\dagger_B-\frac{i}{2}A
(\bar \chi_{\nu BC}\Psi^C)^\dagger))+c.c.\label{coup1}\\
&&+i\epsilon^{\mu\nu\rho}(\bar \chi_\mu ^{AC}\chi_{\nu BC})\tr(Z^B\tilde{D}_\rho Z^\dagger_A)+c.c.\cr
&&-iA(\bar f^{\mu AB}\gamma_\mu \tr(\Psi_AZ^\dagger_B)+\bar f^\mu _{AB}\gamma_\mu \tr(\Psi^{\dagger A}Z^B))\cr
&&-\frac{e}{8}\tilde{R}|Z|^2\label{EHaction}\\
&&+\frac{i}{2}|Z|^2\bar f^\mu_{AB}\chi_\mu ^{AB}\cr
&&+2i\lambda eA\bar \chi_{\mu AB}\gamma^\mu \tr(\Psi^{[B}(Z^{D]}Z^{\dagger}_DZ^A-Z^{|A|}Z^{\dagger}_DZ^{D]})) + c.c.\cr
&&-i\lambda\epsilon^{\mu\nu\rho}(\bar \chi_{\mu AB}\gamma_\nu \chi_\rho ^{CD})\tr(Z^\dagger_DZ^AZ^{\dagger}_CZ^B-Z^{\dagger}_DZ
^BZ^\dagger_CZ^A)\cr
&&+\frac{i}{4}\lambda \epsilon^{\mu\nu\rho}(\bar \chi_{\mu AB}\gamma_\nu \chi_\rho ^{AB})\tr(Z^\dagger_DZ^CZ^\dagger_CZ^D-
Z^\dagger_D Z^DZ^\dagger _C Z^C)\cr
&&-\frac{i}{16}g_M^2e\epsilon^{ABCD}\tr(\bar \Psi_A Z^\dagger_C)\tr(\Psi_BZ^\dagger_D)+c.c.\label{2trace1}\\
&&+\frac{i}{16}g_M^2e\;\tr(\bar \Psi_D^\dagger\Psi^D)|Z|^2-\frac{i}{4}g_M^2e \; \tr(\bar \Psi_D^\dagger\Psi^B)\tr(Z^\dagger_BZ^D)
\label{2trace2}\\
&&+\frac{i}{8}g_M^2e\; \tr(\bar \Psi_D Z^\dagger_B)\tr(\Psi^{\dagger D}Z^B)+\frac{3i}{8}g_M^2e\;\tr(\bar \Psi_DZ^\dagger_B)
\tr(\Psi^{\dagger B}Z^D)\label{2trace3}\\
&&-\frac{i}{16}g_M^2eA\bar \chi_{\mu AB}\gamma^\mu \tr(\Psi^{\dagger B} Z^A)|Z|^2-\frac{i}{4}g_M^2eA\bar \chi_{\mu AB}\gamma^\mu
\tr(\Psi^{\dagger D}Z^B)\tr(Z^AZ^\dagger _D)+c.c.\cr
&&\\
&&-\frac{i}{4}g_M^2\epsilon^{\mu\nu\rho}(\bar \chi_{\nu AB}\gamma_\rho \chi_\mu ^{CD})\tr(Z^AZ^\dagger_C)\tr(Z^BZ^\dagger_D)
+\frac{i}{64}g_M^2|Z|^4\epsilon^{\mu\nu\rho}(\bar\chi_{\nu AB}\gamma_\rho \chi_\mu ^{AB})\cr
&&+\frac{1}{8} g_M^2\lambda e|Z|^2\tr(Z^\dagger _DZ^CZ^\dagger_CZ^D-Z^\dagger_DZ^DZ^\dagger_C Z^C)\cr
&&+\frac{1}{2}g_M^2\lambda e\; \tr(Z^DZ^\dagger_B)\tr(Z^\dagger_DZ^BZ^\dagger_CZ^C-Z^\dagger_DZ^CZ^\dagger_CZ^B)\cr
&&\hspace{8cm}\label{3tracef}\\
&&+\frac{5}{12\cdot 64}g_M^4e |Z|^6-\frac{1}{32} g_M^4e |Z|^2\tr(Z^AZ^\dagger_C)\tr(Z^CZ^\dagger_A)\cr
&&+\frac{1}{48} g_M^4 e\; \tr(Z^AZ^\dagger_C)\tr(Z^BZ^\dagger_A)\tr(Z^CZ^\dagger_B)\label{3trace}\;.
\eea
We do not find it necessary to give the expressions for $L_{sugra}^{conf}$ and $ L^{cov}_{ABJM}$, as the former has no gauge indices and hence stays the same while the latter was already given in this form in \cite{Bagger:2008se}. Note that $\lambda=\tfrac{2\pi}{k}$ multiplied all the three-algebra ${f^{ab}}_{cd}$ terms and hence also naturally  multiplies all $f^{abc}$ terms in the  lagrangian above, obtained when performing the traces. 

\subsubsection{The transformation rules}

In the case of the supersymmetry transformations, everything is straightforwardly done in the same fashion, except for the gauge field which will be dealt with separately. The explicit transformations in the three-algebra language are
\bea
 \delta e_{\mu}{}^{\alpha}&=& i \bar\epsilon_{gAB}\gamma^{\alpha}\chi_{\mu}^{AB},\\
 \delta \chi_{\mu}^{AB}&=&\tilde D_{\mu}\epsilon_{g}^{AB},\\
 \delta B_{\mu ~B}^{~A}&=&\frac{i}{e}(\bar f^{\nu AC}\gamma_{\mu}\gamma_{\nu}\epsilon_{g BC}-
  \bar f^{\nu}_{BC}\gamma_{\mu}\gamma_{\nu}\epsilon_{g}^{AC}
             )\cr
   &&+\tfrac{i}{4}(\bar\epsilon_{BD}\gamma_{\mu}\Psi^{a(D} Z^{A)}_a-\bar\epsilon^{AD}\gamma_{\mu}\Psi_{a(D} \bar Z_{B)}^a)
         \cr
   &&-\tfrac{i}{2}(\bar\epsilon^{AC}_g\chi_{\mu DC}Z^D_a\bar Z^a_B-\bar\epsilon_{g BC}\chi^{DC}_{\mu}Z^A_a\bar Z^a_D)\cr
      && +\tfrac{i}{8}\delta^A_B(\bar\epsilon^{EC}_g\chi_{\mu DC}-
       \bar\epsilon_{gDC}\chi^{EC}_{\mu})Z^D_a\bar Z^a_E
       \cr
   &&+\tfrac{i}{8}(\bar\epsilon^{AD}_g\chi_{\mu BD}-\bar\epsilon_{gBD}\chi^{AD}_{\mu})
   |Z| ^2,\\
 \delta Z^A_a &=&i\bar\epsilon^{AB}\Psi_{Ba}, \\
 \delta\Psi_{Bd}&=&\gamma^\mu\epsilon_{AB}(\tilde D_\mu Z^A_d
 -iA\bar\chi_{\mu}^{AD}\Psi_{Dd})\cr
   &&+ f^{ab}{}_{cd} Z^C{}_a Z^D_b \bar Z_B^c \epsilon_{CD}-
    f^{ab}{}_{cd} Z^A_a Z^C_b \bar Z_C^c\epsilon_{AB}\cr
   &&+\frac{1}{4}Z^C_c Z^D_d \bar Z_B^c \epsilon_{CD}+\frac{1}{16}   |Z| ^2
   Z^A_d\epsilon_{AB},\\
 \delta \tilde A_{\mu~d}^{~c}&=&-i(\bar\epsilon_{AB}\gamma_{\mu}\Psi^{Aa}Z^B_b-
 \bar\epsilon^{AB}\gamma_{\mu}\Psi_{Ab}\bar Z_B^a)f^{bc}{}_{ad}\cr
    &&-2i(\bar\epsilon_g^{AD}\chi_{\mu BD}-\bar\epsilon_{gBD}\chi_{\mu}^{AD})Z^B_b\bar Z^a_A
    f^{bc}{}_{ad},\\
\delta \hat{\hat A}^R_\mu &=&\delta \hat{\hat A}^L_\mu=
\frac{2\pi}{ k'}(\bar\epsilon_{AB}\gamma_{\mu}\Psi^{Aa}Z^B_a-\bar\epsilon^{AB} \gamma_\mu\Psi_{Aa}\bar
Z_B^a)\cr
&&+\frac{4\pi}{k'}(\bar\epsilon_g^{AD}\chi_{\mu BD}-\bar\epsilon_{gBD}\chi_{\mu}^{AD})Z^B_a\bar Z^a_A,\\
\delta C_\mu &=&
-iq(\bar\epsilon_{AB}\gamma_{\mu}\Psi^{Aa}Z^B_a-\bar\epsilon^{AB} \gamma_\mu\Psi_{Aa}\bar
Z_B^a)\cr
&&-2iq(\bar\epsilon_g^{AD}\chi_{\mu BD}-\bar\epsilon_{gBD}\chi_{\mu}^{AD})Z^B_a\bar Z^a_A\;,
\eea
where $ \epsilon_m^{AB}=A\epsilon_g^{AB}=\epsilon^{AB}$. 

Leaving  the gauge transformation law aside for the moment, by multiplying $Z_a^A,\Psi_{Ba}$ with $T^a$ and writing everything in terms of traces and triple brackets according to (\ref{replace}), while also introducing $g_M^2$, we get
\bea
\delta e_{\mu}{}^{\alpha}&=& i \bar\epsilon_{gAB}\gamma^{\alpha}\chi_{\mu}^{AB},\label{1}\\
\delta \chi_{\mu}^{AB}&=&\tilde D_{\mu}\epsilon_{g}^{AB},\label{2}\\
\delta B_{\mu ~B}^{~A}&=&\frac{i}{e}(\bar f^{\nu AC}\gamma_{\mu}\gamma_{\nu}\epsilon_{g BC}-
\bar f^{\nu}_{BC}\gamma_{\mu}\gamma_{\nu}\epsilon_{g}^{AC})\cr
&&+\frac{i}{4}g_M^2(\bar\epsilon_{BD}\gamma_{\mu}\tr(\Psi^{\dagger(D} Z^{A)})-\bar\epsilon^{AD}\gamma_{\mu}\tr(\Psi_{(D} Z^\dagger_{B)}))\cr
&&-\frac{i}{2}g_M^2 (\bar\epsilon^{AC}_g\chi_{\mu DC}\tr(Z^DZ^\dagger_B)-\bar\epsilon_{g BC}\chi^{DC}_{\mu}\tr(Z^AZ^\dagger_D)
)\cr
&&+\frac{i}{8}g_M^2\delta^A_B(\bar\epsilon^{EC}_g\chi_{\mu DC}-\bar\epsilon_{gDC}\chi^{EC}_{\mu})\tr(Z^DZ^\dagger_E)\cr
&&+\frac{i}{8}g_M^2 (\bar\epsilon^{AD}_g\chi_{\mu BD}-\bar\epsilon_{gBD}\chi^{AD}_{\mu})|Z| ^2,\label{3}\\
\delta Z^A &=&i\bar\epsilon^{AB}\Psi_{B}, \label{4}\\
\delta\Psi_{B}&=&\gamma^\mu\epsilon_{AB}(\tilde D_\mu Z^A
-iA\bar\chi_{\mu}^{AD}\Psi_{D})\cr
&&+ \lambda (Z^CZ^\dagger_BZ^D-Z^DZ^\dagger_B Z^C) \epsilon_{CD}-\lambda(Z^AZ^\dagger_CZ^C-Z^CZ^\dagger_CZ^A)\epsilon_{AB}\cr
&&+\frac{1}{4}g_M^2\tr(Z^CZ^\dagger_B)Z^D \epsilon_{CD}+\frac{1}{16} g_M^2  |Z| ^2
Z^A\epsilon_{AB}\cr
\delta C_{\mu}&=&-g_M^2q(\bar\epsilon_{AB}\gamma_{\mu}\tr(\Psi^{\dagger A}Z^B)-
\bar\epsilon^{AB}\gamma_{\mu}\tr(\Psi_{A}Z^\dagger_B))\cr
&&-2g_M^2q(\bar\epsilon_g^{AD}\chi_{\mu BD}-\bar\epsilon_{gBD}\chi_{\mu}^{AD})\tr(Z^BZ^\dagger_A)
\label{5}\;.
\eea
We can now return to the transformation law for the gauge fields. These are obtained by requiring that the  transformation law of the covariant derivative remains unchanged under the replacement\footnote{As advertised, we now work with hermitian gauge fields, hence the field $C_{\mu}$ introduced in \cite{Chu:2009gi} appears as $iC_{\mu}$ in the covariant derivatives {\it etc}.}
\bea
&&T^a D_\mu Z^A_a=T^a ( \d_\mu Z^A_a-{{\tilde{A}_\mu\,}^b}_a Z_b^A)\Rightarrow \nn\\
&&D_\mu Z^A=\d_\mu Z^A +i\hat A_\mu ^{L}Z^A-Z^A i \hat A_\mu^{R}\label{covder}
\eea
used in the pure ABJM case. This implies that
\be
T^d \delta {{\tilde{A}_\mu\,}^c}_dF_c^D=-i\delta \hat A_\mu^{L}F^D+iF^D\delta \hat A_\mu^{R}\label{condition}\;,
\ee
where we used a general object $F^D$ instead of the $Z^A$ in order to facilitate the identification of the transformation
rules. The LHS of (\ref{condition}) gives
\bea
&&-iT^d(\bar\epsilon_{AB}\gamma_{\mu}\Psi^{Aa}Z^B_b-
 \bar\epsilon^{AB}\gamma_{\mu}\Psi_{Ab}\bar Z_B^a)f^{bc}{}_{ad}F_c^D\cr
&& -2i(\bar\epsilon_g^{AD}\chi_{\mu BD}-\bar\epsilon_{gBD}\chi_{\mu}^{AD})T^dZ^B_b\bar Z^a_Af^{bc}{}_{ad}F_c^D\cr
&\rightarrow&-i(\bar \epsilon_{AB}\gamma_\mu [Z^B,F^D;\Psi^A]-\bar\epsilon^{AB}\gamma_\mu [\Psi_A,F^D;\bar Z_B])\cr
&&-2i(\bar\epsilon_g^{AD}\chi_{\mu BD}-\bar\epsilon_{gBD}\chi_{\mu}^{AD})[Z^B,F^D;\bar Z_A]\cr
&\rightarrow &-i\lambda (\bar \epsilon_{AB}\gamma_\mu (Z^B\Psi^{\dagger A}F^D-F^D\Psi^{\dagger A}Z^B)-\bar\epsilon^{AB}
\gamma_\mu (\Psi_AZ^\dagger_BF^D-F^DZ^\dagger_B\Psi_A))\cr
&&-2i\lambda (\bar\epsilon_g^{AD}\chi_{\mu BD}-\bar\epsilon_{gBD}\chi_{\mu}^{AD})(Z^BZ^\dagger_AF^D-F^DZ^\dagger_AZ^B)
\eea
and by matching with the RHS of (\ref{condition}) we obtain 
\bea\label{6}
\delta \hat A_\mu^{L}&=& -\lambda (\bar \epsilon_{AB}\gamma_\mu Z^B\Psi^{\dagger A}-\bar \epsilon^{AB}\gamma_\mu \Psi_A Z^\dagger_B)
-2\lambda(\bar\epsilon_g^{AD}\chi_{\mu BD}-\bar\epsilon_{gBD}\chi_{\mu}^{AD})Z^BZ^\dagger_A\cr
\delta \hat A_\mu^R&=& -\lambda(\bar \epsilon_{AB}\gamma_\mu \Psi^{\dagger A}Z^B-\bar \epsilon^{AB}\gamma_\mu Z^\dagger_B\Psi_A )
-2\lambda(\bar\epsilon_g^{AD}\chi_{\mu BD}-\bar\epsilon_{gBD}\chi_{\mu}^{AD})Z^\dagger_AZ^B\;.
\eea
We can combine this with the abelian factors  
\bea\label{7}
\delta \hat{\hat A}^R_\mu =\delta \hat{\hat A}^L_\mu &=&
\frac{2\pi}{k'}\Big(\bar\epsilon_{AB}\gamma_{\mu}\tr(\Psi^{\dagger A}Z^B)-\bar\epsilon^{AB} \gamma_\mu\tr(\Psi_{A}
Z^\dagger_B)\Big)\cr
&& +\frac{4\pi}{k'}(\bar\epsilon_g^{AD}\chi_{\mu BD}-\bar\epsilon_{gBD}\chi_{\mu}^{AD})\tr(Z^B Z^\dagger_A)\;,
\eea
which, with the definition $k' = k N$, leads to the desired $\U(N)$ gauge fields \cite{Lambert:2010ji}:
\be
 A^{L/R}_\mu = \hat A^{L/R}_\mu -\hat{\hat A}^{L/R}_\mu \one_{N\times N}\;.
\ee

\section{Higgsing and expansion around the VEV solution}\label{Higgsing}

Having recast the theory to our preferred form we now wish to investigate its expansion around a particularly simple background, namely one in which a real  component of the ABJM scalars $Z^A$ develops a VEV, $v$. This triggers the Higgs mechanism  for CS-matter theories introduced in \cite{Mukhi:2008ux}, which turns a parity-invariant pair of CS gauge fields into a single dynamical YM field. In the topologically gauged context there are also chiral CS terms related to the R-symmetry $\U(1)\times \SU(4)$, which include  $C_{\mu}$. These latter fields will also experience a Higgs effect but of a more conventional type. 

During this procedure, following \cite{Aharony:2008ug,Pang:2008hw,Li:2008ya}  for the pure ABJM  case, one would like to keep $v\tfrac{2\pi }{k} = v\lambda  $ fixed, while taking $v\rightarrow \infty$, $\lambda\rightarrow 0$. We will see however that this must be supplemented with a condition on $g_M$, which will be  $g_M v \sqrt N\equiv \bar g_M = \textrm{fixed}$ or $g_M=\tfrac{\bar g_M}{v\sqrt N}\rightarrow 0$. This is the set of limits that will be used for the rest of this section. However, in Section~\ref{interpretation} we will compare them to an alternative set of limits (\ie having a different set of fixed parameters) as mentioned already in the introduction. We work with arbitrary values of $N$ throughout.

For clarity of presentation we split this section into  several parts: We begin with a  review for the higgsing of the pure ABJM theory, then analyze the gravity part of the action and the new features that arise when the gravitational fields are introduced into the covariant derivatives. Finally we address the interactions and supersymmetry transformation rules.

\subsection{M2 to D2 for pure ABJM}

The ABJM model, describing $N$ M2-branes on a $\mathbb C^4/\mathbb Z_k$ singularity, is an $\mathcal N=6$ $\U(N)\times \U(N)$  CS gauge theory with bifundamental matter. However, as discussed in the  previous section, the theory that we here relate to the three-algebra formulation of \cite{Bagger:2008se} is its $\SU(N)\times \SU(N)$ version  \cite{Schnabl:2008wj,Lambert:2010ji}. One  particular example is the $\SU(2)\times \SU(2)$ model that has $\mathcal N=8$ supersymmetry and is equivalent to the BLG-theory, as shown by Van Raamsdonk \cite{VanRaamsdonk:2008ft}. When comparing the situation with and without the two $\U(1)$ factors the higgsing leads to a subtlety  that has not been fully appreciated in the literature and which we explore in the following.

As we will demonstrate later in this section when the full theory is considered, it is possible to find a set of background solutions to the topologically gauged ABJM models parametrized by a real VEV $v$. Thus we would like to see what happens to the lagrangian (\ref{uncoupled})-(\ref{3trace}) under a perturbation schematically of the form
\be
Z^A=v\delta^{A4}+z^A\;,
\ee
with $A=1,...,4$, or more precisely in terms of the real parts\footnote{Here the fields $X^A$ after higgsing are $N\times N$ hermitian matrices which will be expanded in a basis consisting of the unit matrix and a hermitian set of $\SU(N)$ generators.}
\be
Z^A =  v \delta^{A4}\one_{N\times N}+\frac{1}{\sqrt 2}X^A  + i\frac{1}{\sqrt 2} X^{A+4}\;.
\ee

This will cause the theory to break both the $\SU(4)$ R-symmetry and the quiver gauge symmetry
$G\times G$ to the diagonal $G$, with $G$ either $\SU(N)$ or $\U(N)$. In addition there is the abelian gauge symmetry associated with the field $C_{\mu}$ which will mix with the other abelian gauge fields during the higgsing process. These issues will be elaborated upon below when the two different choices for $G$ are discussed one at a time.

As already mentioned, this involves taking the limits $v,k\rightarrow\infty$, with $\tfrac{v}{k}$ fixed. In the spacetime orbifold picture of \cite{Distler:2008mk, Aharony:2008ug}, which is dual to the $\U(N)\times\U(N)$ ABJM model, setting $Z^{1,2,3}=0$ reduces to $\mathbb  C/\mathbb Z_{k\rightarrow \infty}$ with\footnote{The trace parts of the field theory scalars are related to spacetime coordinates by  multiplication with a factor of $l_{11}^{3/2}$.}
\be
 Z^4 \rightarrow  Z^4 e^{\frac{2\pi i}{k}}\simeq  Z^4\Big(1+2\pi i \frac{1}{k}+...\Big)
\simeq Z^4+2\pi i \frac{Z^4}{k}\;.
\ee
Expanding around $Z^4=v+i 0$ with $\frac{v}{k} l_{11}^{3/2}\equiv R$ we obtain that
\be\label{orbifold}
Z^4 l_{11}^{3/2} \rightarrow Z^4 l_{11}^{3/2}+2\pi i R
\ee
is an invariance of the theory, or by decomposing $Z^4=X^4+iX^8$ that $X^8$ is compactified with radius $R$. This is the radius of the M-theory circle.  By letting $\tfrac{v}{k} l_{11}^{3/2} = R\to 0$ one recovers the theory of D2-branes of type IIA string theory in flat space as the direction $X^8$ shrinks to zero.  At low energies this corresponds to an $\mathcal N=8$ SYM theory in 2+1 dimensions, which involves seven, as opposed to eight, scalars. As a result it is natural to expect that the Goldstone modes in the Higgs mechanism, rendering the gauge fields dynamical, should be precisely $X^8$, at least for the diagonal part $X_0^8$ of the general $N\times N$ matrix, corresponding to the center of mass motion of the branes. On the other hand for the BLG-model studied in \cite{Mukhi:2008ux},\footnote{That case corresponds to the $\SU(N)\times   \SU(N)$ model for $N=2$.} which employs real fields, the scalar degree of freedom that disappeared was exactly the one that developed the VEV, as it should due to it not having a complex partner. For our particular choice of VEV this would correspond to $X^4_a$ and $X^4_0$ being singled out as opposed to $X^8_0$ implied by (\ref{orbifold}). This is a sign that the orbifold picture is not an appropriate dual description of the $\SU(N)\times \SU(N)$ theory, which is in line with the fact that there is no known string theory interpretation for ABJM with this choice of gauge group. In fact, we will see that in the $\U(N)\times\U(N)$ case the Goldstone modes that disappear are $X_a^4$ and $X_0^8$, exactly as needed to satisfy both previous pictures.

The pure ABJM lagrangian for both $\SU(N)\times\SU(N)$ and $\U(N)\times\U(N)$ gauge groups is  given by \cite{Terashima:2008sy}:
\bea
S &=& \int d^3 x \Big[\frac{k}{4 \pi} \varepsilon^{\mu\nu\lambda}\tr\Big( A^L_\mu\partial_\nu A^L_\lambda +\frac{2i}{3} A^L_\mu A^L_\nu A^L_\lambda - A^R_\mu\partial_\nu A^R_\lambda -\frac{2i}{3}   A^R_\mu A^R_\nu A^R_\lambda\Big)\cr
&&\qquad \qquad\qquad - \tr \hat D_\mu Z^\dagger_A \hat D^\mu Z^A - i \tr \bar \Psi^{\dagger A}\gamma^\mu \Psi \hat D_\mu\Psi_A - V_{bos}- V_{Yukawa}\Big]\;,
\eea
where
\bea
V_{bos} &=& -\frac{4 \pi^2}{3k^2}\tr\Big( Z^AZ^\dagger_A Z^B Z^\dagger_B Z^C Z^\dagger_C +Z^\dagger_A Z^A  Z^\dagger_B Z^B  Z^\dagger_C Z^C \nn\\
&& + 4 Z^AZ^\dagger_B Z^C Z^\dagger_A Z^B Z^\dagger_C - 6 Z^\dagger_A Z^B  Z^\dagger_B Z^A  Z^\dagger_C Z^C\Big)\\
V_{Yukawa}&=& -\frac{2\pi i}{k}\tr \Big( Z^{\dagger}_ A Z^A \bar \Psi^{\dagger B }\Psi_B - \bar \Psi^{\dagger B } Z^A Z^{\dagger}_ A   \Psi_B - 2 Z^{\dagger}_ A Z^B \bar \Psi^{\dagger A }\Psi_B + 2 \bar \Psi^{\dagger B } Z^A Z^{\dagger}_ B   \Psi_A \cr
&& - \epsilon^{ABCD}Z^\dagger_A \bar\Psi_B Z^\dagger_C \Psi_D+\epsilon^{ABCD}Z^A \bar\Psi^{^\dagger B} Z^C \Psi^{^\dagger D}\Big)
\eea
and
\be
\hat D_\mu Z^A = \partial_\mu Z^A + i A^L_\mu Z^A - i Z^A A^R_\mu\;.
\ee

For the higgsing it is appropriate to define
\be
A^+_\mu = \frac{1}{2}(A^L_\mu +  A^R_\mu) \;, \qquad A^-_\mu = \frac{1}{2}(A^L_\mu - A^R_\mu)\;,
\ee
which, observing that $A^L_\mu Z^A - Z^A A^R_\mu =[A^+_\mu , Z^A]+\{ A^-_\mu, Z^A\}$,  translates into
\bea
\hat D_\mu Z^A &=& D_\mu Z^A+i\{ A^-_\mu, Z^A\}\cr
D_\mu Z^A  &=& \partial_\mu Z^A + i[A^+_\mu , Z^A]\cr
F^+_{\mu\nu} &=& \partial_\mu A^+_\nu +i [A^+_\mu , A^+_\nu]\;.
\eea
Note that the abelian gauge fields do not appear in the covariant derivative $D_{\mu}$.
In terms of these new variables the CS part of the lagrangian becomes
\be
S_{CS}= \int d^3x \frac{k}{2\pi}\varepsilon^{\mu\nu\lambda}\tr \Big( A^-_\mu F^+_{\nu\lambda}+\frac{2i}{3}A^-_\mu A^-_\nu A^-_\lambda\Big)\;.
\ee
The analysis will be performed for the bosonic sector but the fermions follow suit.

\subsubsection{The $\SU(N)\times \SU(N)$ case}

We start by looking at the BLG case in Van Raamsdonk's $\SU(2)\times \SU(2)$  formulation, before generalizing  to $\SU(N)\times \SU(N)$.
The identification between the three-algebra and the conventional gauge theory formulations is (adopting the notation of \cite{VanRaamsdonk:2008ft} to that used in this paper)
\bea
X^I=X_0^I+i\sigma_iX^I_i\;, && \Psi^I=\psi_0^I+i\sigma_i\psi_i^I\cr
A^L_\mu=A_{\mu 4i}^L\sigma_i\;,&& A^R_\mu=A_{\mu 4i}^R\sigma_i\;,
\eea
with
\bea
A_{\mu ab}&=&-\frac{1}{2f}(A_{\mu ab}^L+A_{\mu ab}^R)\;,\cr
A_{\mu ab}^{L(R)}&=&\pm\frac{1}{2}\epsilon_{abcd}A^{L(R)}_{\mu cd}
\eea
and the three-algebra structure constants $f^{abcd}=f\epsilon^{abcd}$, while the $\SU(2)$ generators are normalized by $\tr(\sigma_i\sigma_j)$ $
=2\delta_{ij}$. Note that we choose
the old and new covariant derivatives to differ by a factor of $i$.
Still in real notation ($I=1,...,8$) we thus make the identification:
\bea
&& D_\mu X^{Ia}=\d_\mu X^{Ia}+{f^{abc}}_dA_{\mu bc} X^{Id}\cr
&&D_\mu X^I=\d_\mu X^I+iA^L_\mu X^I-iX^I A^R_\mu\;.
\eea
The generalization to $\SU(N)\times \SU(N)$ was given in \cite{Pang:2008hw}. However,  there it was applied to the $\U(N)\times \U(N)$  ABJM case although the $\U(1)$ factors seem not to have been included.

In order to proceed in the general $\SU(N)\times \SU(N)$ case one has  to expand the complex ABJM scalars as follows:\footnote{We normalize the $\SU(N)$ generators as $\tr(T^aT^b)=\delta^{ab}$, while $T^0 = \one_{N\times N}$.}
\be
Z^A=z^A_0T^0+iz^A_aT^a;\;\;\;
\Psi^A=\psi^A_0T^0+i\psi^A_aT^a;\;\;\;
A_\mu^R=A_\mu^{aR}T_a;\;\;\; A^L_{\mu}=A_\mu^{aL}T_a
\ee
and define the $A_{\mu}^{\pm}$ as above.
Then one can  split the fields into real and imaginary parts, while  giving a VEV to the real scalar component along $T^0 = \one_{N\times N}$:\footnote{Note that compared to \cite{Pang:2008hw} we use a real and not an imaginary VEV. As a result the roles of $X^4$ and $X^8$ will be interchanged in the final result compared to that reference.}
\bea
&&Z^A=\left(\frac{X_0^A}{\sqrt{2}}+ v\delta^{A,4}\right)T^0+i\frac{X_0^{A+4}}{\sqrt{2}}T^0+i\frac{X^A_a}{\sqrt{2}}T^a-
\frac{X_a^{A+4}}{\sqrt{2}}T^a\cr
&&\Psi^A=\psi^A_0T^0+i\psi^A_aT^a+i\psi^{A+4}_0T^0-\psi^{A+4}_aT^a\label{scalarexp}\;.
\eea
With the definitions $[T^a,T^b]=i{f^{ab}}_cT^c, \{ T^a,T^b\}={d^{ab}}_cT^c$ one gets
\bea
&&\hat D_\mu Z^A=\frac{\d_\mu X_0^A}{\sqrt{2}}T^0-\frac{D_\mu X_a^{A+4}}{\sqrt{2}}T^a+\frac{i\d_\mu X^{A+4}_0}{\sqrt{2}}
T^0+\frac{iD_\mu X_a^A}{\sqrt{2}}T^a\nn\\
&&\qquad\qquad
+2ivA_{\mu a} ^-T^a\delta^{A4}+i\sqrt{2}A_{\mu a}^-X_0^AT^a-\sqrt{2}A^-_{\mu a}X_0^{A+4}T^a\nn\\
&&\qquad\qquad-\frac{i}{\sqrt{2}}A_\mu^{-a}d_{abc}T^cX_b^{A+
4}-\frac{1}{\sqrt{2}}A_\mu^{-a}d_{abc}T^c X_b^A\nn\\
&&\qquad\qquad +i[v({B_\mu^A}_4+qC_\mu\delta^{A4})+{\rm subleading}]\;,
\eea
where we note that only the first of the $A^-_{\mu}$ terms involves the VEV $v$. Note also that in the last line we wrote also the $B$ and $C$ terms in the covariant derivative, since they combine with the rest. Indeed, we see that $[v({B^4_\mu}_4+qC_\mu)+\tfrac{1}{\sqrt 2}\d_\mu X_0^8]$, $[v{\rm Re}({B^{A'}_\mu}_4)+\tfrac{1}{\sqrt 2}\d_\mu X_0^{A'+4}]$ and $[-v{\rm Im}({B^{A'}_\mu}_4)+\tfrac{1}{\sqrt 2}\d_\mu X_0^{A'}]$ appear together signalling the Higgs mechanism, as we will discuss in more detail in a later section. However, since in this subsection we look only at the pure ABJM part we will  ignore $B$ and $C$ in the following discussion. Then
\bea
\tr|\hat D_\mu Z^A|^2&=&N\frac{(\d_\mu X_0^A)^2}{2}+N\frac{(\d_\mu X_0^{A+4})^2}{2}
+\left(\frac{(D_\mu X)_c^{A+4}}{\sqrt{2}}\right)^2+\left(2vA_{\mu c}^-\delta^{A4}+\frac{(D_\mu X)_c^A}{\sqrt{2}}\right)^2\cr
&&\qquad +{\rm subleading}\;,
\eea
where all the terms with $A^-_\mu$, except the one shown explicitly, are subleading in the large $v$ limit.
Adding for free a term that gives zero under the integral by the Bianchi identity
\be\label{free}
\frac{k}{2\pi}\varepsilon^{\mu\nu\lambda} \frac{1}{v}\frac{1}{2\sqrt  2} (D_\mu X)_a^4 F^{+a}_{\nu\lambda}
\ee
we get, together with the CS terms,
\bea
S &=& \int d^3x \Big[\frac{k}{2\pi}\varepsilon^{\mu\nu\lambda}\tr \Big( A^-_\mu F^+_{\nu\lambda}+\frac{2i}{3}A^-_\mu A^-_\nu A^-_\lambda\Big)
- \tr |\hat D_\mu Z^A|^2\Big]\cr
&=& \int d^3 x \Big[ \frac{k}{2\pi} \varepsilon^{\mu\nu\lambda}  (A^-_{\mu a}+ \frac{1}{2v}\frac{1}{\sqrt 2}
(D_\mu X)_a^4) F^+_{\nu\lambda a} - (2v A_{\mu a}^- +  \frac{1}{\sqrt 2}(D_\mu X)^4_a)^2\cr
&&\qquad\qquad -   \frac{1}{2}(D_\mu X)_a^{I'} (D^\mu X)_a^{I'} -\frac{1}{2} ND_\mu X_0^I D^\mu X_0^I+
\textrm{higher order}  \Big]\;,
\eea
where $I = 1,...,8$ and $I'=\{I\neq 4\}$.
The higher order terms also include a contribution proportional to  $(A_\mu^-)^3$. However, these terms are  subleading  in the limit $k,v\to \infty$  and can hence be ignored.

Now one can perform a shift involving $X^4_a$:
\be\label{shift}
A^-_{a\mu} \to A^-_{a\mu}  -  \frac{1}{2v}\frac{1}{\sqrt 2} D_\mu X^4_a
\ee
that leads to the simple expression
\bea
S &=& \int d^3 x   \Big(\frac{k}{2\pi} \varepsilon^{\mu\nu\lambda} A^-_{\mu a} F^+_{\nu\lambda a}
-4 v^2 A_{\mu a}^- A_a^{-\mu}-  \frac{1}{2}(D_\mu X)_a^{I'} (D^\mu X)_a^{I'} \cr
&&-\frac{1}{2}N\d_\mu X_0^I\d^\mu X_0^I
+ \textrm{higher order}  \Big).
\eea
In the above the $X_a^4$ component has vanished. This is the Goldstone mode that will render the $\SU(N)$ gauge field $A^+_{\mu a}$ dynamical. Note that the $X_0$'s are decoupled.

We finally integrate out $A^-_{\mu a}$  to obtain
\be
A^{-}_{\mu a} = \frac{k}{16\pi v^2}\varepsilon_{\mu\nu\lambda} F_a^{+\nu\lambda} + \textrm{higher order}
\ee
and upon plugging this into the above expression for the action we get
\be
S = \int d^3 x   \Big(-\frac{k^2}{32\pi^2 v^2} F^{+\mu\nu a}F_{\mu\nu}^{+ a}  -\frac{1}{2}(D_\mu X)_a^{I'} (D^\mu
X)_a^{I'} -\frac{1}{2}N\d_\mu X_0^I\d^\mu X_0^I+ \textrm{higher order}  \Big).
\ee
Then using the definition
\be
\frac{k^2}{32\pi^2 v^2} = \frac{1}{4 g^2_{YM}} \label{gidentif}
\ee
and taking the limit $k,v\to\infty$, with $\tfrac{k}{v} = \textrm{fixed}$, the higher order terms drop out. Furthermore, performing an abelian duality transformation on the free scalar degree of freedom $X^4_0$, one recovers an additional $\U(1)$ that combines with the $\SU(N)$ factor. The result is nothing but the bosonic part for the kinetic terms of $\U(N)$ SYM theory in three dimensions, including all the correct coefficients.

The situation can be summarized by referring to the coset $\SU(N)\times \SU(N)/\SU(N)$ where the group in the denominator is the diagonal subgroup of those in the numerator. By standard Higgs arguments one can use the elements in the coset to rotate \eg the $N\times N$ matrix $X^4$ into its diagonal part $X^4_0$. Due to the  symmetry of the lagrangian the corresponding gauge transformation will trivially cancel out in all non-kinetic terms proving that $X^4_a$ will not appear anywhere in those terms. This will be explicitly demonstrated later in this section. What remains to be analyzed is the kinetic terms which were dealt with above. We now turn to the $\U(N) \times \U(N)$ case where also  the $\U(1)$ factors will experience a Higgs effect.

\subsubsection{The $\U(N) \times \U(N)$ case}

In this case, which is the one that is relevant for the physics of M2-branes,  we will obtain some important differences compared to the previous result.

The expansion of the fields now includes the $\U(1)$ factors for the gauge fields
\bea
Z^A=z^A_0T^0+iz^A_aT^a\;, && \Psi^A=\psi^A_0T^0+i\psi^A_aT^a\,,\cr
A^L_\mu=A^{L0}_{\mu}T^0 + A_\mu^{La}T^a\;, &&  A^R_\mu = A^{R0}_\mu T^0+  A_\mu^{Ra}T^a\,,
\eea
and
\bea
&&Z^A=\left(\frac{X_0^A}{\sqrt{2}}+v\delta^{A,4}\right)T^0+i\frac{X_0^{A+4}}{\sqrt{2}}T^0+i\frac{X^A_a}{\sqrt{2}}T^a-
\frac{X_a^{A+4}}{\sqrt{2}}T^a\,,\cr
&&\Psi^A=\psi^A_0T^0+i\psi^A_aT^a+i\psi^{A+4}_0T^0-\psi^{A+4}_aT^a\,,\label{scalarexp2}
\eea
with the rest of the definitions remaining as for $\SU(N)\times \SU(N)$. As a result one gets some extra terms involving $A_{\mu 0}^-$ in
\bea
&&\hat D_\mu Z^A=\frac{\pd_\mu X_0^A}{\sqrt{2}}T^0-\frac{D_\mu X_a^{A+4}}{\sqrt{2}}T^a+\frac{i\pd_\mu X^{A+4}_0}{\sqrt{2}}
T^0+\frac{iD_\mu X_a^A}{\sqrt{2}}T^a\cr
&&\qquad\qquad
+2ivA_{\mu a} ^-T^a\delta^{A4}+i\sqrt{2}A_{\mu a}^-X_0^AT^a-\sqrt{2}A^-_{\mu a}X_0^{A+4}T^a+A_\mu^{-a}d_{abc}T^cX_b^{A+
4}\cr
&& \qquad\qquad -iA_\mu^{-a}d_{abc}T^c X_b^A +2ivA_{\mu 0} ^-T^0\delta^{A4}+i\sqrt{2}A_{\mu 0}^-X_0^AT^0-\sqrt{2}A^-_{\mu 0}X_0^{A+4}T^0\cr
&& \qquad\quad-\sqrt{2}A_{\mu 0}^-X_a^AT^a-i\sqrt{2}A^-_{\mu 0}X_a^{A+4}T^a+i[v({B_\mu^A}_4+qC_\mu\delta^{A4})+{\rm subleading}]\,.\label{covaderi}
\eea
As in the $\SU(N)\times \SU(N)$ case, we introduced the $B$ and $C$ fields since they combine with 
$X_0^{A'},X_0^{A'+4},X_0^8$ and now additionally with $A_0^-$. Specifically, we have 
$[v({B^4_\mu}_4+qC_\mu+A_{0\mu}^-)+\tfrac{1}{\sqrt{2}}\d_\mu X_0^8]$, $[v{\rm Re}({B^{A'}_\mu}_4)+\tfrac{1}{\sqrt{2}}\d_\mu X_0^{A'+4}]$ and $[-v{\rm Im}({B^{A'}_\mu}_4)+\tfrac{1}{\sqrt{2}}\d_\mu X_0^{A'}]$ appearing together, once again signalling the Higgs mechanism. However the  $B,C$ fields will be dropped for the moment and dealt with in later subsections. Then we obtain
\bea
\tr|\hat D_\mu Z^A|^2&=&N\frac{(\pd_\mu X_0^A)^2}{2}
+\left(\frac{(D_\mu X)_c^{A+4}}{\sqrt{2}}\right)^2+\left(2vA_{\mu c}^-\delta^{A4}+\frac{(D_\mu X)_c^A}{\sqrt{2}}\right)^2\cr
&&+N\left(\frac{1}{\sqrt 2}\pd_\mu X_0^{A+4}+ 2 v A^-_{\mu 0}\delta^{A4}\right)^2+{\rm subleading}\;.
\eea
Adding for free the two terms, of which the second is new compared to the $\SU(N)\times \SU(N)$ case,
\be
\frac{k}{2\pi}\varepsilon^{\mu\nu\lambda} \frac{1}{v}\frac{1}{2\sqrt  2} (D_\mu X)_a^4 F^{+a}_{\nu\lambda}+\frac{Nk}{2\pi}\varepsilon^{\mu\nu\lambda} \frac{1}{v}\frac{1}{2\sqrt  2} (\pd_\mu X_0^8) F^{+0}_{\nu\lambda}
\ee
and with the inclusion of the CS terms, we get that the lagrangian is
\bea
S &=& \int d^3x \Big[\frac{k}{2\pi}\varepsilon^{\mu\nu\lambda}\tr \Big( A^-_\mu F^+_{\nu\lambda}+\frac{2i}{3}A^-_\mu A^-_\nu A^-_\lambda\Big)
- \tr |\hat D_\mu Z^A|^2\Big]\cr
&=& \int d^3 x \Big[ \frac{k}{2\pi} \varepsilon^{\mu\nu\lambda}  \left(A^-_{\mu a}+ \frac{1}{2v}\frac{1}{\sqrt 2}
(D_\mu X)_a^4\right) F^{+a}_{\nu\lambda }+\frac{Nk}{2\pi} \varepsilon^{\mu\nu\lambda} \left(A^-_{\mu 0}+ \frac{1}{2v}\frac{1}{\sqrt 2}
(\pd_\mu X)_0^8\right) F^{+0}_{\nu\lambda } \cr
&&\qquad\qquad- \left(2v A_{\mu a}^- +  \frac{1}{\sqrt 2}(D_\mu X)^4_a\right)^2-
N\left(\frac{1}{\sqrt 2}\pd_\mu X_0^{A+4}+ 2 v A^-_{\mu 0}\delta^{A4}\right)^2\cr
&&\qquad\qquad -   \frac{1}{2}(D_\mu X)_a^{I'} (D^\mu X)_a^{I'} -\frac{1}{2}N \pd_\mu X_0^A \pd^\mu X_0^A+
\textrm{higher order}  \Big]\;,
\eea
where $I'=\{I\neq 4\}$. The higher order terms will once again include   $(A_\mu^-)^3$, which will lead to subleading expressions in the limit $k,v\to \infty$.

Note that in this case we can perform a double shift; both  the $A^-_{\mu a}$ shift of (\ref{shift}) as well as one for the new abelian component $A^-_{\mu 0}$:
\be
A^-_{\mu a} \to A^-_{\mu a}  -  \frac{1}{2v}\frac{1}{\sqrt 2} (D_\mu X)^4_a \qquad\textrm{and}\qquad A^-_{\mu 0} \to A^-_{\mu 0}  -  \frac{1 }{2v}\frac{1}{\sqrt 2} (\pd_\mu X^8_0)\label{higgseating}
\ee
and that leads to
\bea\label{above}
S &= &\int d^3 x   \Big(\frac{k}{2\pi} \varepsilon^{\mu\nu\lambda} (A^-_{\mu a} F^{a+}_{\nu\lambda }+N A^-_{\mu 0} F^{+}_{\nu\lambda \;0})
-4 v^2 A_{\mu a}^- A_a^{-\mu}-4 N v^2  A_{\mu 0}^- A_0^{-\mu}\cr
&&\qquad -  \frac{1}{2}(D_\mu X)_a^{I'} (D^\mu X)_a^{I'} -\frac{1}{2}N\pd_\mu X_0^{\tilde I'} \pd^\mu X_0^{\tilde I'} + \textrm{higher order}  \Big)\;,
\eea
where $\tilde I'=\{I\neq 8\}$.

It is important to observe that in the above both the $X_a^4$ and the $X^8_0$ components vanish to leading order, as opposed to only $X^4_a$ for $\SU(N)\times\SU(N)$.  These fields make up the Goldstone modes that render, respectively, $A^+_{\mu a}$ and $A^+_{\mu 0 }$ dynamical to give back a $\U(N)$ gauge field.  This situation corresponds to the coset $\U(N)\times \U(N)/\U(N)$.  Without the vanishing of the $X_0^8$ one would have ended up with excessive degrees of freedom compared to the initial theory.  This result is also in line with the spacetime orbifold picture, as it should apply to the trace component $X_0^8$, as opposed to the non-trace $X_a^4$.

We can now integrate $A^-_{\mu }$ out to obtain
\be\label{integrate}
A^-_\mu = \frac{k }{16\pi v^2}\varepsilon_{\mu\nu\lambda} F^{+\nu\lambda} + \textrm{higher order}
\ee
and upon plugging into the above this gives rise to
\be
S = \int d^3 x  \Big[ \tr \Big(-\frac{k^2 }{32\pi^2 v^2} F^{+\mu\nu}F_{\mu\nu}^+\Big)  -  \frac{1}{2}(D_\mu X)_a^{I'} (D^\mu X)_a^{I'} -\frac{1}{2}N\pd_\mu X_0^{\tilde I'} \pd^\mu X_0^{\tilde I'} + \textrm{higher order}  \Big]\;.
\ee
As for $\SU(N)\times \SU(N)$, along with the definition (\ref{gidentif}) for the YM coupling
and taking the limit $k,v\to\infty$, with $\frac{v}{k} \to \textrm{fixed}$, the higher order terms drop out and upon combining the remaining traceless part of $X_a^8$ with the  trace part of $X_0^4$ this is nothing but the bosonic part for the kinetic terms of $\U(N)$, three-dimensional SYM theory.

Similarly to the discussion at the end of the previous subsection, the Higgs phenomenon can in this case be related to the coset $\U(N)\times \U(N)/\U(N)$ where global factors in the definition of $\U(N)$ are not considered. Thus a full $N\times N$ matrix is involved in the higgsing which we saw above corresponded to $X_0^8$ and $X_a^4$. When gravity is added to this picture the situation becomes a bit more interesting, as we will discuss next.

\subsection{Higgsing including the gravity sector}\label{backgrounds}

We will next investigate what happens when the gravitational sector is added to the theory. Explicitly, the bosonic subsector of the lagrangian relevant for this analysis is\footnote{The other set of limits discussed in this paper in Section~\ref{interpretation} are based on the same lagrangian but with an extra overall factor of $g_M^2$.}
\be\label{grav}
L=g_M^{-2}L_{CS(\tilde\omega)}-e\partial Z^A\partial \bar Z_A -\tfrac{e}{8}R\vert Z\vert^2-eV_{pot}(Z,\bar Z)\,,
\ee
where $V_{pot}$ is the full bosonic potential. We remind the reader that apart from the one-trace term coming from the original ABJM lagrangian the potential also contains double- and triple-trace terms arising from the topological gauging. The terms in the potential are scaled in such a way that they pick up one factor of $g_M^2$ for each trace beyond the first, following the relevant discussion in Section~\ref{3andbi}. 

We note that  due to  the gauge-gravity interaction terms the introduction of the VEV, $Z^A = v\delta^{A4} + z^A$, leads to a nonzero cosmological term in the lagrangian
\be
V_{pot}(v)=V_{pot}^{3-trace}(v)=\tfrac{1}{4^4}N^3 g_M^4v^6\;.
\ee
It is also immediately clear that the third term in the above  lagrangian will lead to an Einstein-Hilbert term that will play an important role later.

\subsubsection{Background: solution,  decoupling limit}\label{background}

From (\ref{grav}) one obtains the following equations of motion, by  varying with respect to the scalars and  the metric respectively:
\bea
&&\Box Z^D -\frac{1}{8}RZ^D+
\frac{5}{4\cdot 64}g_M^4|Z|^4Z^D-\frac{g_M^4}{32}Z^D\tr(Z^AZ^\dagger_C)\tr(Z^CZ^\dagger_A)\nn\\
&&-\frac{g_M^4}{16}|Z|^2Z^A\tr(Z^DZ^\dagger_A)+\frac{g_M^4}{16}Z^A\tr(Z^BZ^\dagger_A)\tr(Z^DZ^\dagger_B)=0\label{newscalareq}
\eea
and
\bea\label{einsteincotton}
&&-e(\partial_{\mu}Z^A\partial_{\nu}\bar Z_A-\tfrac{1}{2}g_{\mu\nu}\partial^{\rho}Z^A\partial_{\rho}\bar Z_A)+
g_M^{-2}{\mathcal C}_{\mu\nu}+\tfrac{e}{2}g_{\mu\nu}V_{pot}\cr
&&-\tfrac{e}{8}(R_{\mu\nu}-\tfrac{1}{2}g_{\mu\nu}R)\vert Z\vert^2+\tfrac{e}{8}(\nabla_{(\mu}\nabla_{\nu)}\vert Z\vert^2-g_{\mu\nu}\Box\vert Z\vert^2)=0\;,
\eea
where
\be
\mathcal C_{\mu\nu} \equiv \epsilon_\mu^{\phantom{\mu}\alpha\beta}\nabla_\alpha (R_{\beta\nu}- \frac{1}{4}g_{\beta\nu}R)\label{Cottontensor}
\ee
is the Cotton tensor.

It is now straightforward to verify that a constant VEV $v$ solves the scalar equations of motion: The $D=1,2,3$ and the imaginary part of the $D=4$ equation are satisfied by the VEV, while
the real part of the $D=4$ equation gives
\be\label{R1}
R^{BG}=-\tfrac{3}{32}(g_Mv\sqrt N)^4\,,
\ee
which suggests an AdS background geometry. Although this looks like an Einstein equation, it comes from the {\em scalar} equations of motion, so {\it a priori} it is a constraint on top of the graviton field equations which we next check for consistency.

For AdS solutions the Cotton tensor vanishes, since the geometry is conformal, reducing the gravitational field equation to 
\be\label{BGeom}
\tfrac{1}{4}(R_{\mu\nu}-\tfrac{1}{2}g_{\mu\nu}R)\vert_{AdS}\,Nv^2 =V_{pot}(v)g_{\mu\nu}\;,
\ee
which in turn gives 
\be\label{R2}
R_{\mu\nu}^{BG}=2\Lambda g_{\mu\nu}\qquad\textrm{with}\qquad\Lambda=-\tfrac{1}{64}(g_Mv\sqrt N)^4\;.
\ee
The trace gives exactly (\ref{R1}), as expected for $\textrm{AdS}_3$ with $R = 6 \Lambda$.

Since one should be able to recover the pure ABJM lagrangian for {\em generic} values of the ABJM fields and parameters but {\em small perturbations} around the (AdS) gravitational background, we would next like to understand whether it is possible to decouple gravity from the gauge theory in the lagrangian (\ref{uncoupled})-(\ref{3trace}) using the gravitational coupling $g_M$. 

This decoupling limit is in fact  $g_M\rightarrow 0$ with all other parameters fixed. In this limit the terms in the lagrangian that couple gravity to matter become weak and the background becomes flat, as seen from (\ref{R1}).  The double-trace and triple-trace interaction terms in (\ref{3trace}) vanish trivially. Since the $R^{BG}|Z|^2$ background term vanishes, and  we  are only considering perturbations much smaller than the background, $R|Z|^2$ also vanishes in the limit.\footnote{We note however that it was essential that we only considered small perturbations around the background, otherwise for generic perturbations 
and generic nonzero VEV the $R|Z|^2$ term gives gravity with finite gravitational coupling $\kappa^2\sim \tfrac{1}{v}$.} The remaining single-trace interaction terms involve at least one gravitino. By rescaling the Rarita-Schwinger action to the canonical form through multiplying the gravitini by $g_M$ cancelling the $\tfrac{1}{g_M^2}$ coefficient, all the single-trace interaction terms pick up $g_M$ factors and vanish as well.  Therefore $g_M\rightarrow 0$  decouples gravity completely and we are left with the pure ABJM lagrangian in a flat background.

In summary, the VEV  is indeed a consistent solution of the equations of motion, giving an anti-de Sitter space as a background. The limit  $g_M\rightarrow 0$, with all other parameters fixed and for small perturbations around the background, decouples gravity from the ABJM part of the action.

\subsubsection{Higgsed supergravity action and the chiral point}\label{relations}

We now  write down the relevant bosonic part of the higgsed gravitational lagrangian, keeping in mind that this is in the presence of the  $\textrm{AdS}_3$ background we have just found. This can be readily determined and reads
\bea\label{N6sugra}
L_{sugra} =-e\frac{v^2N}{8}(R-2\Lambda) +\frac{1}{2g_M^2}\epsilon^{\mu\nu\rho}
\Tr_{\alpha}(\tilde\omega_{\mu}\partial_{\nu}\tilde\omega_{\rho}+ \frac{2}{3}\tilde\omega_{\mu}\tilde\omega_{\nu}\tilde\omega_{\rho}) \;.
\eea

In order for it to be compatible with the ABJM part of the higgsing process the above is valid in the limit $v\to\infty, \lambda\to 0$, with $v\lambda = \textrm{fixed}$. However, in the presence of gravity one needs to  additionally supplement the former with a condition on their  scalings relative to the coupling $g_M$. This can be seen as follows: In Eq.~(\ref{R1}) we discovered that the curvature for the background solution was given by
\be
R^{BG}=-\frac{3}{32}g_M^4v^4N^2\label{rvalue}\nn\;.
\ee
For a background space with fixed and finite curvature,
the RHS of this equation must not scale with $v$. This can be achieved by letting $g_M\to 0$ in such a way that 
\be
\bar g_M^2= g_M^2 v^2 N\label{gsrescaling}= \textrm{fixed}.
\ee
For the background this means that, since $R=6\Lambda$,
\be
\Lambda= -\frac{1}{64}\bar g_M^4. \label{rvalue2}
\ee
The Newton constant can also be read off from the above lagrangian:
\be
\kappa^2=\frac{8}{v^2N}\,.
\ee
If we then adopt the definition of the constant $\mu$ in front of the gravitational CS-term as done in the work of Li, Song and Strominger \cite{Li:2008dq} we have $\kappa^2\mu=g_M^2$ and hence
\be
\mu=\frac{\bar g_M^2}{8}\,.
\ee
Finally, expressing the cosmological constant $\Lambda$ in terms of a length $\ell$ as $\Lambda=-\frac{1}{\ell^2}$ one  finds that\footnote{Here we define $\ell$ to be positive. The sign of $\ell$ does not enter the bosonic part of the theory, but is related to the chirality of the supersymmetries when the fermionic sector is included.}
\be
\mu \ell= 1\;,
\ee
\ie, as was first noticed in \cite{Chu:2009gi}, the theory sits at a `chiral point' in the sense of \cite{Li:2008dq}. Note, however, that the signs of the Einstein-Hilbert term and the cosmological term in (\ref{N6sugra}) are opposite to the ones used in \cite{Li:2008dq}, although they coincide with the ones normally used in TMG. We will come back to this point in Section~\ref{TMG}.

The limit with $\bar g_M^2 = \rm{fixed}$ looks similar to the weak gravity limit discussed at the end of the previous section although now one has $v\rightarrow \infty$, $g_M\rightarrow 0$, as well as a nontrivial gravitational background.  As a result it is not the same as the trivial limit that decouples gravity and gives back pure ABJM.  In fact, we will next find that at the end of the higgsing process one obtains finite interaction terms between gravity and ABJM in addition to the finite curvature gravitational piece. We will however postpone a full discussion of the physics in the gravity sector post-higgsing to Section~\ref{interpretation}.

\subsection{The interaction terms}\label{interaction}

We now turn to the interaction terms of the full action. These do not show any qualitative differences  between $\SU(N)\times\SU(N)$ and $\U(N)\times\U(N)$ so for the remainder of this section we will assume the latter, which is the case of direct physical interest.

For the initial purpose of finding the leading contribution in $v$ we can refrain from using the full expansion  (\ref{scalarexp}), as this is rather involved and in fact some of the terms might end up canceling. Instead we first perform  the expansion around the VEV $v$ using the expression
\be
Z^A = v\delta^{A,4} + z^A\;
\ee
and  derive the result for all the interaction  terms as an order-by-order expansion in $v$ .

\subsubsection{Bosonic terms}\label{bosonic}

The bosonic terms can be organized as single-, double- and triple-trace. The single-trace 6-scalar structures appear only in the ABJM part of the potential $V_6$,
\bea
-V_6&=&\tr(Z^AZ^\dagger_ZZ^BZ^\dagger_BZ^CZ^\dagger_C+Z^\dagger_ZZ^AZ^\dagger_BZ^BZ^\dagger_CZ^C\cr
&&+4Z^AZ^\dagger_BZ^CZ^\dagger
_AZ^BZ^\dagger_C-6Z^AZ^\dagger_BZ^BZ^\dagger_AZ^CZ^\dagger_C),
\eea
which was already analyzed in \cite{Pang:2008hw}. Here we simply wish to observe that all the terms scaling like $v^6,...,v^3$ vanish and that one is left with a
potential which is of order $v^2\lambda^2\propto g^2_{YM}$, and hence fourth order in the scalar fields, as expected \cite{Pang:2008hw}. The remaining terms are subleading in $v$ and will vanish in the $v\rightarrow \infty$ limit. The term surviving in the limit is
\be
- V_6\to-\frac{ g_{YM}^2}{4}\tr \Big([X^{I'},X^{J'}][X^{J'}, X^{I'}]\Big)
\ee
but since this term is part of the ordinary Yang-Mills theory it will not be mentioned further  when the scalar field interaction terms are discussed below.

For the rest of the interaction terms the results can be obtained using the formulas from  Appendix~\ref{expansion}. For the double-trace expressions we find that the $v^6, v^5$ and $v^4$ terms are identically zero, while for lower orders we obtain 
\bea
-V_{2-trace}&=&g_M^2\frac{\lambda}{8}|Z|^2\tr(Z^\dagger_DZ^CZ^\dagger_C Z^D-Z^\dagger_DZ^DZ^\dagger_CZ^C)\cr
&&+g_M^2\frac{\lambda}{2}\tr(Z^DZ^\dagger_B)\tr(Z^\dagger_DZ^BZ^\dagger_CZ^C-Z^\dagger_DZ^CZ^\dagger_CZ^B)\cr
&=& g_M^2\lambda\Big[\frac{1}{4}v^3N\tr((z^4+\zd_4)[\zd_D,z^D])+\frac{1}{2}v^3N\tr((z^4+\zd_4)[\zd_4,z^4])\cr
&&-\frac{v^2N}{8}\tr(\zd_Dz^D\zd_Cz^C-z^D\zd_Dz^C\zd_C)+\frac{v^2N}{2}\tr(\zd_4z^4\zd_Cz^C-z^4\zd_4z^C\zd_C)\cr
&&+\frac{v^2}{4}\tr(z^4-\zd_4)\tr((z^4-\zd_4)[\zd_D,z^D])
+\frac{v^2}{2}\tr\; z^D\tr(\zd_D[\zd_C,z^C]+\zd_D[z^4,\zd_4])\cr
&&+\frac{v^2}{2}\tr\;\zd_D\tr(z^D[\zd_C,z^C]+z^D[z^4,\zd_4])\Big]+{\cal O}(v)\,.
\eea
Here we should note that  the coefficients for the leading terms in $v$  are $\sim g_M^2\l v^3$ which is just $\bar g_M^2 g_{YM}$ and thus fixed in the scaling limit. As in the previous case, any remaining contributions are of  lower order in $v$ and will vanish as $v\rightarrow \infty$.

For the triple-trace part the situation is more intricate.  In this case no terms will vanish identically and, due to the special properties of the set of limits used here, we must keep not only the terms that are multiplied by parameters fixed under this scaling but also terms that blow up. These latter terms will determine the background geometry and are crucial for the interpretation of the results.  We find 
\bea
-V_{3-trace}&=&g_M^4\Big[\frac{5}{12\cdot 64}|Z|^6-\frac{1}{32}|Z|^2\tr(Z^AZ^\dagger_C)\tr(Z^CZ^\dagger_A)\cr
&&+\frac{1}{48}
\tr(Z^AZ^\dagger_C)\tr(Z^BZ^\dagger_A)\tr(Z^CZ^\dagger_B)\Big]\cr
&=&g_M^4\Big\{-\frac{1}{256}(v^2N+v\tr(z^4+\zd_4))^3-\frac{3}{256}\tr(z^A\zd_A)(v^2N+v\tr(z^4+\zd_4))^2\cr
&&+v^2N\Big[\frac{5}{256}(\tr(z^A\zd_A))^2-\frac{1}{32}\tr(z^A\zd_B)\tr(z^B\zd_A)-\frac{1}{32}\tr(z^4\zd_4)
\tr(z^A\zd_A)\cr
&&+\frac{1}{16}\tr(z^4\zd_A)\tr(\zd_4z^A)\Big]
+\frac{v^2}{16}\Big[\tr \zd_A(\tr z^B \tr(z^A\zd_B)\cr
&&-\tr z^4 \tr(z^A\zd_4))+\tr z^A(\tr \zd_A\tr(z^4\zd_4)
-\tr \zd_4 \tr(z^4\zd_A))\Big]\Big\}+{\cal O}(v)\label{3tracepart}\;.
\eea

There appear three types of terms that survive the scaling limit and must be analyzed: First, the cosmological term is blowing up since it is multiplied by $g_M^4v^6\sim \bar g_M^4 v^2$ but will nevertheless result in a fixed AdS geometry (also the EH term is $\sim v^2$). Second, there are mass terms $\sim \bar g_M^4$ that do not scale. In addition there are terms linear in the scalar fields which are $\sim \bar g_M^4 v$. Fortunately this latter kind of term arises also from $R|Z^2|$  and as we will see shortly the two contributions cancel each other.

Collecting the results, the  leading terms in $v$ are
\bea
-V_B&=&-\frac{g_M^2v^3N \lambda}{2\sqrt{2}}f^{abc}X_a^{A+4}X_b^A X_c^8-\frac{1}{256}g_M^4v^3N^3(v+\sqrt{2}X_0^4)^3\cr
&&-\frac{3}{2\cdot 256}g_M^4v^2N^2(v+\sqrt{2}X_0^4)^2(NX_0^IX_0^I+X_a^IX_a^I)+{\cal O}(v^2)
\label{higherorder}\;.
\eea
The remaining  terms of ${\cal O}(v^2)$ are presented in Appendix~\ref{subleading}. As we have already explained under Eq.~(\ref{3trace}), all terms with $f^{abc}$  also come multiplied by a power of $\lambda=\tfrac{2\pi}{k}$.  This is sensible, since one has from (\ref{gidentif}) that
\be
g_{YM}=\frac{2\pi}{k}v\sqrt{2}=\lambda v \sqrt{2}\;.
\ee

At this point the reader might be alarmed by the fact that the leading contribution from the non-ABJM interaction terms is $\mathcal O (v^3)$, while the one for ABJM is $\mathcal O(v^2)$. However, one needs to remember that it is the quantity $\bar g_M=g_M v$ from (\ref{gsrescaling}) that we would like to keep fixed in our limit. Additionally, it is necessary to perform the rescaling
\be
X_0^I=\frac{\tilde X_0^I}{\sqrt{N}}\, ,\;\;\;\;
\psi_0^I=\frac{\tilde \psi_0^I}{\sqrt{N}}
\ee
in order to make these of the same order as $X_a^I,\psi_a^I$. After doing so  we obtain that $V_{2-trace,v^2}\sim \mathcal O(\lambda)\rightarrow 0$, while
$V_{3-trace,v^2}\sim \mathcal O(g_M^2)\rightarrow 0$ and indeed the nonzero terms come from the $\mathcal O (v^3)$ contributions, giving
\bea\label{358}
-V_B&=&-\frac{\bar g_M^2}{4}g_{YM}f^{abc}X_a^{A'}X_b^{A'}X_c^8-\frac{\bar g_M^4v^2N}{256}-\frac{3\sqrt{2}}{256}\bar g_M^4v\sqrt{N}\tilde
X_0^4-\frac{3}{128}\bar g_M^4(\tilde X_0^4)^2\nn\\
&&-\frac{3}{2\cdot 256}\bar g_M^4(\tilde X_0^I\tilde X_0^I+X_a^IX_a^I)+{\rm subleading}\;.
\eea
In the summary of the final results in Section~\ref{explicit}, these terms appear together with the related terms arising from the expansion of the conformal scalar-curvature term:
\be\label{359}
-\frac{R}{8}|Z|^2=-\frac{R}{8}\left[Nv^2+v\sqrt{N}\sqrt{2}\tilde X_0^4+\frac{\tilde X_0^I\tilde X_0^I}{2}+\frac{X_a^IX_a^I}{2}\right].
\ee
It is important to note here that there are unwanted terms linear in the scalar field $\tilde X^4_0$ in both of \eqref{358} and \eqref{359}. However,  by using the background curvature relation (\ref{rvalue2}) the last equation reads
\be
-\frac{R_{BG}}{8}|Z|^2=\frac{3}{256}\bar g_M^4\left[Nv^2+v\sqrt{N}\sqrt{2}\tilde X_0^4+\frac{\tilde X_0^I\tilde X_0^I}{2}+\frac{X_a^IX_a^I}{2}\right]
\ee
and we find that these linear terms cancel in an expansion around the AdS background. This  also affects the values of scalar masses and the  bosonic interaction terms as seen from 
\be
-V_B-\frac{R_{BG}}{8}|Z|^2=\frac{1}{128}\bar g_M^4v^2N-\frac{3}{128}\bar g_M^4(\tilde X_0^4)^2
-\frac{\bar g_M^2}{4}g_{YM}f^{abc}X_a^{A'+4}X_b^{A'}X_c^8+{\rm subleading}\label{bosonicHiggs}\;.
\ee

As noted already in the previous subsection, the first term in this result is related to the cosmological constant which is proportional to 
$ \bar g_M^4 v^2N$
and thus diverges in the limit used here.\footnote{The second scaling procedure described in the next section gives rise to a finite cosmological constant as well as a freely tunable Newton constant.} The remaining terms are all finite as they should.
Moreover, there is no dependence on  $X_a^4, X_0^8$, exactly as in the pure ABJM part, while $X_0^4$ is now massive. However, in addition to this pure ABJM-like result there is also no dependence on any of the $X_0$ scalars except $X_0^4$. This indicates that in this topologically gauged case there is an extended Higgs effect at work, where the local $\SU(4)$ R-symmetry has also  been broken. The coset $\SU(4)/\SU(3)$ indeed accounts for the missing seven scalars.

In conclusion, we obtain  an AdS background geometry that is strongly enforced: Both the Einstein  and the cosmological constant terms  are  multiplied by $v^2$, which diverges in the limit used here. Apart from this fact, all the interaction terms between gravity and ABJM are finite.

\subsubsection{Fermionic terms}

We quote the leading fermionic interaction terms from (\ref{fermionterms}), while referring the reader who is interested in the details  to Appendix~\ref{subleading}:
\bea
-V_F&&=-\frac{2 \sqrt 2}{\kappa} iA\bar{ f}^{\mu A4}\gamma_\mu (\tilde \psi_0^A+i\tilde \psi_0^{A+4})
-\frac{2 \sqrt 2}{\kappa}\frac{i}{16}\bar g_M^2eA\bar{\chi}_{\mu 4A}\gamma^\mu(\tilde\psi_0^A-i\tilde\psi_0^{A+4})+c.c.\cr
&&+\frac{i\bar g_M^2}{16}e\Big[3\bar{\tilde{\psi}}_0^{I''}\tilde\psi_0^{I''}+\bar\psi_a^{I''}\psi_a^{I''}\Big]
+\frac{i\bar g_M^2}{16}e\Big[5(\bar{\tilde{\psi}}_0^4\tilde\psi_0^4+\bar{\tilde{\psi}}_0^8
\tilde\psi_0^8)-3(\bar\psi_a^4\psi_a^4+\bar\psi_a^8\psi_a^8)\Big]\cr
&&+\frac{4i\bar g_M^2}{\kappa^2}\bar{ f}^\mu_{AB} \chi_\mu^{AB}+\frac{i\bar g_M^2}{8\kappa^2}\epsilon^{\mu\nu\rho}\bar{\chi}_{\nu AB}\gamma_\rho
 \chi_\mu^{AB}+{\rm subleading}\;,
\label{fermionicHiggs}
\eea
with $I'' = \{I\neq 4,8\}$. These contributions are also all finite. We note that $\tilde \psi_0^{I''},\psi_a^{I''},\tilde\psi_0^{4,8}$, 
$\psi_a^{4,8}$ as well as the gravitini have acquired mass terms and that in extracting the leading contributions we have used $A^-_\mu\sim \tfrac{F_\mu^+}{v^2}$ from (\ref{integrate}). Also note that the kinetic term for the gravitini has been included in this expression and that, after the breaking to the ${\bf 3}+\bar{\bf 3}$ of the R-symmetry group $\SU(3)$, all six gravitini are still AdS-massless. This fact was incorporated into the supergravity lagrangian quoted in the introductory section, which was claimed to have $\mathcal N=6$ supersymmetry.

\subsubsection{The $B^A_\mu{}_B,C_\mu$ terms}

The gravity sector also contains the fields $B_{\mu}^A{}_B$ and $C_{\mu}$, which are used to gauge the global symmetries $\SU(4)_R$ and $\U(1)$ respectively. Although $C_{\mu}$ does not couple to any of the other gravitational fields including $B_{\mu }^A{}_B$ \cite{Chu:2009gi}, it is natural to discuss both of these two chiral CS fields in the gravity context since both groups are obtained when breaking $\SO(8)$.

On top of their kinetic terms, in the presence of a scalar VEV there are also mass terms coming from the ABJM piece, $-e\tr[\tilde D_\mu Z^A\tilde D^\mu \bar Z_A]$. We have already seen in (\ref{covaderi}) that the fields $B_{\mu}^A{}_B$ and $C_{\mu}$ appear in the covariant derivative as 
\bea
\tilde DZ^4&=&i[v({B^4_\mu}_4+qC_\mu+2A_{0\mu}^-)+\frac{1}{\sqrt{2}}\d_\mu X_0^8]+...\cr
\tilde DZ^{A'}&=&[-v{\rm Im}({B^{A'}_\mu}_4)+\frac{1}{\sqrt{2}}\d_\mu X_0^{A'}]+i[v{\rm Re}({B^{A'}_\mu}_4)
+\frac{1}{\sqrt{2}}\d_\mu X_0^{A'+4}]+...\;.
\eea
Then, in the absence of a  potential for $X_0^{A'},X_0^{A'+4}, X_0^8$, one can indeed implement the usual Higgs mechanism to have the fields $X_0^{A'}$ and $X_0^{A'+4}$ `eaten' through the redefinition
\bea
&&{\rm Re}({B^{A'}_\mu}_4)+\frac{1}{\sqrt 2 v}\d_\mu X_0^{A'+4}\rightarrow{\rm Re}({B^{A'}_\mu}_4)\cr
&&{\rm Im}({B^{A'}_\mu}_4)-\frac{1}{\sqrt 2 v}\d_\mu X_0^{A'}\rightarrow{\rm Im}({B^{A'}_\mu}_4)\label{ba4redef}\;.
\eea
The gauge fields ${B^{A'}_\mu}_4$ obtain the usual mass terms, while $X_0^{A'},X_0^{A'+4}$ disappear from the action. 

On the other hand, $X_0^8$ was already `eaten' by $A^-_{0\mu}$  in (\ref{higgseating}), which allowed for the $\U(1)$ part of the nondynamical $\U(N)\times \U(N)$  CS field to turn into the $\U(1)$ part of the dynamical $\U(N)$  Yang-Mills field. However, in that pure gauge theory discussion the effects of $B_{\mu}^A{}_B$ and $C_{\mu}$ were ignored. Accounting for them involves first redefining
\be
{B^4_\mu}_4+qC_\mu+2A^-_{0\mu}=\tilde A_{0\mu}^-\label{bcredef}
\ee
so that 
\be\label{bceating}
\tilde A_{0\mu}^--\frac{1}{\sqrt{2} v}\d_\mu X_0^8\to \tilde A_{0\mu}^-\;.
\ee
This means that the relevant terms in the action (\ref{above}) become
\bea
&&\frac{kN}{2\pi} \varepsilon^{\mu\nu\lambda}  A^-_{\mu 0} F^{+0}_{\nu\lambda }
-4 N v^2  (\tilde A_{\mu 0}^-  +\frac{1}{\sqrt{2}v}\d_\mu X_0^8)^2\cr
&\to&\frac{kN}{4\pi} \varepsilon^{\mu\nu\lambda}  \tilde A^-_{\mu 0} F^{+0}_{\nu\lambda }
-4 N v^2  \tilde A_{\mu 0}^- \tilde A_0^{-\mu}-\frac{kN}{4\pi}\varepsilon^{\mu\nu\lambda}(qC_\mu+{B^4_\mu}_4)F^{+0}_{\nu
\lambda}\;.
\eea
As was the case before the $\tilde A_{0\mu}^-$ terms give the $\U(1)$ part of the Yang-Mills action, while the extra one is
\be
-\frac{kN}{4\pi}\varepsilon^{\mu\nu\lambda}(qC_\mu+{B^4_\mu}_4)F^{+0}_{\nu
\lambda}=-\frac{2\sqrt{N}}{g_{YM}\kappa}\varepsilon^{\mu\nu\lambda}(qC_\mu+{B^4_\mu}_4) F^{+0}_{\nu\lambda}\;.
\ee
Rewriting everything in terms of $\kappa$ and $\mu$, we finally get the $B,C$ action
\begin{multline}
\frac{1}{\kappa^2\mu^2}\int d^3 x \Big[-\frac{\mu}{2}\varepsilon^{\mu\nu\rho}C_\mu \d_\nu C_\rho
-2\mu\varepsilon^{\mu\nu\rho}\Tr_A
(B_{\mu}\partial_{\nu}B_{\rho}+\frac{2}{3}B_{\mu}B_{\nu}B_{\rho})\\-8\mu^2 |{B^{A'}_\mu}_4|^2
-\frac{4\sqrt{N\mu}\kappa^2 \mu^2}{g_{YM}}\varepsilon^{\mu\nu\lambda}(qC_\mu+{B^4_\mu}_4)F^{+0}_{\nu\lambda}
\Big]\;,
\end{multline}
where the $\SU(4)$ (anti-)hermitian traceless field ${B_\mu^A}_B$ has been split into a massless
$\SU(3)$ (anti-)hermitian traceless part ${B_\mu^{A'}}_{B'}$, the massless $3\times 3$ trace $iB_{0\mu}\equiv \sum_{A'}{B^{A'}_\mu}_{A'}=-{B_\mu^4}_4$, and the 
massive complex vector ${B_\mu^{A'}}_4$ (which ate the $X_0^{A'},X_0^{A'+4}$ fields). Note that ${B_\mu^4}_{A'}$ is related to ${B_\mu^{A'}}_4$ by the (anti-)hermiticity condition.

Under this decomposition the CS term for the $\SU(4)$ field ${B_\mu^A}_B$ splits into $\SU(3)$ components as
\bea
-\frac{2}{\mu}\varepsilon^{\mu\nu\rho}\Tr_{A}
(B_{\mu}\partial_{\nu}B_{\rho}+\frac{2}{3}B_{\mu}B_{\nu}B_{\rho})&=&-\frac{2}{\mu}\varepsilon^{\mu\nu\rho}\Tr_{A'}
(B_{\mu}\partial_{\nu}B_{\rho}+\frac{2}{3}B_{\mu}B_{\nu}B_{\rho})
\cr 
&&-\frac{4}{\mu}\varepsilon^{\mu\nu\rho}B^{A'}_{\mu\,4} \check{D}_{\nu}B^4_{\rho\,A'} +\frac{2}{\mu}\varepsilon^{\mu\nu\rho}B_{0\mu}\partial_{\nu}B_{0\rho}\;,
\eea
where
\be
\check{D}_{\nu}B^4_{\rho\,A'} \equiv \pd_{\nu}B_\rho^{4}{}_{A'} + i B_{0\nu} B_\rho^{4}{}_{A'}-\frac{2}{3}B_\nu^{B'}{}_{A'}B_\rho^{4}{}_{B'}\;.
\ee
It is evident that the $\SU(4)$ R-symmetry has been broken to $\SU(3) \times \U(1)$. At the same time one combination  of the abelian vector fields becomes massive. This field is still denoted $A_{\mu}^+$ although it is a sum of several fields. Hence there is  just one scalar degree of freedom being absorbed in the higgsing of the abelian sector. Note that despite the appearance of the coupling term between $F^+$ and $(qC+B)$ there is no further breaking of the $\U(1)$ R-symmetry. The counting of degrees of freedom is then easily understood in terms of the $6+N^2$  dimensional coset $\SU(4)\times \U(1)\times \U(N) \times \U(N)/\SU(3)\times \U(1)\times \U(1) \times \U(N)$.

The above discussion  relied on the fact that massive CS-theories in three dimensions have been analyzed in detail  and seen to have one degree of freedom. Thus the action for the ${B_\mu^{A'}}_4$ field is the `self-dual action in odd dimensions' of \cite{Townsend:1983xs}.  For an abelian field $A_\mu$ one could use the transformation of \cite{Deser:1984kw}, which allows to express an abelian self-dual field as a Maxwell-CS (topologically massive) one, \ie
 \be
 \frac{1}{\kappa^2\mu^2}\int d^3 x\Big[-\frac{1}{4} F_{\mu\nu}^2-\frac{\mu}{2}\varepsilon^{\mu\nu\rho}A'_\mu \d_\nu A'_\rho\Big]\;,
 \ee
where $F_{\mu\nu}=\d_{[\mu}A'_{\nu]}$. 

For nonabelian, coupled fields like our ${B_\mu^{A'}}_4$ this transformation cannot be performed \cite{Townsend:1983xs2}.\footnote{The action for ${B_\mu^{A'}}_4$ contains the terms   ${B_\mu^{A'}}_4{B_\nu^{4}}_{B'}{B_\rho^{B'}}_{A'}$.} However, there is nothing unusual with having self-dual actions in odd dimensions in the context of supergravity. In fact, such an action appears \eg in 7 dimensional gauged supergravity  and it was shown in \cite{Nastase:1999cb,Nastase:1999kf} that it can arise by dimensional reduction of the action of 11 dimensional supergravity.\footnote{In odd dimensions the usual Klein-Gordon kinetic   operator splits into two self-dual parts of opposite relative sign
 \be
 \Box -m^2\sim (\epsilon   \d +m)(\epsilon \d-m)\;,
\ee 
which means that a field with the Klein-Gordon equation of motion   $(\Box -m^2)A=0$ can be split, by further adding an auxiliary field $B$ with action $B^2$, into two   equations of motion of self-dual type
 \be
 (\epsilon\d +m)P=0\;\;\;{\rm and}\;\;\;   (\epsilon\d-m)Q=0\;.
 \ee
These equations are in turn coming from an action of the type $\sim [\epsilon P\d P+mP^2-\epsilon Q\d   Q+mQ^2]$. In this sense, the self-dual kinetic operator is the `square root' of the Klein-Gordon  one.}

\subsection{Supersymmetry rules}

We finally turn our attention to the effect of the higgsing on the supersymmetry transformation rules. Since both the $\U(N)\times \U(N)$ gauge fields $A_\mu$ and  the ${B_\mu^{A'}}_4$ fields have `eaten' scalars to become dynamical, it is important  to clarify how the transformations rules organize the physical states into ${\cal N}=6$ supermultiplets. The answer to this question will, however, require additional information about the exact form of the special superconformal transformations leaving the unhiggsed theory invariant. 

As we will see below there is an interplay between the supersymmetry and special conformal supersymmetry transformations that needs to be understood in order to determine the form of the supersymmetry transformations that are left unbroken  after higgsing of the theory. The introduction of a VEV with non-zero mass dimension breaks the conformal and dilation dependent symmetries, in particular the special superconformal symmetry. However, this does not imply that the supersymmetry that remains after higgsing coincides  exactly with that called supersymmetry in the conformal theory. Instead, in the analysis below we find several facts that point towards one and the same solution to the question of how to define supersymmetry after higgsing. This solution involves forming a particular linear combination of the original supersymmetry and the special conformal supersymmetry transformations.

\subsubsection{The effect of the VEV on the supersymmetry of the original fields}

In this subsection we will first derive the rules in terms of the original fields, \ie before the `eating' has been implemented. In fact, this latter step would mean identifying the final supermultiplets and that we will not be able to do in this work since we lack a full control of the special superconformal transformations in the ABJM theory coupled to conformal gravity, \ie of the topologically gauged theory. The special superconformal transformations are easily constructed in both the pure ${\cal N}=6$ superconformal gravity theory, as demonstrated in \cite{Gran:2008qx,Chu:2009gi}, as well as in the flat space ABJM theory \cite{Bandres:2008ry}. However, the validity of these expressions have not yet been verified in the coupled theory.

By  replacing the Higgs ansatz in the supersymmetry rules (\ref{1})-(\ref{4}) we obtain
\bea
\delta e_\mu^\a&=&i \bar \epsilon_{gAB}\gamma^\a\chi_\mu ^{AB}\nn\\
\delta\chi_\mu^{AB}&=&\tilde D_\mu \epsilon_g^{AB}\nn\\
\delta B_\mu^A\;
&=&\frac{i}{e}(\bar{f}^{\nu AC}\gamma_\mu \gamma_\nu \epsilon_{g BC}-\bar{f}^\nu_{BC} \gamma_\mu\gamma_\nu \epsilon_g^{AC})\cr
&&+\frac{i}{4 \kappa}\Big[\delta^{A4}\bar\epsilon_{BD}\gamma_\mu (\tilde \psi_0^D-i\tilde \psi_0^{D+4})+\bar \epsilon_{B4}\gamma_\mu (\tilde
\psi_0^A-i\tilde\psi_0^{A+4})\cr
&&-\delta_{B4}\bar\epsilon^{AD}\gamma_\mu
(\tilde\psi_0^D+i\tilde\psi_0^{D+4})-\bar\epsilon^{A4}\gamma_\mu(\tilde\psi_0^B+i\tilde\psi_0^{B+4})\Big]\cr
&&-\frac{4i}{\kappa^2}(\bar\epsilon_g^{AC}\chi_{\mu 4C}\delta_{B4}-\bar \epsilon_{gBC} \chi_\mu^{4C}\delta^{A4})
+\frac{i}{\kappa^2}(\bar \epsilon_g^{AD}\chi_{\mu BD}-\bar\epsilon_{gBD} \chi_\mu ^{AD})\cr
\delta C_\mu&=&-vq(\bar \epsilon_{A4}\gamma_\mu \tr(\Psi^{\dagger A})-\bar \epsilon^{A4}\gamma_\mu \tr(\Psi_A))
-\frac{16}{\kappa^2}q(\bar\epsilon_g^{4D}\chi_{\mu 4D}-\bar \epsilon_{g4D} \chi_\mu ^{4D})\cr
&=&-\frac{2q}{\kappa}\Big[\bar\epsilon_{A4}\gamma_\mu (\tilde \psi_0^A-i\tilde \psi_0^{A+4})-
\bar\epsilon^{A4}\gamma_\mu (\tilde\psi_{0}^A+i\tilde\psi_{0}^{A+4})\Big]\cr
&&-\frac{16}{\kappa^2}q(\bar\epsilon_g^{4D}\chi_{\mu 4D}-\bar \epsilon_{g4D} \chi_\mu ^{4D})\cr
\delta(\tilde X_0^A+i\tilde X_0^{A+4})&=&i\bar \epsilon^{AB}(\tilde \psi_0^B+i\tilde\psi_0^{B+4})\cr
\delta( X_a^A+iX_a^{A+4})&=&i\bar\epsilon^{AB}(\psi_a^B+i\psi_a^{B+4})\;.
\eea

For $\Psi_B$ of (\ref{5}) and the combined $A_\mu^{L/R}$ of (\ref{6})-(\ref{7}) we only write the new terms (gravitational interaction terms)
\bea
\delta\Psi_B-\delta_0\Psi_B 
&=&\frac{\bar g_M^2 v}{16}\epsilon_{4B}+\frac{\bar g_M^2}{4\sqrt{2}}\delta_{B4}\Big[\frac{(\tilde X_0^A+i\tilde X_0^{A+4})}{\sqrt{N}}
(\epsilon_{A4}+\epsilon_{4A})+(iX_a^A-X_a^{A+4})T^a\epsilon_{4A}\Big]\cr
&&+\frac{\bar g_M^2}{16\sqrt{2}}\Big[\frac{(\tilde X_0^A+i\tilde X_0^{A+4})}{\sqrt{N}}\epsilon_{AB}+\frac{2\tilde X_0^4}{\sqrt{N}}\epsilon_{AB}
+(iX_a^A-X_a^{A+4})\epsilon_{AB}T^a\Big]\cr
\delta A_\mu^{L}-\delta_0 A_\mu^{L}&=&
\delta A_\mu^{R}-\delta_0 A_\mu^{R}=
\frac{8i}{\kappa}\Big[-\frac{ \tilde X_0^{A+4}}{\sqrt{N}}+X_a^AT^a\Big]
(\bar \epsilon_g^{4D}\chi_{\mu AD}-\bar \epsilon_{g AD} \chi_\mu ^{4D})\;,
\eea
noting that in the first equation the interaction term $iA\gamma^\mu \epsilon_{AB}\bar\chi^{BD}\Psi_D$ vanishes under rescaling and higgsing. Expanding  $\Psi_B$ we get
\bea
\delta(\tilde \psi_0^B+i\tilde \psi_0^{B+4})-\delta_0(\tilde \psi_0^B+i\tilde \psi_0^{B+4})&=&\frac{\sqrt{2N}\bar g_M^2}{16}\Big(v+\frac{\sqrt{2}\tilde X_0^4}{\sqrt{N}}\Big)
\epsilon_{4B}+\frac{\bar g_M^2}{16}(\tilde X_0^A+i\tilde X_0^{A+4})\epsilon_{AB}\cr
\delta(i\psi_a^B-\psi_a^{B+4})-\delta_0(i\psi_a^B-\psi_a^{B+4})&=&\frac{\bar g_M^2}{16}(iX_a^A-X_a^{A+4})(\epsilon_{AB}+4\epsilon_{4A}\delta_{B4})\;.
\label{fermivaria}
\eea
In the above, $\delta_0$ denote the  set of supersymmetry transformations for the pure ABJM theory. Pre-higgsing these were given by
\bea
\delta_0\Psi_A&=& -\gamma_\mu \epsilon_{AB}\tilde D_\mu Z^B+\lambda(-\epsilon_{AB}(Z^CZ^\dagger_CZ^B-Z^BZ^\dagger_CZ^C)+2\epsilon_{CD}Z^CZ^\dagger_A
Z^D)\cr
\delta_0 A^{L}_\mu&=&\lambda(\bar\epsilon_{AB}\gamma_\mu Z^B \Psi^{A\dagger} -\bar\epsilon^{AB}\gamma_\mu \Psi_A Z^\dagger_B)\cr
\delta_0 A^{R}_\mu &=& \lambda(\bar\epsilon_{AB}\gamma_\mu\Psi^{A\dagger}Z^B -\bar\epsilon^{AB}\gamma_\mu Z^\dagger_B\Psi_A)\nn\;.
\eea
Post-higgsing, the $v$ contribution cancels between the two terms in the gauge field transformations, hence
\bea
\delta_0 A^{L}_\mu&=&\lambda(\bar\epsilon_{AB}\gamma_\mu z^B \Psi^{A\dagger} -\bar\epsilon^{AB}\gamma_\mu \Psi_A z^\dagger_B)\nn\\
\delta_0 A^{R}_\mu &=& \lambda(\bar\epsilon_{AB}\gamma_\mu\Psi^{A\dagger}z^B -\bar\epsilon^{AB}\gamma_\mu z^\dagger_B\Psi_A)\;.
\eea
For the fermions we have
\bea
\delta_0 \Psi_A&=&iv\gamma_\mu \epsilon_{BA}[2\delta^{B4}A_\mu^-+({B^B_\mu}_4+qC_\mu \delta^{B4})]
-\gamma_\mu \epsilon_{AB} \tilde D_\mu z^B\cr
&&+\sqrt{2}g_{YM}(\delta_{A4}\epsilon_{CD}z^Cz^D+\epsilon_{C4}[z^C,z^\dagger_A])
-\frac{g_{YM}}{\sqrt{2}}(\epsilon_{A4}[z^C,z^\dagger_C]+\epsilon_{AB}[z^4+z^\dagger_4,z^B])\cr
&=&\Big[\frac{2 \sqrt 2 i}{\kappa \sqrt{N}}({ B^{B}_\mu\;}_4-{ B^4_\mu\;}_4\delta^{B4})
+\frac{i}{2\sqrt{2}g_{YM}}\varepsilon_{\mu\nu\rho}F^{+\nu\rho}\delta^{B4}\Big]\gamma^\mu\epsilon_{BA}\cr
&&-\frac{1}{\sqrt 2}\gamma_\mu \epsilon_{AB'} \d_\mu [\frac{1}{\sqrt N}(X_0^{B'} + i X_0^{B'+4}) + (iX_a^{B'} - X_a^{B'+4})T^a]\cr
&&-\frac{1}{\sqrt 2}\gamma_\mu \epsilon_{A4} \d_\mu [\frac{1}{\sqrt N}X_0^4  - X_a^{8}T^a]\cr
&&+\sqrt{2}g_{YM}[\frac{1}{4}\delta_{A4}\epsilon_{CD}(iX_a^C-X_a^{C+4})(iX_b^D-X_b^{D+4})\cr
&&+\frac{1}{2}\epsilon_{C4}(iX_a^C-X_a^{C+4})(-iX_b^A-X_b^{A+4})]{f^{ab}}_cT^c\cr
&&-\frac{g_{YM}}{\sqrt{2}}(\epsilon_{A4}X_a^CX_b^{C+4}+i\epsilon_{AB}
(iX_a^B-X_a^{B+4})X_b^8{f^{ab}}_cT^c\;.\label{psiafermi}
\eea

This last set of higgsed pure ABJM transformations should be compared to  the ones for three-dimensional SYM in the low-energy limit of $N$ D2-branes in flat space.

\subsubsection{Supersymmetry for redefined higgsed fields}

We can now use \eqref{higgseating}, (\ref{ba4redef}) and (\ref{bceating}) to obtain the transformations for the higgsed fields after the `eating' redefinition. Since the $A^-_{\mu a}$ and $ A_{\mu 0}^-$ fields get integrated out, the bosonic on-shell degrees of freedom are contained in $X^{I'}_a$, $X^4_0$,  $A^+_{a\mu}$, $A^+_{0\mu}$ and $B^{A'}_\mu{}_4$. The above redefinitions affect the supersymmetry rules through the change
\be
 {\delta B^{A'}_\mu}_4-\frac{1}{\sqrt 2 v}\d_\mu \delta(X_0^{A'}+i X_0^{A'+4})={\delta \hat B^{A'}_\mu}_4\;
\ee
and one also needs to drop the $X_0^{A'},X_0^{A'+4},X_0^8$ transformation rules altogether.\footnote{For notational simplicity  we will drop the hat on ${B^{A'}_\mu}_4$ after the 
redefinition, with the hope that this will not cause any confusion.} The transformations for the bosonic fields then read
\bea
\delta e_\mu^\a&=&i \bar \epsilon_{gAB}\gamma^\a \chi_\mu ^{AB}\cr
\delta\chi_\mu^{AB}&=&\tilde D_\mu \epsilon_g^{AB}\cr
\delta {B_\mu^{A'}\;}_{B'}
&=&\frac{i}{e}(\bar{f}^{\nu A'C}\gamma_\mu \gamma_\nu \epsilon_{g B'C}-\bar{f}^\nu_{B'C} \gamma_\mu\gamma_\nu \epsilon
_g^{A'C})+\frac{i}{\kappa^2}(\bar \epsilon_g^{A'D}\chi_{\mu B'D}-\bar\epsilon_{gB'D} \chi_\mu ^{A'D})\cr
\delta {B_\mu^{A'}\;}_4
&=&\frac{i}{e}(\bar{f}^{\nu A'C}\gamma_\mu \gamma_\nu \epsilon_{g 4C}-\bar{f}^\nu_{4C} \gamma_\mu\gamma_\nu \epsilon
_g^{A'C})+\frac{\kappa\sqrt N}{4}\d_\mu \Big[\bar \epsilon^{AB}(\tilde \psi_0^B+i\tilde\psi_0^{B+4})\Big]\cr
&&+\frac{i}{4\kappa}\Big[-\bar\epsilon^{A'D}\gamma_\mu (\tilde\psi_0^D+i\tilde\psi_0^{D+4})-\bar\epsilon^{A'4}\gamma_\mu(\tilde\psi_0^4+i\tilde\psi_0^{8})\Big]\cr
&&-\frac{4i}{\kappa^2}\bar\epsilon_g^{A'C}\chi_{\mu 4C}+\frac{i}{\kappa^2}(\bar \epsilon_g^{A'D}\chi_{\mu 4D}-\bar\epsilon_{g4D} \chi_\mu ^{A'D})\cr
\delta B_{0\mu}\equiv i\delta {B^4_\mu\;}_4
&=&-\frac{1}{e}(\bar{f}^{\nu 4C}\gamma_\mu \gamma_\nu \epsilon_{g 4C}-\bar{f}^\nu_{4C} \gamma_\mu\gamma_\nu \epsilon
_g^{4C})\cr
&&-\frac{4}{\kappa}\Big[\bar\epsilon_{4D}\gamma_\mu (\tilde \psi_0^D-i\tilde \psi_0^{D+4})-\bar\epsilon^{4D}\gamma_\mu 
(\tilde\psi_0^D+i\tilde\psi_0^{D+4})\Big]\cr
&&-\frac{1}{\kappa^2}(\bar \epsilon_g^{4D}\chi_{\mu 4D}-\bar\epsilon_{g4D} \chi_\mu ^{4D})\cr
\delta C_\mu&=&-\frac{2q}{\kappa}\Big[\bar\epsilon_{A4}\gamma_\mu (\tilde \psi_0^A-i\tilde \psi_0^{A+4})-
\bar\epsilon^{A4}\gamma_\mu (\tilde\psi_{0}^A+i\tilde\psi_{0}^{A+4})\Big]\cr
&&-\frac{16}{\kappa^2}q(\bar\epsilon_g^{4D}\chi_{\mu 4D}-\bar \epsilon_{g4D}\chi_\mu ^{4D})\cr
\delta\tilde X_0^4&=&i\bar \epsilon^{4B'} \tilde \psi_0^{B'}\cr
\delta X_a^{A'} &=&i\bar\epsilon^{A'B}\psi_a^{B}\cr
\delta X_a^{A+4}&=&i\bar\epsilon^{AB}\psi_a^{B+4}\cr
\delta A_{0\mu}^+ &= &-\frac{ 8}{\kappa\sqrt{N}} \tilde X_0^{A+4}
(\bar \epsilon_g^{4D}\chi_{\mu AD}-\bar \epsilon_{g AD}\chi_\mu ^{4D})\cr
&& + \frac{\lambda}{2}(\bar\epsilon_{AB}\gamma_\mu \frac{1}{N}\tr[\{z^B, \Psi^{A\dagger}\}] -\bar\epsilon^{AB}\gamma_\mu \frac{1}{N}
\tr[\{\Psi_A, z^\dagger_B\}])\cr
\delta A_{a\mu}^+ &= & \frac{8}{\kappa} X_a^A
(\bar \epsilon_g^{4D} \chi_{\mu AD}-\bar \epsilon_{g AD} \chi_\mu ^{4D})\cr
&& + \frac{\lambda}{2}(\bar\epsilon_{AB}\gamma_\mu \{z^B, \Psi^{A\dagger}\}_a -\bar\epsilon^{AB}\gamma_\mu \{\Psi_A, z^\dagger_B\}_a) \;.
\eea
The above set of transformations, along with the ones for the fermionic fields presented in the previous subsection,\footnote{Note that the term $\frac{\bar g_M^2}{16}(\tilde X_0^A+i\tilde   X_0^{A+4})\epsilon_{AB}$ in the fermion variation (\ref{fermivaria}) is the one needed to cancel   the variation of the background term $-\frac{1}{8}R_{BG}|Z|^2$. Also, the $X_0^{A'},X_0^{A'+4}$ terms in   (\ref{psiafermi}) get absorbed in ${B^{A'}_\mu\;}_4$ by the `eating' redefinition.} should leave the final action invariant. The situation should be completely analogous to the case of pure ABJM in the limit of large $v$, where the supersymmetry transformations post-higgsing can be seen to exactly reduce to the ones for three-dimensional SYM.

However, the more intriguing properties of the above transformation rules are found by looking at the transformation rules for the spinors. The gravitini transformation differs from the one expected in a theory with an AdS background in that the covariant derivative in $\delta\chi_\mu^{AB}=\tilde D_\mu \epsilon_g^{AB}$ does not contain a term proportional to a three-dimensional gamma matrix.
Such a term appears however in the special superconformal transformations of the pure ${\cal N}=6$ supergravity theory \cite{Gran:2008qx, Chu:2009gi} and by combining this with the original supersymmetry transformations given above the correct form of the variation of the gravitini could be obtained. By correct we mean that the resulting Killing spinor equation should have six solutions corresponding to the number of expected supersymmetries in the AdS background.

There is another fact pointing in the same direction, namely the variation of the spinor field from the ABJM theory. As seen above, the ordinary supersymmetry in the conformal theory has picked up a constant shift after higgsing. Such a term would normally indicate that supersymmetry is broken and that the fermion in question is a Goldstone fermion. However, in our case we can eliminate this constant shift by adding a special superconformal transformation in the same way as described in the previous paragraph. 

The conclusion is that the supersymmetry that survives the higgsing of the superconformal theory is a linear combination of the supersymmetry $Q$ and the special conformal supersymmetry $S$ in the superconformal symmetry algebra of the unhiggsed theory, \ie the original topologically gauged ABJM theory. To make this explicit, the special superconformal transformations identified in \cite{Gran:2008qx, Chu:2009gi} read
\be
\delta\chi_{\mu AB}=\gamma_{\mu}\eta_{AB}
\ee
and the corresponding transformation for the ABJM spinor field is \cite{Bandres:2008ry}
\be
\delta\Psi_A=Z^B\eta_{AB}\,,
\ee
which in fact is the unique expression for this transformation.
The arguments presented above translate into the following result:
\be
Q(\epsilon_{AB}^{\textrm{AdS}})=Q(\epsilon_{AB}^{\textrm{conf}})+S(\eta_{AB}^{\textrm{conf}}= \frac{\bar g_M^2}{16}\epsilon_{AB}^{\textrm{conf}})\,.
\ee
Recalling now that $\frac{\bar g_M^4}{64}=\frac{1}{\ell^2}$ we conclude that the supersymmetry transformation of the gravitini in  AdS becomes
\be
\delta\chi_{\mu AB}=(\tilde D_{\mu}+\frac{1}{2\ell}\gamma_{\mu})\epsilon_{AB}\,,
\ee
where we have chosen the positive root of $\bar g_M^4$. This choice will be further justified in Section~\ref{TMG}. With this variation of the gravitini we get a Killing spinor equation that has six solutions, that is, the theory has six supersymmetries as expected (see \eg \cite{Duff:1986hr}). This result also fits well with the corresponding formula in ${\cal N}=1$ chiral supergravity \cite{Becker:2009mk}.

We end this section with an observation on the multiplet structure. Before performing the higgsing, we started with two interacting ${\cal N}=6$ supermultiplets: the ABJM one, with 8 bosonic and 8 fermionic on-shell degrees of freedom, and the topological conformal supergravity one, with no on-shell degrees of freedom. After the higgsing, due to the R-symmetry fields of the topological multiplet eating scalars from the abelian part of the ABJM multiplet, we have a single entangled ${\cal N}=6$ supermultiplet with the same on-shell degrees of freedom as the original ABJM multiplet.\footnote{In the language of the simple WZ analogy, after the   redefinition one of the scalar degrees of freedom in the WZ multiplet gets turned into a degree   of freedom for the gauge field in the vector multiplet.}  We are not aware of such examples being discussed in the literature, so we will not try to fit this final multiplet into a general analysis. The fact that the various fields in this multiplet have different masses is typical of AdS backgrounds \cite{Kim:1985ez,Duff:1986hr}.

\section{Interpreting the supergravity action}\label{interpretation}

Let us summarize our findings thus far: After higgsing the topologically gauged ABJM theory we obtained an ${\cal N}=6$ action for three-dimensional SYM/CS theory coupled to supergravity. The gravity sector is interesting since it consists of an EH kinetic term plus a cosmological constant on top of the conformal gravity part that we started off with. As we will discuss shortly this is a supersymmetric extension of the chiral gravity of \cite{Li:2008dq} coupled to a SYM/CS theory.

We begin with a few more comments on the effects of the nontrivial gravitational background on the physics. 
In general dimension $d$, a term in the action of the type
\be
\frac{1}{16\pi G}\int (R-2\Lambda)
\ee
leads to Einstein equations
\be
R_{\mu\nu}-\frac{1}{2}g_{\mu\nu}R=-\Lambda g_{\mu\nu}\;.
\ee
Tracing the above we get
\be
R=\frac{2d}{d-2}\Lambda\;,
\ee
which  gives rise to the relation 
\be
\Lambda=-\frac{(d-1)(d-2)}{2\ell^2}
\ee 
between  the $\textrm{AdS}_d$ radius $\ell$ defined by $R_{\mu\nu}{}^{\rho\sigma}=-\frac{1}{\ell^2}(\delta_{\mu}^{\rho}\delta_{\nu}^ {\sigma}-\delta_{\mu}^{\sigma}\delta_{\nu}^{\rho})$ and  the cosmological constant.

Because of the presence of the gravitational CS part in the three-dimensional theory studied here, the  Einstein equations will in general also involve the Cotton tensor $\mathcal C_{\mu\nu}$, as in  (\ref{einsteincotton}), which however vanishes on the conformally flat  $\textrm{AdS}_3$ background. Moreover, since then $R=6\Lambda$ one has from (\ref{R1}) that (recalling the definition $\bar g_M^2=g_M^2v^2\,N$)
\be
\Lambda =-\frac{1}{\ell^2}=-\frac{\bar g_M^4}{64} = \textrm{fixed}\label{Lambda}\;,
\ee
which means we are in a finite curvature AdS space and confirms our result (\ref{R2}). 

In $\textrm{AdS}_d$ space the Breitenlohner-Freedman (BF) bound for scalars allows  to a certain extent for a negative mass squared. The bound relates the mass to the curvature through
\be
m^2\geq -\frac{1}{\ell^2}\frac{(d-1)^2}{4}\;,
\ee
which in our case of $d=3$ gives
\be\label{mu}
m^2\geq -(\frac{\bar g_M^2}{8})^2 \equiv - \mu^2\
\ee
and we would like to note the condition $\mu \ell = 1$ found in the previous section. Hence the BF bound is always respected by the scalars in the higgsed theory, since they are either massless or massive ($\tilde X_0^4$) but never `tachyonic', that is, never have a negative mass squared.

\subsection{Comparison with the chiral gravity  and chiral ${\cal N}$=1 supergravity literature}\label{TMG}

By adding to the lagrangian (\ref{N6sugra}) the necessary fermionic terms coming from the third line of (\ref{fermionicHiggs}), we recover the full pure gravity sector of the $\mathcal N=6$ action, which was given explicitly in (\ref{finalsugra}). Let us however restrict ourselves to the graviton plus single gravitino terms  and put the rest of the fields to their VEVs. This leads to an $\mathcal N=1$ supergravity which can be expressed as
\bea
{\cal L}&=&\Big[-e \frac{v^2N}{8}(R-2\Lambda)+\frac{v^2N}{8}\frac{1}{2\mu}\Tr_\alpha(\tilde\omega\wedge d\tilde\omega+\frac{2}{3}\tilde\omega\wedge\tilde \omega\wedge \tilde\omega)\Big]\nn\\
&&-2i \frac{v^2N}{8}\Big[\frac{e}{2\mu}\tilde D_\mu \bar\chi_\nu\gamma^{\rho\sigma}\gamma^{\mu\nu}\tilde D_\rho\chi_{\sigma }
-\epsilon^{\mu\nu\rho}\bar \chi_\mu^{}\Big(\tilde D_\nu +\frac{1}{2\ell}\gamma_\nu\Big)\chi_{\rho }\Big]\;.
\eea

A similar action was derived  by Becker {\it et al.} in \cite{Becker:2009mk} as  a (1,0) supersymmetric extension of the chiral pure gravity case considered in \cite{Li:2008dq}. The lagrangian of  \cite{Becker:2009mk} reads in our conventions\footnote{Reference \cite{Becker:2009mk} uses a different   conventions for the fermions: $\bar\psi=\psi^\dagger \gamma_0=-\psi^\dagger \gamma^0$ and for   $\epsilon^{012}=-1$, while our conventions are $\bar \psi=\psi^\dagger \gamma^0$ and   $\epsilon^{012}=+1$.} 
\bea
{\cal L}^{\prime}&=&\frac{1}{16\pi G}\Big[e(R-2\Lambda)+\frac{1}{2\mu}\Tr_\alpha(\tilde\omega\wedge d\tilde\omega+\frac{2}{3}\tilde\omega\wedge \tilde\omega\wedge\tilde \omega)\Big]\nn\\
&&-\frac{i}{16\pi G}\Big[\frac{e}{2\mu}\tilde D_\mu \bar\chi_\nu \gamma^{\rho\sigma}\gamma^{\mu\nu}\tilde D_\rho\chi_{\sigma}
+\epsilon^{\mu\nu\rho}\bar \chi_\mu^{} \Big(\tilde D_\nu -\frac{1}{2\ell}\gamma_\nu\Big)\chi_{\rho}]\;.
\eea
One immediately sees that by rescaling the gravitino by a factor of $1/\sqrt 2$, and identifying 
\be
\frac{1}{\kappa^2}\equiv \frac{1}{16\pi G}=\frac{v^2N}{8}\label{kappa}
\ee
the two actions are  identical up to some signs. The relative sign between the covariant derivative and the gamma matrix differ in the two cases but this is just a matter of whether we have $(6,0)$ or $(0,6)$ supersymmetry and is not important  here.

Another sign  difference with important physical implications is, however, expected in view of an observation made already in \cite{Chu:2009gi} relšated to the sign of $R$. This sign was taken to be $+$ in \cite{Li:2008dq,Becker:2009mk}, which is opposite to the usual conventions in the literature on TMG and the one appearing in the theory discussed here. The purpose of choosing the $+$ sign was to have positive energy BTZ black holes but negative energy massive gravitons. However, \cite{Li:2008dq} made probable that when the theory sits at the chiral point $\mu\ell =1$ the massive gravitons become zero energy states and one is led to a more sensible theory with a conjectured chiral-CFT dual.  This point of view generated some controversy in the literature which is summarized in \cite{Maloney:2009ck}, where the reader can find an extensive list of references. In fact, it is still a matter of debate whether the boundary dual to the above theory is a chiral or logarithmic CFT, depending on the subtle choice of gravitational boundary conditions discussed also in \cite{Skenderis:2009nt,Skenderis:2009kd,Grumiller:2010rm}.

Our theory on the other hand is a TMG theory which appears at the chiral point $\mu \ell=1$  as an automatic consequence of higgsing the topologically gauged ABJM theories. This suggests a reversal of the signs of the energies for the gravity states and thus leads to zero energy gravitons together with  negative energy (unstable) black holes, as is customary from the three-dimensional TMG literature. These black holes are potentially unpleasant and attempts have been made to argue that they can never be formed in physical processes, see for instance \cite{Deser:2010df}. Chirality might in our case, as in the pure chiral gravity case, eliminate the propagating gravitational bulk modes although this analysis has not been carried out for the $\mathcal N=6$ supersymmetric situation found here.  In any case,  our construction provides an $\mathcal N=6$ generalization of previous  models  \cite{Li:2008dq,Becker:2009mk}  and a setup for investigating these issues from other angles.

A final question concerns  the supersymmetry transformation rules for the gravitini and the form  of the AdS covariant derivative. Recall that we concluded in the previous section that after the VEV is introduced we get
\be
\delta\chi_\mu^{AB}= \Big(\tilde D_\mu+\frac{1}{2\ell}\gamma_\mu\Big) \epsilon_g^{AB},
\ee
appropriate for a gravitational theory with a (negative) cosmological constant. However, the relative sign between the covariant derivative and the gamma matrix was  determined by the analysis of the transformation rules in the previous section and must be consistent with  the terms in the action containing the same kind of AdS combination of a derivative and gamma matrix. 

First, of course, the sign is given by the gravitini kinetic terms presented in the beginning of this subsection and we find also in this case a relative $+$ sign between the covariant derivative and the gamma matrix.
There are, however, other terms containing generalized AdS derivatives. By
looking at our full $\mathcal N=6$ lagrangian, we see that there exists a mixing term coming from the first line of (\ref{fermionicHiggs}) and involving the spin-$\frac{1}{2}$ fermions:
\be
\Big[\bar{  f\,}^{\mu A4}-e\frac{ \bar g_M^2 }{16}\bar{ \chi\,}^{\mu A4}\Big]\gamma_\mu (\psi_0^A+i\psi_0^{A+4})\;,
\ee
which can be rewritten by spinor flipping and seen to contain
\be
\gamma^\mu\Big[ f_{\mu}^{AB}-\frac{ e}{2\ell} \chi_{\mu}^{AB}\Big] =\frac{e}{2}\gamma^{\nu\mu}\Big(\tilde D_\nu+\frac{1}{2\ell}\gamma_\nu
\Big)\chi_{\mu}^{ AB},
\ee
thus confirming that the relative sign is +. Having firm evidence of the form of the supersymmetry transformation rules we conclude that  this AdS theory has six chiral supersymmetries\footnote{We have not established an exact  dictionary between our conventions and those of \cite{Becker:2009mk} and can therefore not compare our definition of $(p,q)$ to theirs.} which we might denote as $(0,6).$
 The above terms in the action  are included in the final form of the higgsed action presented in equations (\ref{final}) to (\ref{interactions}).

\subsection{Gravity interaction with matter}

We next try to understand how gravity interacts with matter. The theory has two gravi\-tational-type coupling parameters: The usual Newton coupling $\kappa$ from (\ref{kappa}), which goes to zero when the limits are taken but is kept non-zero in most of our analysis, and the finite $\bar g_M^2 \sim \mu$, which multiplies all  interaction terms between the gravitational and ABJM sectors, given in  (\ref{bosonicHiggs}) and (\ref{fermionicHiggs}) for the bosonic and fermionic contributions respectively. The same $\bar g_M$ also defines $\Lambda=-\mu^2$ through (\ref{Lambda}).

Because of (\ref{kappa})  gravity is weakly interacting as can be straightforwardly seen: One can define as usual the perturbative graviton $h_{\mu\nu}$ in terms of
\be
g_{\mu\nu}=g_{\mu\nu}^{(0)}+\kappa h_{\mu\nu}
\ee
so that its kinetic term is canonical\footnote{That is, proportional to $\frac{1}{2}h\Box h+...$ , without any additional coupling in front.} while the interactions, defined by $\kappa$, are weak. This is the case even though the background ${\rm AdS}_3$ metric $g_{\mu\nu}^{(0)}$ satisfies an equation of motion involving the finite coupling $\bar g_M$ in (\ref{BGeom}) and implying a finite cosmological constant. On the other hand, the pure ABJM part will couple to $h_{\mu\nu}$ only through $\kappa$.

Since $\kappa\to 0$ in the higgsing limit, the graviton $h_{\mu\nu}$ will be weakly coupled to matter. However, one still has a bit of freedom: By increasing $\bar g_M$ it is possible to increase the coupling of $h_{\mu\nu}$ to (\ref{bosonicHiggs}) and (\ref{fermionicHiggs}), though in doing so one also increases the curvature of the background AdS space. If we instead keep $\bar g_M$ of $\mathcal O(1)$, then the pure ABJM and interaction terms will be of the same order and both will only generate weak gravitational perturbations.

\subsection{Higgsing summary and an alternative procedure}\label{summary}

It is possible to envision an alternative higgsing procedure that  leads to a set of tunable coupling constants for the gauge theory as well as for the gravity sector of the theory. In order to see how this might happen, let us concisely summarize the main points of the higgsing process for the topologically gauged ABJM theory of \cite{Chu:2009gi} so that the possibility of the alternative  procedure involving a slightly different set of parameters and  limits becomes clear. 

\subsubsection{The  higgsing procedure discussed so far}
The higgsing discussed in this paper rely on the introduction of a new parameter\footnote{This parameter plays a similar role to   the level introduced in three-dimensional conformal gravity by Horne and Witten in   \cite{Horne:1988jf}.} $g_M$ in the conformal supergravity sector of the topologically gauged ABJM theory and lead to relations between the fundamental parameters of the theory and the final gravitational and Yang-Mills coupling constants.\footnote{For details how this is done see Section~\ref{relations}.} Starting from the following schematically written part of the bosonic lagrangian where, however, all parameters appear explicitly,
\bea
L=&&\tfrac{1}{g^2_M}L_{CS(\omega)}-R\vert Z \vert^2+\tfrac{1}{\lambda}(AdA+A^3)-DZD\bar Z \cr &&-\lambda^2V^{\rm 1-trace}(Z)-\lambda g^2_M V^{\rm 2-trace}(Z)-g^4_M V^{\rm 3-trace}(Z)\;,
\eea
where $\lambda=\tfrac{2\pi}{k}$ with $k$ the Chern-Simons level, higgsing is straightforward and works as follows: First the vector potential $A$ is divided into two fields $A^+$ and $A^-$
in the above lagrangian which then appear through a term $\tfrac{1}{\lambda}A^-F^+$. Inserting the scalar VEV $\langle Z^4 \rangle=v$ into the scalar field kinetic term one obtains a term $v^2(A^-)^2$, which together with the previous result leads to the equation $v^2A^-\sim \tfrac{1}{\lambda}F^+$. Together these  generate a proper Yang-Mills kinetic term, namely $\tfrac{1}{\lambda^2v^2}(F^+)^2$, and hence we conclude that
the Yang-Mills coupling constant is determined by
\be
g^2_{YM}\sim\lambda^2 v^2.
\ee
Since we want to keep $g^2_{YM}$ finite when the limits are taken, we find that these must involve
\be\label{limits}
v \rightarrow \infty,\,\, \lambda \rightarrow 0\,.
\ee
It is also clear from the $R|Z^2| = v^2 R +... $ term in the lagrangian above that it is necessary to identify the gravitational coupling constant as
\be
\kappa^2\sim\tfrac{1}{v^2}\;,
\ee
 which then goes to zero when taking the limits (\ref{limits}). The lagrangian now reads
schematically, to lowest order in the expansion around the VEV and with $z$ the deviation from the VEV,
\bea
L=&&\tfrac{1}{g^2_M}L_{CS(\omega)}-\tfrac{1}{\kappa^2}R+\tfrac{1}{g_{YM}^2}(F^+)^2-D z D\bar z \nn\\ &&-(\lambda^2v^2(z)^4+{\rm subleading})-(\lambda g^2_M v^3(z)^3+{\rm subleading})\nn\\
&&-(g^4_M v^6+g^4_M v^5(z)+g^4_M v^4(z)^2+{\rm subleading})\;,
\eea
where the sub-leading terms become small in the $v\to\infty$ limit. For the single- and double-trace potential terms (second line  above) one of course also needs to show that,  in this limit, any contributions that  would grow  relative to the explicitly given Yang-Mills terms vanish. For the single-trace term of the original ABJM theory this was shown in \cite{Pang:2008hw}, while for the double-trace terms this was demonstrated in Section~\ref{interaction} of this paper. As also seen there, the linear term in the 
triple-trace potential (last line) cancels against a term coming from the expansion of $R|Z|^2$.
Finally, we identified the coefficients in front of the three potential terms and the gravitational Chern-Simons term and found
\bea
L=&&\tfrac{1}{\kappa^2\mu}L_{CS(\omega)}-\tfrac{1}{\kappa^2}R+\tfrac{1}{g_{YM}^2}(F^+)^2-D z D\bar z \cr &&-(g_{YM}^2(z)^4+{\rm subleading})-(\mu g_{YM}(z)^3+{\rm subleading})\cr
&&-(\tfrac{1}{\kappa^2 \ell^2}+...+{\rm subleading})\;,
\eea
where $subleading$\footnote{The terms displayed are the non-vanishing ones with lowest number of scalar fields. Terms with more scalar fields have thus fewer powers of the VEV $v$.} refers to positive powers of $\kappa$ or $\frac{1}{v^2}$, and where
\be
\mu\sim\tfrac{g_M^2}{\kappa^2}\;,\;\ell=\tfrac{1}{\mu}
\ee
with $\ell$ the AdS radius as in the previous sections. This confirms the prediction of \cite{Chu:2009gi} that this theory is automatically sitting at a chiral point in the sense of  Li, Song and Strominger \cite{Li:2008dq}.

\subsubsection{An alternative higgsing procedure}
We have thus far described the higgsing procedure for the interacting ABJM plus gravitational theory, presented as an extension of the M2 to D2 procedure of \cite{Mukhi:2008ux} for pure ABJM. In this context, the starting point was uniquely determined by the requirement to obtain a pure supergravity part that comes with the usual  $\tfrac{1}{g_M^2}$ coefficient, as well as the wish to recover pure ABJM at $g_M=0$, \ie  to have no factors of $g_M$ in the pure ABJM part of the lagrangian.

However if one forgets for a moment about the M2-brane origin of the theory and its associated ABJM interpretation, and only considers higgsing some field theory coupled to gravity, it is possible to consider a slightly different starting point: 
 \bea L=&&L_{CS(\omega)}-g^2_MR\vert Z \vert^2+\tfrac{g^2_M}{\lambda}(AdA+A^3)-g^2_MDZD\bar Z \cr &&-\lambda^2g^2_MV^{1-trace}(Z)-\lambda g^4_M V^{2-trace}(Z)-g^6_M V^{3-trace}(Z)\;,
 \eea
 which is essentially the lagrangian used in the previous discussion multiplied by a factor of  $g^2_M$. As we will see below this minor alteration  nevertheless has  an impact on the behavior  of the various coefficients when taking the limits.

It can be easily  verified that performing the higgsing based on the same VEV as above, $\langle Z^4\rangle=v$, now leads to the identifications
\be
g^2_{YM}\sim\frac{\lambda^2 v^2}{g^2_M}\,,\,\, \kappa^2\sim\frac{1}{v^2g_M^2}\;,
\ee
resulting in the lagrangian
\bea
L&=&L_{CS(\omega)}-\tfrac{1}{\kappa^2}R+\tfrac{1}{g_{YM}^2}(F^+)^2-g^2_MD z D\bar z \cr
&&-(\lambda^2v^2g^2_M(z)^4+{\rm subleading})-(\lambda g^4_M v^3(z)^3+{\rm subleading})\cr
&&-(g^6_M v^6+...+{\rm subleading})\;.
\eea
At this point one may rescale the scalar fields 
\be
z= \frac{\tilde  z}{g_{YM}g_M}\;,
\ee
which will turn all coefficients in $L$ into combinations of only $\kappa$
 and $g_{YM}$:
 \bea
L&=&L_{CS(\omega)}-\tfrac{1}{\kappa^2}R+\tfrac{1}{g_{YM}^2}(F^+)^2-\tfrac{1}{g^2_{YM}}D\tilde z D\bar{\tilde z} 
\nn\\
&&-(\tfrac{1}{g^2_{YM}}(\tilde z)^4+{\rm subleading})
-(\tfrac{1}{\kappa^{2}g^2_{YM} }(\tilde z)^3+{\rm subleading})\nn\\
&&-(\tfrac{1}{\kappa^{6}}+
\tfrac{1}{\kappa^{5}g_{YM} }(\tilde z)+\tfrac{1}{\kappa^{4}g^2_{YM} }(\tilde z)^2+{\rm subleading})\;.
\eea
From this form of the lagrangian we conclude that 
\be
\mu=\kappa^{-2},\,\,\ell=\kappa^2
\ee
and hence that the theory is at the `chiral point' just as we found in the first analysis. 

One also discovers that the sub-leading terms all appear with more than two inverse powers of $g_{YM}$ and thus become negligible close to the IR fixed point, \ie for large Yang-Mills coupling. This is not the same procedure as that done in the pure ABJM case, where we wanted $v\rightarrow\infty$, $\lambda\to 0$ and we obtained three-dimensional SYM at {\em finite} $g_{YM}\sim v\lambda$. However, the current starting point may not have an M2/ABJM interpretation, in which case this is not an  issue.

The analysis of the sub-leading terms is  done as follows: The one-trace (ABJM) potential $V$ (proportional to $Z^6$) has an expansion around the VEV $v$ that gives the term at order $v^2\tilde z^4/g^{2}_{YM}$. Replacing one $\tilde z$ with a VEV $v$ means acquiring an extra factor of $vg_Mg_{YM}=\tfrac{g_{YM}}{\kappa}$.  As is known from \cite{Mukhi:2008ux,Distler:2008mk} this and the terms resulting after two and more such replacements all vanish. Replacing a VEV by $\tilde z$ on the other hand will produce extra factors of $g_{YM}$ in the denominator and yield sub-leading expressions in the IR limit.

For the double-trace potential we get a cubic contribution with the remaining terms in its expansion behaving in a similar way.  Finally, the triple-trace potential starts with the cosmological constant term $\kappa^{-6}$ and by replacing VEVs with $\tilde z$'s on gets a linear term that vanishes and a bilinear contribution that is a mass term with coefficient $g^{-2}_{YM}\kappa^{-4}$. One also has a term linear in the scalar field with a non-vanishing coefficient. However, this term is needed since it will cancel a similar one coming from the expansion of the conformal $R|Z|^2$ in the lagrangian. The rest of the terms are all sub-leading in the $g_{YM}\to\infty$ limit.

The main novelty of this alternative approach is  that  the  three  limits
\be
v \rightarrow \infty,\,\,\lambda \rightarrow 0,\,\, g_M \rightarrow 0
\ee
can  be used to keep the gravitational coupling
\be
\kappa^2=\tfrac{8}{g_M^2v^2}
\ee
fixed and finite. However, it does not seem possible to completely switch off gravity since then the AdS radius goes to zero, \ie in that case  one is dealing with a background of infinite curvature. We should keep in mind that in this case the M2/ABJM interpretation may be lost but one can nevertheless study the model in its own right.

Hence the two pictures present different virtues: While for the usual ABJM interpretation, considered in the bulk of this paper, we could not avoid to decouple gravity (since $v\sim\tfrac{1}{\kappa}$ and $v\rightarrow \infty$). This, however, happens at a fixed AdS radius, which is welcome. On the other hand, in the alternative approach we have a fixed gravitational coupling constant with an AdS radius that is proportional to the gravitational coupling. Both features are somewhat unusual but should be viewed as consequences of the fact that the theory is tied to the chiral point. It seems that the physical effects of the chiral point in this model need to be further studied.

\section{Conclusions and Outlook}\label{conclusions}

In this paper we  analyzed the higgsing \cite{Mukhi:2008ux} of the topologically gauged ABJM model constructed in \cite{Chu:2009gi}. We first re-expressed the model of \cite{Chu:2009gi} in a bifundamental Lie algebra formulation and introduced a coupling constant $g_M$ for the conformal supergravity sector and  the matter-gravity interactions. We proceeded to briefly review the higgsing for the original ABJM case. In doing so we took some care to clarify the role of U(1) factors that were previously omitted in the literature.

 We then presented the higgsing procedure for the theory of \cite{Chu:2009gi} as an extension to ABJM.  For the latter one needs to take  the VEV $v$ and CS level $k$ to infinity, with $\tfrac{v}{k}\equiv g_{YM}={\rm fixed}$, while in the theory studied in this paper one must also take $g_M\rightarrow 0$, with $\tfrac{g^2_M v^2 N}{8}\equiv \mu={\rm fixed}$, where $\frac{1}{\mu}$ is the coefficient of the gravitational Chern-Simons term. This leads to an ${\cal N}=6$ supersymmetric theory involving an ${\rm AdS}_3$ background of finite radius $\ell$ with $\mu\ell =1$. All matter coupled to gravity through a weak Newton coupling, $\kappa^2\sim \tfrac{1}{v^2}\rightarrow 0$.

The corresponding action was given in Eq.~(\ref{final}), with the gravity part being a supersymmetric generalization of topologically massive gravity (TMG) at the `chiral point' in the sense of \cite{Li:2008dq}. 
The supersymmetries appearing in the higgsed theory can be seen to arise as  linear combinations of supersymmetries and special superconformal symmetries of the unhiggsed theory. One should note, however, that we here find  the sign of the Einstein-Hilbert term reversed  compared to \cite{Li:2008dq} as is more customary in the three-dimensional TMG literature. It would be interesting to further elucidate the role of this sign issue and to see whether our construction can provide a different angle on the subtle questions of gravitational boundary conditions that determine the nature of the physical states and the dual CFT for these theories.

Besides the standard ABJM Higgs effect we also find that the R-symmetry is broken to $\SU(3)$ with the result that all but one of the eight Goldstone scalars $X_0^I$ are `eaten' by gauge fields. This makes it hard to find a D-brane interpretation for the final theory unless the branes experience some kind of position stabilization in analogy to the discussion of \cite{Goldberger:1999uk} in the context of brane-world models.  Since in string theory independent world-volume Einstein gravity on branes is not allowed, world-volume gravity would most likely arise from the space-time fields. Indeed, in brane-world models, world-volume gravity does arise from brane-restriction of space-time gravity as in the Randall-Sundrum model \cite{Randall:1999vf}, and position stabilization generates an effective action for the space-time fields restricted to the world-volume \cite{Goldberger:1999uk}.

Similar questions remain in connection with the interpretation of the theory in the conformal phase. Since the ABJM theory captures the dynamics of $N$ M2-branes on a $\mathbb C^4/\mathbb Z_k$ orbifold, the coupling to conformal world-volume supergravity might be thought of as a construction resembling the string action with world-sheet gravity, used \eg when quantizing the string on higher genus Riemann surfaces, or perhaps as a first step towards coupling the theory to more general backgrounds.

In this context one could also ask what the string/M-theory origin of a topologically gauged theory might be. There exist examples for which the low energy limit has been known to fail in decoupling gravity, \eg systems involving D3-branes intersecting D7/O7 configurations on a 1+1d subspace.  This special property is due to the particular way $\alpha'$ appears in front of the different terms in the lagrangian \cite{Harvey:2007ab,Harvey:2008zz}. Perhaps the model of \cite{Chu:2009gi} arises in a similar way from some yet to be determined M-theory configuration. It would be interesting to investigate further if such a picture can be realized.

\section*{Acknowledgments}

We would like to thank Bobby Ezhuthachan, Alex Maloney, Andy Royston and Kostas Skenderis for discussions and comments. CP is supported by the STFC grant ST/G000395/1 and XC
by the Belgian Fonds de la Recherche Scientifique (FNRS) and Belgian Science Policy IAP.

\begin{appendix}

\section{Expansion and trace formulas}\label{expansion}

In this appendix we list various formulas that were used for the higgsing of the topologically gauged ABJM action.  In particular, we find for the bosonic single-trace 2-scalar structures:
\bea
|Z|^2&=&\sum_A\tr(Z^AZ^\dagger_A)= \tr \Big( v^2 + v(z^4 + z^\dagger_4) + z^A z^\dagger_A\Big) \cr
\tr(Z^BZ^\dagger_A)&=&  \tr \Big( v^2 \delta_{A,4}  \delta^{B,4}+ v(z^B \delta_{A,4} + z^\dagger_A \delta^{B,4}) + z^B z^\dagger_A\Big) \cr
\tr(Z^B\tilde D_\rho Z^\dagger_A)&=& \tr \Big(v\tilde D_\rho v \delta_{A,4}  \delta^{B,4}+ z^B \tilde D_\rho v \delta_{A,4} + v \tilde D_\rho z^\dagger_A \delta^{B,4}+ z^B \tilde D_\rho z^\dagger_A\Big)\;,
\eea
which then also implies
\bea
\tr(Z^BZ^\dagger_A)\tr(Z^AZ^\dagger_B)&=&(v^2N+v\tr(z^4+\zd_4))^2+2v^2N\tr(z^4\zd_4)\cr
&&+2v^2(\tr z^A\tr \zd_A-\tr z^4\tr \zd_4)\cr
&&+2v(\tr z^A\tr(z^4\zd_A)+\tr \zd_A\tr(\zd_4 z^A))+\tr(z^A\zd_B)\tr(z^B\zd_A)\cr
\tr(Z^AZ^\dagger_C)\tr(Z^BZ^\dagger_A)\tr(Z^CZ^\dagger_B)  
&=&(v^2N+v\tr(z^4+\zd_4))^3+3v^2(v^2N+v\tr(z^4+\zd_4))\times \cr
&&\times (\tr z^A\tr \zd_A-\tr z^4\tr \zd_4)+3v^4N^2\tr(z^4+\zd_4)\cr
&&+3v^3N[\tr(z^4+\zd_4)\tr(z^4\zd_4)+\tr(z^4\zd_A)\tr z^A\cr
&&+\tr(z^A\zd_4)\tr \zd_A]+\mathcal O(v^2)\;.
\eea
The bosonic single-trace 4-scalar structures are
\bea
\tr(Z^\dagger_DZ^BZ^\dagger_CZ^C-Z^BZ^\dagger_DZ^CZ^\dagger_C)&=& \tr\Big( v [z^\dagger_C, z^C] (\delta^{B,4} z^\dagger_D +\delta_{D,4} z^B )+ v z^\dagger_D ( [z^B, z^4] + [z^B, z^\dagger_4])\cr
&& + z^\dagger_D z^B z^\dagger_C z^C-z^B z^\dagger_Dz^C z^\dagger_C\Big)\cr
\tr(Z^\dagger_DZ^DZ^\dagger_CZ^C-Z^DZ^\dagger_DZ^CZ^\dagger_C)&=& \tr\Big(2 v (z^4 + z^\dagger_4) [z^\dagger_D , z^D] 
+ z^\dagger_Dz^Dz^\dagger_C z^C-z^D z^\dagger_D z^C z ^\dagger_C\Big)\;.\nn\\
\eea
Combining the above we get
\bea
\tr(Z^DZ^\dagger_B)\tr(Z^\dagger_DZ^BZ^\dagger_CZ^C-Z^BZ^\dagger_DZ^CZ^\dagger_C)&=&  \tr(v^2) \tr\Big( v [z^\dagger_C, z^C](z^\dagger_4 + z^4) + v (z^4+z^\dagger_4)[z^\dagger_4, z^4]\Big)\cr
&& + \tr(v^2)\tr(z^\dagger_4 z^4 z^\dagger_C z^C - z^4 z^\dagger_4 z^C z^\dagger_C)\cr
&& + \tr(v z^D)\tr\Big(v[\zd_C, z^C] \zd_D + v \zd_D[z^4, \zd_4]\Big)\cr
&& + \tr(v \zd_B)\tr\Big(v[\zd_C, z^C] z^B + v z^B[z^4,\zd_4])\Big)\cr
&&+ \tr(v z^4)\tr\Big(v [\zd_C, z^C]z^4)\cr
&&\qquad\qquad\qquad + \tr(v \zd_4)\tr(v [\zd_C, z^C]\zd_4\Big)\cr
&&+\tr(z^D \zd_B)\tr\Big(v \zd_D ([z^B, z^4] + [z^B, \zd_4])\Big)\cr
&&+ \tr(z^D \zd_4)\tr\Big(v [\zd_C, z^C]\zd_D)\cr
&&\qquad\qquad\qquad + \tr(z^4\zd_B)\tr(v[\zd_C, z^C]z^B\Big)\cr
&&+\tr(vz^A)\tr(\zd_A z^4\zd_C z^C-z^4\zd_Az^C\zd_C)\cr
&&+\tr(v\zd_A)\tr(\zd_4z^A\zd_Cz^C-z^A\zd_4z^C\zd_C)\cr
&&+\tr(z^Dz^\dagger_B)\tr(z^\dagger_Dz^B z^\dagger_C z^C-z^B z^\dagger_Dz^C z^\dagger_C)\;.
\eea
Finally we get the combinations:
\bea
|Z^6| &=& \tr(v^2)^3 + \tr(v (z^4 + \zd_4))^3 + \tr(z^A\zd_A)^3\cr
&& + 3\Big(\tr(v^2)^2 [\tr(v(z^4 + \zd_4)) + \tr(z^A \zd_A)]\cr
&& + \tr(v(z^4 + \zd_4))^2[\tr(v^2) + \tr(z^A \zd_A)]\cr
&& + \tr(z^A \zd_A)^2 [\tr(v^2) + \tr(v(z^4+\zd_4))]\Big)\cr
&&+ 6\tr(v^2)\tr(v(z^4+\zd_4))\tr(z^A \zd_A)\\
|Z^2|\tr(Z^A Z^\dagger_B)\tr(Z^B Z^\dagger_A) &=&\tr(v^2)^3 + \tr(v^2)^2\Big(3 \tr(v z^4) +3 \tr(v\zd_4)\cr
&&\qquad\qquad\qquad\qquad\qquad + \tr(z^A \zd_A)+2\tr(z^4\zd_4)\Big)\cr
&& + \tr(v^2)\Big( 3(\tr(v z^4) + \tr(v\zd_4))^2 \cr
&& + 2 \tr(v\zd_A) \tr(z^A\zd_4)  + \tr(z^A\zd_B)\tr(z^B\zd_A)\cr
&&+ 2 \tr(z^4 \zd_4)\tr(v z^4) +2 \tr(z^4 \zd_4)\tr(v\zd_4)\cr
&& + 2 \tr(v z^A) \tr(z^4\zd_A) + \tr(z^4\zd_4)\tr(z^A\zd_A)\cr
&& + 2 \tr(v z^4) \tr(z^A\zd_A) + 2 \tr(v \zd_4) \tr(z^A\zd_A) \Big)\cr
&& + \tr(z^C \zd_C) \tr(z^A z^\dagger_B)\tr(z^B z^\dagger_A)+ \Big(\tr(v z^4)+ \tr(v \zd_4)\Big)^3\cr
&& + 2\Big(\tr(v z^A)\tr(z^4\zd_A) + \tr(v \zd_A)\tr(z^A\zd_4)\Big)\times \cr
&& \times \Big(\tr(v z^4) + \tr(v \zd_4) + \tr(z^A \zd_A) \Big)\cr
&& + \tr(z^A\zd_A) \Big(\tr(v z^4 ) + \tr(v\zd_4) \Big)^2\cr
&& + \tr(z^A \zd_B)\tr(z^B \zd_A)\tr(v z^4)\cr
&&+ \tr(z^A \zd_B)\tr(z^B \zd_A)\tr(v \zd_4)\\
|Z^2|\tr(Z^\dagger_DZ^DZ^\dagger_CZ^C-Z^DZ^\dagger_DZ^CZ^\dagger_C) &=&  \Big(2\tr( v (z^4 + z^\dagger_4) [z^\dagger_D , z^D])\cr
&&\qquad\qquad +\tr( z^\dagger_Dz^Dz^\dagger_C z^C-z^D z^\dagger_D z^C z ^\dagger_C)\Big)\nonumber\\
&&\times\Big( \tr(v^2) + \tr(v(z^4 + z^\dagger_4)) +\tr( z^A z^\dagger_A)\Big)\;.
\eea

Working with the normalisation $T^0={\one}_{N\times N}$, considering  $[T^a,T^b]=i{f^{ab}}_cT^c$ and $\tr(\{T^a,T^b\}T^c)=d^{abc}$, a few trace formulas relevant for the bosonic terms  are
\bea
&&[z^A,\zd_A]=-X_a^{A+4}X_b^A{f^{ab}}_cT^c\cr
&&\tr(z^4+\zd_4)=N\sqrt{2}X_0^4\cr
&&\tr(z^4-\zd_4)=i\sqrt{2}NX_0^8\cr
&&\tr(z^A)\tr(\zd_A)-\tr(z^4)\tr(\zd_4)=\frac{N^2}{2}\Big(X_0^{I}X_0^{I} - X_0^{4}X_0^{4} -X_0^{8}X_0^{8}\Big)\cr
&& z^A\zd_A = \frac{1}{2}\Big( X_0^I X_0^I + X_a^I X_b^I T^a T^b + 2 X^A_a X_0^{A+4} T^a - 2 X_0^A X_a^{A+4}T^a
-X_a^{A+4}X_b^A{f^{ab}}_cT^c\Big)\cr
&&\zd_A z^A=\frac{1}{2}\Big( X_0^I X_0^I + X_a^I X_b^I T^a T^b + 2 X^A_a X_0^{A+4} T^a - 2 X_0^A X_a^{A+4}T^a
+X_a^{A+4}X_b^A{f^{ab}}_cT^c\Big)\cr
&&\tr(z^A\zd_A)=\frac{1}{2}(N\:X_0^IX_0^I+X_a^IX_a^I)\cr
&&\tr((z^4+\zd_4)[\zd_D,z^D])=-\sqrt{2}f^{abc}X_a^{A+4}X_b^AX_c^8\cr
&&\tr(\zd_4[z^4,\zd_4])=\frac{1}{\sqrt{2}}f^{abc}X_a^8X_b^4(X_c^8+iX_c^4)\cr
&& \tr((z^4-\zd_4)[\zd_D,z^D])=i\sqrt{2}f^{abc}X_a^{A+4}X_b^AX_c^4\cr
&&\tr(\zd_Dz^D\zd_C z^C-z^D\zd_Dz^C\zd_C)=X_a^{A+4}X_b^Af^{abc}(2X_c^BX_0^{B+4}-2X_0^BX_c^{B+4}
+\frac{1}{2}{d^{ef}}_cX^I_eX^I_f)\cr
&&\tr(\zd_4z^4\zd_Cz^C-z^4\zd_4z^C\zd_C)=\frac{1}{2}f^{abc}\Big\{X_a^{A+4}X_b^A(2X_c^4X_0^8-2
X_0^4X_c^8+\frac{1}{2}{d^{ed}}_c(X_e^4X_f^4+X_e^8X_f^8))\cr
&&\qquad\qquad\qquad\qquad\qquad\qquad+X_a^8X_b^4(2X_c^BX_0^{B+4}-2X_0^BX_c^{B+4}+\frac{1}{2}{d^{ef}}_c(X^I_eX^I_f))\Big\}\cr
&&\tr(z^D)\tr(\zd_D[\zd_C,z^C])=-\frac{N}{2}f^{abc}X_a^{C+4}X_b^C(X_c^{D+4}+iX_c^D)(X_0^D+iX_0^{D+4})\cr
&&\tr(z^D)\tr(\zd_D[z^4,\zd_4])=\frac{N}{2}f^{abc}X_a^{8}X_b^4(X_c^{D+4}+iX_c^D)(X_0^D+iX_0^{D+4})\cr
&&\tr(\zd_D)\tr(z^D[\zd_C,z^C])=-\frac{N}{2}f^{abc}X_a^{C+4}X_b^C(X_c^{D+4}-iX_c^D)(X_0^D-iX_0^{D+4})\cr
&&\tr(\zd_D)\tr(z^D[z^4,\zd_4])=\frac{N}{2}f^{abc}X_a^{8}X_b^4(X_c^{D+4}-iX_c^D)(X_0^D-iX_0^{D+4})\cr
&&\tr(z^A)\tr(\zd_A)\tr(z^4\zd_4)=\frac{N^2}{4}X_0^IX_0^I\Big[N((X_0^4)^2+(X_0^8)^2)+(X_a^4)^2+(X_a^8)^2\Big]\cr
&&\tr(z^A)\tr(\zd_4)\tr(z^4\zd_A)  = \frac{N^2}{4}\Big[N X_0^I X_0^I((X_0^4)^2+(X_0^8)^2)+(X_0^A X_0^4 + X_0^{A+4}X_0^8)(X_a^A X_a^4 + X_a^{A+4}X_a^8)\cr
&&\qquad\qquad\qquad\qquad+ (X_0^A X_0^8 - X_0^{A+4}X_0^4)(X_a^A X_a^8 - X_a^{A+4}X_a^4) \Big]\cr
&&\tr(\zd_A)\tr(z^4)\tr(z^A\zd_4)=\frac{N^2}{4}\Big[NX_0^IX_0^I((X_0^4)^2+(X_0^8)^2)\cr
&&\qquad\qquad\qquad\qquad\qquad+(X_0^A-iX_0^{A+4})(iX_a^A-
X_a^{A+4})(X_0^4+iX_0^8)(-iX_a^4-X_a^8)\Big]\cr
&&\qquad\qquad\qquad\qquad \qquad= \tr(z^A)\tr(\zd_4)\tr(z^4\zd_A)\cr
&&\tr(\zd_A)\tr(z^B)\tr(z^A\zd_B)= \frac{N^2}{4}\Big[N(X_0^IX_0^I)^2 + (X_0^A X_0^B + X_0^{A+4}X_0^{B+4}) (X_a^A X_a^B + X_a^{A+4}X_a^{B+4})\cr
&&\qquad\qquad\qquad\qquad + (X_0^A X_0^{B+4} - X_0^{A+4}X_0^{B}) (X_a^A X_a^{B+4} - X_a^{A+4}X_a^{B})\Big]\cr
&& \tr(z^A \zd_B) = \frac{1}{2}\Big(N(X_0^A + i X_0^{A+4})(X_0^B - i X_0^{B+4}) - (iX_a^A - X_a^{A+4})(X_a^{B+4} + i X_a^B)\Big)\cr
&& \tr(z^A \zd_B)\tr(z^B \zd_A) 
 =\frac{1}{4}\Big[N^2(X_0^IX_0^I)^2+(X_a^IX_b^I)^2+(X_a^AX_b^{A+4}-X_a^{A+4}X_b^A)^2\cr
&& \qquad\qquad\qquad\qquad +2N[(X_0^IX_a^I)^2+
 (X_0^{A+4}X_a^A-X_0^AX_a^{A+4})^2]\Big]\cr
&&\tr(z^A \zd_4)\tr(z^4 \zd_A)= \frac{1}{4}\Big[ N^2 \Big(X_0^I X_0^I (X_0^4 X_0^4 +X_0^8 X_0^8)\Big)  + |(X_a^{8}+ i X_a^4)( i X_a^A - X_a^{A+4} )|^2\cr
 && \qquad\qquad\qquad\qquad\qquad\qquad - N (X_0^4 - i X_0^{8})(X_0^A + i X_0^{A+4})(  X_a^{A+4}+iX_a^A)( i X_a^4-X_a^{8}) - c.c.\Big]\cr &&\qquad\qquad\qquad\qquad
 =\frac{1}{4}\Big[N^2X_0^IX_0^I(X_0^4X_0^4+X_0^8X_0^8)
 \cr
&&\qquad\qquad\qquad\qquad+(X_a^{A+4}X_b^4+X_a^AX_b^8)(X_b^{A+4}X_a^4+X_b^AX_a^8)\cr &&\qquad\qquad\qquad\qquad+(X_a^AX_b^4-X_a^{A+4}X_b^8)(X_b^AX_a^4-X_b^{A+4}X_a^8)\cr
&&\qquad\qquad\qquad\qquad+2N[(X_0^AX_0^4+X_0^{A+4}X_0^8) (X_a^{A+ 4} X_a^8+X_a^AX_a^4)\cr
&&\qquad\qquad\qquad\qquad+(X_0^4X_0^{A+4}-X_0^8X_0^A)(X_a^{A+4}X_a^4
-X_a^{A}X_a^8)]\Big]\cr
&&\tr(z^4 \zd_4)\tr(z^A \zd_A) = \frac{1}{4}(NX_0^4 X_0^4 + NX_0^8 X_0^8 +X_a^4 X_a^4 + X_a^8 X_a^8 )(NX_0^I X_0^I + X_a^I X_a^I)\;,
\eea
where in the first identity we do not sum over $A$.

Formulas relevant for the fermionic terms are
\bea
&&[z^A,z^\dagger_D] = \frac{1}{2}{f^{ab}}_c(i X_a^A X_b^D - X_a^{A+4}X_b^D + X_a^A X_b^{D+4}+ i X^{A+4}X_b^{D+4})T^c\cr
&&[z^A,z^D] = \frac{1}{2}{f^{ab}}_c(-i X_a^A X_b^D + X_a^{A+4}X_b^D + X_a^A X_b^{D+4}+ i X^{A+4}X_b^{D+4})T^c\cr
&&\tr(Z^\dagger_DZ^CZ^\dagger_CZ^D-Z^\dagger_DZ^DZ^\dagger_CZ^C)=2\sqrt{2}vf^{abc}X_a^{A+4}X_b^AX_c^8-f^{abc}X_a^{A+4}X_b^A\cr
&&\qquad\qquad\qquad\qquad\qquad\qquad\qquad(2X_c^BX_0^{B+4}-2
X_0^BX_c^{B+4}+\frac{1}{2}{d^{ef}}_cX_e^IX_f^I)\cr
&&\tr(Z^BZ^\dagger_A)=v^2N\delta_{A,4}\delta^{B,4}+\frac{vN}{\sqrt{2}}((X_0^A-iX_0^{A+4}) \delta^{B,4} +\delta_{A,4}(X_0^B+iX_0^{B+4}))\cr
&&\qquad\qquad\qquad\qquad+\frac{N}{2}(X_0^B+iX_0^{B+4})(X_0^A-iX_0^{A+4})+\frac{1}{2}(X_a^{B+4}-iX_a^B)(X_a^{A+4}+iX_a^A)\cr
&&\tr(Z^B\tilde{D}_\rho Z^\dagger_A)=\frac{vN}{\sqrt{2}}\delta^{B,4}\pd_\rho (X_0^A-iX_0^{A+4})+\sqrt{2}v(X_a^B+iX_a^{B+4})A_\rho^{a-}\delta^{A,4}\cr
&&\qquad\qquad\qquad+\frac{N}{2}(X_0^B+iX_0^{B+4})\pd_\rho(X_0^A-iX_0^{A+4})+\frac{1}{2}(X_a^{B+4}-iX_a^B)D_\rho(X_a^{A+4}+iX_a^A)\cr
&&\qquad\qquad\qquad +{\rm subleading}-i N \delta^{A,4} A_{\mu 0}^- \Big [2 v^2  \delta_{B,4} + \sqrt 2  (X_0^B + i X_0^{B+4})  \Big]\cr
&&\tr(Z^\dagger_{[D}Z^{[A}Z^\dagger_{C]}Z^{B]}-Z^\dagger_{[D}Z^{B]}Z^\dagger_{C]}Z^{A]})=\frac{1}{\sqrt{2}}vif^{abc}\Big[(iX_a^{[C}+X_a^{[C+4})(iX_b^{D]}+X_b^{D]+4}) \times\cr
&&\qquad\qquad\qquad \qquad\qquad\qquad \times(iX_c^{[A}-X_c^{[A+4}) \delta^{B],4}\cr
&&\qquad\qquad\qquad \qquad\qquad\qquad +(iX_a^{[A}-X_a^{[A+4})(iX_b^{B]}-X_b^{B]+4}) (iX_c^{[C}+X_c^{[C+4})\delta^{D],4}\Big]\cr
&&\qquad\qquad\qquad\qquad\qquad\qquad+{\rm subleading}\cr
&&\tr(\bar\Psi_D^\dagger \Psi^B)=\frac{N}{\sqrt{2}}(\psi_0^{[A}+i\psi_0^{[A+4})\pd_\mu (X_0^{B]}-iX_0^{B]+4})\cr
&&\qquad\qquad\qquad+(i\psi_a^{[A}-\psi_a^{[A+4})\Big[\frac{D_\mu(-iX_a^{B]}-X_a^{B]+4})}{\sqrt{2}}-2ivA_{\mu a}^- \delta^{B,4}\Big]\cr
&&\qquad\qquad\qquad- 2i N v A_{\mu 0}^- \delta^{A,4}(\psi_0^A + i \psi_0^{A+4})+{\rm subleading} \cr
&&\tr(\Psi_A\Psi^{\dagger C})=N(\psi_{0}^{A}+i\psi_{0}^{A+4})(\psi_0^{\dagger C}-i\psi_0^{\dagger C+4})+(i\psi_{a}^{A}-\psi_{a}^{A+4})
(-i\psi_a^{\dagger C}-\psi_a^{C+4})\cr
&&\tr(\Psi_A Z^\dagger_B)=\frac{N}{\sqrt{2}}(\psi_0^A+i\psi_0^{A+4})(X_0^B-iX_0^{B+4})\cr
&& \qquad\qquad\qquad+\frac{1}{\sqrt{2}}(i\psi_a^A-\psi_a^{A+4})(-iX_a^B-X_a^{B+4}) + v N \delta_{B,4}(\psi_0^A + i \psi_0 ^{A+4})\cr
&&\tr(\Psi^B(Z^DZ^\dagger_DZ^A-Z^AZ^\dagger_DZ^D))=
-iv(i\psi_c^B-\psi_c^{B+4})f^{abc}X_a^8(iX_b^A-X_b^{A+4})\cr
&&\qquad\qquad\qquad\qquad\qquad\qquad-v\delta^{A,4}X_a^{D+4}X_b^D(i\psi_c^B-\psi_c^{B+4})+{\rm subleading}\cr
&&\tr(\Psi^D(Z^{[B}Z^\dagger_DZ^{A]}-Z^{[A}Z^\dagger_DZ^{B]}))=
-\frac{i}{2}v(i\psi_c^4-\psi_c^8)(iX_a^{[A}-X_a^{[A+4})(iX_b^{B]}-X_b^{B]+4})f^{abc}\cr
&&\qquad\qquad\qquad\qquad\qquad-iv(i\psi_c^D-\psi_c^{D+4})\delta^{[A,4}(-iX_{a,D}-X_{a,D+4})(iX_b^{B]}-X_b^{B]+4})f^{abc}\cr
&&\qquad\qquad\qquad\qquad\qquad\qquad+{\rm subleading}
\;.
\eea

\section{Subleading terms in $v$}\label{subleading}

Here we present formulas for the Higgsed action down to ${\cal O}(v)$ for both the bosonic and fermionic pieces. 

\subsection{Bosonic terms}

The bosonic 2-trace terms give
\bea
-V_{2-trace,v^2}&=&-\frac{g_M^2v^2N \lambda}{8}X_a^{A+4}X_b^Af^{abc}(2X_c^BX_0^{B+4}-2X_0^BX_c^{B+4}
+\frac{\lambda}{2}{d^{ef}}_cX^I_eX^I_f)\cr
&&+\frac{g_M^2v^2N \lambda}{4}f^{abc}\Big\{X_a^{A+4}X_b^A(2X_c^4X_0^8-2
X_0^4X_c^8+\frac{1}{2}{d^{ed}}_c(X_e^4X_f^4+X_e^8X_f^8))\cr
&&+X_a^8X_b^4(2X_c^BX_0^{B+4}-2X_0^BX_c^{B+4}+\frac{1}{2}{d^{ef}}_c(X^I_eX^I_f))\Big\}\cr
&&-\frac{g_M^2v^2N \lambda}{2}f^{abc}X_a^{A+4}X_b^AX_c^4X_0^8\cr
&&+\frac{g_M^2v^2N \lambda}{2}f^{abc}(X_c^{D+4}X_0^D-X_c^DX_0^{D+4})(X_a^8X_b^4-X_a^{C+4}X_b^C)\;.
\eea

The bosonic 3-trace terms give
\bea
-V_{3-trace,v^2}&=&\frac{5g_M^4v^2N}{4\cdot 256}(N^2 (X_0^IX_0^I)^2+(X_a^IX_a^I)^2 + 2 X_0^I X_0^I X_a^J X_a^J)\cr
&&-\frac{g_M^4v^2N}{2\cdot64}\Big[N^2X_0^IX_0^I(X_0^JX_0^J-(X_0^4)^2-(X_0^8)^2)+N[2(X_0^IX_a^I)^2 \cr
&&
+2(X_0^{A+4}X_a^A-X_0^AX_a^{A+4})^2 +X_0^IX_0^I((X_a^4)^2+(X_a^8)^2)
\cr
&&+X_a^IX_a^I((X_0^4)^2+(X_0^8)^2)-4(X_0^AX_0^4+X_0^{A+4}X_0^8)(X_a^{A+4}X_a^8+X_a^A X_a^4)\cr
&&-4(X_0^4X_0^{A+4}-X_0^8X_0^A)(X_a^{A+4}X_a^4-X_a^{A}X_a^8)] +X_a^IX_a^I((X_b^4)^2+(X_b^8)^2) \cr
&&-2(X_a^{A+4}X_b^4+X_a^AX_b^8)(X_b^{A+4}X_a^4+X_b^AX_a^8)+(X_a^IX_b^I)^2
\cr
&&-2(X_a^AX_b^4-X_a^{A+4}X_b^8)(X_b^AX_a^4-X_b^{A+4}X_a^8)+ (X_a^A X_b^{A+4}-X_a^{A+4}X_b^A)^2\Big]\cr
&&+\frac{g_M^4v^2N^2}{64}\Big[NX_0^IX_0^I(X_0^JX_0^J-(X_0^4)^2-(X_0^8)^2)+X_0^IX_0^I((X_a^4)^2+(X_a^8)^2)\cr
&& +(X_0^A X_0^B + X^{A+4}_0 X_0^{B+4}) (X_a^A X_a^B + X^{A+4}_a X_a^{B+4})\cr
&&+ (X_0^A X_0^{B+4} - X^{A+4}_0 X_0^{B}) (X_a^A X_a^{B+4}- X^{A+4}_a X_a^{B})\cr
&&-2 (X_0^A X_0^4 + X^{A+4}_0 X_0^{8}) (X_a^A X_a^4 + X^{A+4}_a X_a^{8})\cr
&&-2 (X_0^A X_0^{8} - X^{A+4}_0 X_0^{4}) (X_a^A X_a^{8}- X^{A+4}_a X_a^{4})\Big]\;.
\eea

Note that if we exclude the contributions containing $X_0$'s, which decoupled during the higgsing of  the pure ABJM theory, we are left with
\bea
-V_{v^2}^{\SU(N)}&=&-\frac{g_M^2v^3N \lambda}{2\sqrt{2}}f^{abc}X_a^{A+4}X_b^AX_c^8\cr
&&+\frac{g_M^2v^2N \lambda}{8}f^{abc}{d^{ef}}_c\Big[X_a^{A+4}X_b^A(-\frac{1}{2}X_e^IX_f^I+X_e^4X_f^4+X_e^8X_f^8)
+X_a^8X_b^4X_e^IX_f^I\Big]\cr
&&-\frac{g_M^4v^2N}{128}\Big[(X_a^IX_b^I)^2+(X_a^{A}X_b^{A+4}-X_a^{A+4}X_b^A)^2 -\frac{5}{8}(X_a^IX_a^I)^2\cr
&&+X_a^IX_a^I((X_b^4)^2+(X_b^8)^2)-2(X_a^{A+4}X_b^4+X_a^AX_b^8)(X_b^{A+4}X_a^4+X_b^AX_a^8)\cr
&&-2(X_a^AX_b^4-X_a^{A+4}X_b^8)(X_b^AX_a^4-X_b^{A+4}X_a^8)\Big]\;.\nn\\
\eea
If we further restrict to pieces not involving the `special' directions  4 and 8, which we will denote with indices $I''$, the ${\cal O} (v^2)$ potential looks very simple:
\bea
V_{v^2}^{\SU(N),I''}&=&-\frac{ g_M^4v^2N}{128}
[X_a^{I''}X_b^{I''}X_a^{J''}X_b^{J''}+(X_a^{A'}X_b^{A'+4}-X_a^{A'+4}X_b^{A'})^2 -\frac{5}{8}(X_a^{I''}X_a^{I''})^2]\cr 
&&-\frac{ g_M^2v^2N \lambda}{16}f^{abc}{d^{ef}}_cX_a^{A'+4}X_b^{A'}X_e^{I''}X_f^{I''}\;,
\eea
where $A'=1,2,3$.

\subsection{Fermionic terms}

The full fermionic terms, including the ones of the same order as the above bosonic terms, are
\bea
&&\frac{i}{2}f^\mu_{AB}\chi_\mu^{AB}(v^2N+vN\sqrt{2}X_0^4+\frac{1}{2}(NX_0^IX_0^I+X_a^IX_a^I))\cr
&&
+\frac{iv}{\sqrt{2}}\lambda\epsilon^{\mu\nu\rho}(\bar \chi_{\mu AB}\gamma_\nu \chi_\rho^{AB})
f^{abc}X_a^{C'+4}X_b^{C'}X_c^8\cr
&&+\frac{i}{64}g_M^2\epsilon^{\mu\nu\rho}(\bar \chi_{\nu AB}\gamma_\rho\chi_\mu^{AB})(v^2N+vN\sqrt{2}X_0^4)
\Big[v^2N+vN\sqrt{2}X_0^4+\frac{1}{2}(NX_0^IX_0^I+X_a^IX_a^I)\Big]\cr
&&+\frac{i}{16}g_M^2e\Big[N(\bar \psi_{0}^{D}-i\bar \psi_{0}^{ D+4})(\psi_0^D+i\psi_0^{D+4})
\cr
&&\qquad\qquad\qquad\qquad+(i\bar \psi_{a}^{D}+\bar \psi_{a}^{ D+4})(-i\psi_a^D+\psi_a^{D+4})\Big](v^2N+vN\sqrt{2}X_0^4)\cr
&&+i\epsilon^{\mu\nu\rho}\bar\chi_\mu^{AC}\chi_{\nu BC}\Big[\frac{vN}{\sqrt{2}}\delta^{B,4}\pd_\rho (X_0^A-iX_0^{A+4})
+\sqrt{2}v(X_a^B+iX_a^{B+4})A_\rho^{a-}\delta^{A,4}\cr
&&+\frac{N}{2}(X_0^B+iX_0^{B+4})\pd_\rho(X_0^A-iX_0^{A+4})+\frac{(X_a^{B+4}-iX_a^B)D_\rho(X_a^{A+4}+iX_a^A)}{2}\cr
&&- i N \delta^{A,4} A_{\mu 0}^-[2v^2 \delta_{B,4}+ \sqrt 2 (X_0^B + i X_0^{B+4})\Big]+c.c\cr
&&-\frac{i}{4}g_M^2e \Big[N(\bar \psi_{0}^D-i\bar \psi_{0}^{ D+4})(\psi_0^B+i\psi_0^{B+4})+(i\bar \psi_{a}^{D}+\bar \psi_{a}^{ D+4})(-i\psi_a^B+\psi_a^{B+4})\Big]\times\cr
&&\times \Big[v^2N\delta_{B,4}\delta^{D,4}+\frac{vN}{\sqrt{2}}((X_0^B-iX_0^{B+4})\delta^{D,4}+\delta_{B,4}(X_0^D+iX_0^{D+4}))\Big]\cr
&&-\frac{i}{4}g_M^2\epsilon^{\mu\nu\rho}(\bar \chi_{\nu AB}\gamma_\rho \chi_\mu^{CD})
\Big[v^2N\delta_{C,4}\delta^{A,4}+\frac{vN}{\sqrt{2}}((X_0^C-iX_0^{C+4})\delta^{A,4}+\delta_{C,4}(X_0^A+iX_0^{A+4}))\Big]\times\cr
&&\times\Big[v^2N\delta_{D,4}\delta^{B,4}+\frac{vN}{\sqrt{2}}((X_0^D-iX_0^{D+4})\delta^{B,4}+\delta_{D,4}(X_0^B+iX_0^{B+4}))\Big]\cr
&&+iAe\bar\chi_\mu^{BA}\gamma^\mu\gamma^\nu\Big\{\frac{N}{\sqrt{2}}
(\psi_0^{[A}+i\psi_0^{[A+4})D_\mu (X_0^{B]}-iX_0^{B]+4})\cr
&&+(i\psi_a^{[A}-\psi_a^{[A+4})\Big[\frac{D_\mu(-iX_a^{B]}-X_a^{B]+4})}{\sqrt{2}}-2ivA_{\mu a}^- \delta^{B,4}\Big]\Big\}+c.c.\cr
&&+\frac{A^2}{2}e\bar\chi_\mu ^{BA}\gamma^\nu \gamma^\mu(\bar \chi_{\nu BC})^\dagger \Big[
N(\psi_{0}^{A}+i\psi_{0}^{A+4})(\psi_0^{ C}-i\psi_0^{ C+4})\cr
&&\qquad\qquad\qquad\qquad\qquad\qquad+(i\psi_{a}^{A}-\psi_{a}^{A+4})(-i\psi_a^{ C}
-\psi_a^{C+4})\Big]+c.c.\cr
&&-iA\bar f^{\mu AB}\gamma_\mu 
\Big[vN\delta_{B4} (\psi_0^A+i\psi_0^{A+4})+\frac{N}{\sqrt{2}}(\psi_0^A+i\psi_0^{A+4})(X_0^B-iX_0^{B+4})\cr
&&
+\frac{1}{\sqrt{2}}(i\psi_a^A-\psi_a^{A+4})(-iX_a^B-X_a^{B+4})\Big]+c.c.\cr
&&-\frac{i}{16}g_M^2eA\bar\chi_{\mu AB}\gamma^\mu \Big[vN\delta^{4A}(\psi_0^ B-i\psi_0^{B+4})+
\frac{N}{\sqrt{2}}(\psi_0^B-i\psi_0^{B+4})(X_0^A+iX_0^{A+4})\cr
&&+\frac{1}{\sqrt{2}}
(-i\psi_a^B-\psi_a^{B+4})(iX_a^A-X_a^{A+4})\Big] (v^2N+vN\sqrt{2}X_0^4)+c.c.\cr
&&-\frac{ie}{4}g_M^2A\bar\chi_{\mu AB}\gamma^\mu \Big[vN\delta^{B4}(\psi_0^D-i\psi_0^{D+4})+
\frac{N}{\sqrt{2}}(\psi_0^D-i\psi_0^{D+4})(X_0^B+iX_0^{B+4})\cr
&&+\frac{1}{\sqrt{2}}
(-i\psi_a^D-\psi_a^{D+4})(iX_a^B-X_a^{B+4})\Big] \Big[v^2N\delta_{D,4}\delta^{A,4}+ \frac{vN}{\sqrt{2}}((X_0^D-iX_0^{D+4})\delta^{A,4}\cr
&&\qquad\qquad\qquad\qquad\qquad\qquad\qquad\qquad+\delta_{D,4}(X_0^A+iX_0^{A+4}))\Big]+c.c.\cr
&&-\frac{i}{4}g_M^2\epsilon^{\mu\nu\rho}(\bar \chi_{\nu AB}\gamma_\rho \chi_\mu^{CD})
\Big[v^2N\delta_{C,4}\delta^{A,4}+\frac{vN}{\sqrt{2}} ((X_0^C-iX_0^{C+4})\delta^{A,4}\cr
&&\qquad\qquad\qquad\qquad\qquad\qquad\qquad\qquad+\delta_{C,4}(X_0^A+iX_0^{A+4}))\Big]\times\cr
&&\times \Big[v^2N\delta_{D,4}\delta^{B,4}+\frac{vN}{\sqrt{2}}((X_0^D-iX_0^{D+4})\delta^{B,4}+\delta_{D,4}(X_0^B+iX_0^{B+4}))\Big]\cr
&&+\lambda \frac{v}{\sqrt{2}}\epsilon^{\mu\nu\rho}(\bar\chi_{\mu AB}\gamma_\nu\chi_\rho^{CD})
f^{abc}\Big[(iX_a^{[C}+X_a^{[C+4})(iX_b^{D]}+X_b^{D]+4})(iX_c^{[A}-X_c^{[A+4})\delta^{B],4}\cr
&&+(iX_a^{[A}-X_a^{[A+4})(iX_b^{B]}-X_b^{B]+4})(iX_c^{[C}+X_c^{[C+4})\delta^{D],4}\Big]\cr
&&+\lambda eA\bar\chi_{\mu AB}\gamma^\mu f^{abc}
\Big[v(i\psi_c^B-\psi_c^{B+4})X_a^8(iX_b^A-X_b^{A+4})-iv\delta^{A,4}X_a^{D+4}X_b^D(i\psi_c^A-\psi_c^{A+4})
\Big]\cr
&&-\lambda eA\bar\chi_{\mu AB}\gamma^\mu f^{abc}\Big[\frac{1}{2}v(i\psi_c^4-\psi_c^8)(iX_a^{[A}-X_a^{[A+4})(iX_b^{B]}-X_b^{B]+4})\cr
&&+v(i\psi_c^D-\psi_c^{D+4})\delta^{[A,4}(-iX_{a,D}-X_{a,D+4})(iX_b^{B]}-X_b^{B]+4})\Big]+{\rm subleading}\label{fermionterms}\;.
\eea

Restricting to the terms of the same order as the bosonic ${\cal O}(v^2)$ terms, and also to the $\SU(N)$ component, we obtain
\bea
&&\frac{i}{4}f^\mu_{AB}\chi_\mu^{AB}X_a^IX_a^I
+\frac{iv}{\sqrt{2}}\lambda\epsilon^{\mu\nu\rho}(\bar \chi_{\mu AB}\gamma_\nu \chi_\rho^{AB})
f^{abc}X_a^{C'+4}X_b^{C'}X_c^8\cr
&&+i\epsilon^{\mu\nu\rho}\bar\chi_\mu^{AC}\chi_{\nu BC}\Big[
\sqrt{2}v(X_a^B+iX_a^{B+4})A_\rho^{a-}\delta^{A,4}
+\frac{1}{2}(X_a^{B+4}-iX_a^B)D_\rho(X_a^{A+4}+iX_a^A)\Big]+c.c\cr
&&+iAe\bar\chi_\mu^{BA}\gamma^\mu\gamma^\nu
(i\psi_a^{[A}-\psi_a^{[A+4})\Big[\frac{1}{\sqrt 2}D_\mu(-iX_a^{B]}-X_a^{B]+4})-2ivA_{\mu a}^- \delta^{B,4}\Big]+c.c.\cr
&&+\frac{A^2}{2}e\bar\chi_\mu ^{BA}\gamma^\nu \gamma^\mu(\bar \chi_{\nu BC})^\dagger (i\psi_{a}^{A}-\psi_{a}^{A+4})(-i\psi_a^{ C}
-\psi_a^{C+4})+c.c.\cr
&&-iA\bar f^{\mu AB}\gamma_\mu
\frac{1}{\sqrt{2}}(i\psi_a^A-\psi_a^{A+4})(-iX_a^B-X_a^{B+4})+c.c.\cr
&&+\lambda \frac{v}{\sqrt{2}}\epsilon^{\mu\nu\rho}(\bar\chi_{\mu AB}\gamma_\nu\chi_\rho^{CD})
f^{abc}\Big[(iX_a^{[C}+X_a^{[C+4})(iX_b^{D]}+X_b^{D]+4})(iX_c^{[A}-X_c^{[A+4})\delta^{B],4}\cr
&&+(iX_a^{[A}-X_a^{[A+4})(iX_b^{B]}-X_b^{B]+4})(iX_c^{[C}+X_c^{[C+4})\delta^{D],4}\Big]\cr
&&+\lambda veA\bar\chi_{\mu AB}\gamma^\mu f^{abc}
\Big[(i\psi_c^B-\psi_c^{B+4})X_a^8(iX_b^A-X_b^{A+4})-i\delta^{A,4}X_a^{D+4}X_b^D(i\psi_c^A-\psi_c^{A+4})
\Big]\cr
&&-\lambda veA\bar\chi_{\mu AB}\gamma^\mu f^{abc}\Big[\frac{1}{2}(i\psi_c^4-\psi_c^8)(iX_a^{[A}-X_a^{[A+4})(iX_b^{B]}-X_b^{B]+4})\cr
&&+(i\psi_c^D-\psi_c^{D+4})\delta^{[A,4}(-iX_{a,D}-X_{a,D+4})(iX_b^{B]}-X_b^{B]+4})\Big]+{\rm subleading}\;.
\eea
If we further restrict only to the $I''$ indices, that is we again  exclude contributions from the directions  $4$ and $8$, we get
\bea
&&\frac{i}{4}f^\mu_{A'B'}\chi_\mu^{A'B'}X_a^{I''}X_a^{I''}
+i\epsilon^{\mu\nu\rho}\bar\chi_\mu^{A'C'}\chi_{\nu B'C'}
\frac{1}{2}(X_a^{B'+4}-iX_a^{B'})D_\rho(X_a^{A'+4}+iX_a^{A'})+c.c\cr
&&+iAe\bar\chi_\mu^{B'A'}\gamma^\mu\gamma^\nu
(i\psi_a^{[A'}-\psi_a^{[A'+4})\Big[\frac{1}{\sqrt 2}D_\mu(-iX_a^{B']}-X_a^{B']+4})\Big]+c.c.\cr
&&+\frac{A^2}{2}e\bar\chi_\mu ^{B'A'}\gamma^\nu \gamma^\mu(\bar \chi_{\nu B'C'})^\dagger (i\psi_{a}^{A'}-\psi_{a}^{A'+4})(-i\psi_a^{ C'}
-\psi_a^{C'+4})+c.c.\cr
&&-iA\bar f^{\mu A'B'}\gamma_\mu
\frac{1}{\sqrt{2}}(i\psi_a^{A'}-\psi_a^{A'+4})(-iX_a^{B'}-X_a^{B'+4})+c.c.+{\rm subleading}\;.
\eea

\end{appendix}

\bibliographystyle{JHEP}
\bibliography{abjmgrav}

\end{document}